\documentclass[a4paper,12pt]{article}
\pdfoutput=1
\usepackage{graphicx}
\usepackage{mathtools}
\usepackage{amssymb}
\usepackage{amsfonts}
\usepackage[footnotesize]{caption}
\usepackage[font=scriptsize]{subcaption}
\usepackage{color}
\usepackage{braket}
\usepackage{cite}
\usepackage{hyperref}
\usepackage{url}
\usepackage{multirow}
\usepackage{relsize}
\usepackage{fullpage}
\usepackage{makecell}
\usepackage{blkarray}
\usepackage{fullpage}

\newcommand{\newc}{\newcommand}
\newc{\be}{\begin{equation}}
\newc{\ee}{\end{equation}}
\newc{\beq}{\begin{equation}}
\newc{\eeq}{\end{equation}}
\newc{\bea}{\begin{eqnarray}}
\newc{\eea}{\end{eqnarray}}
\newc{\simlt}{~\mbox{\smaller\(\lesssim\)}~}
\newc{\simgt}{~\mbox{\smaller\(\gtrsim\)}~}

\newcommand{\pmatr}[1]{\begin{pmatrix} #1 \end{pmatrix}}

\def\vev#1{\left\langle #1\right\rangle}

\begin{document}

\begin{titlepage}
\begin{center}
{\bf\Large
\boldmath{
Twin Pati-Salam theory of flavour \\ with a TeV scale vector leptoquark
}
} \\[12mm]
Stephen~F.~King$^{\star}$%
\footnote{E-mail: \texttt{king@soton.ac.uk}}
\\[-2mm]
\end{center}
\vspace*{0.50cm}
\centerline{$^{\star}$ \it
Department of Physics and Astronomy, University of Southampton,}
\centerline{\it
SO17 1BJ Southampton, United Kingdom }
\vspace*{1.20cm}

\begin{abstract}
{\noindent
We propose a twin Pati-Salam (PS) theory of flavour broken to the $G_{4321}$ gauge group at high energies, then to the Standard Model at low energies,
yielding a TeV scale vector leptoquark $U^{\mu}_1(3,1,2/3)$ which has been suggested to address the 
lepton universality anomalies $R_{K^{(*)}}$ and $R_{D^{(*)}}$ in $B$ decays.
Quark and lepton masses are mediated by vector-like fermions, with personal Higgs doublets for the second and third families,
which may be replaced by a two Higgs doublet model (2HDM). The twin PS theory of flavour successfully accounts for all quark and lepton (including neutrino) masses and mixings, and predicts a dominant coupling of 
$U^{\mu}_1(3,1,2/3)$ to the third family left-handed doublets.
However the predicted mass matrices, assuming natural values of the parameters,
are not consistent with the single vector leptoquark
solution to the $R_{D^{(*)}}$ anomaly, given its current value.
}
\end{abstract}
\end{titlepage}

\section{Introduction}
The Standard Model (SM) despite its many successes leaves the flavour puzzle unanswered.
The low energy quark and lepton masses may be expressed approximately as~\cite{Zyla:2020zbs}
\begin{align}
\label{tcu}
m_t &\sim v,\ \ \ \ \ \  
m_c \sim \lambda^{3.3} v,\ \ 
m_u\sim \lambda^{7.5} v \\
\label{bsd}
 m_b &\sim \lambda^{2.5} v,\ \ 
m_s \sim \lambda^{5.0} v, \ \ 
m_d\sim \lambda^{7.0} v\\
\label{taumue}
m_{\tau} &\sim \lambda^{3.0} v, \ \ 
m_{\mu}\sim \lambda^{4.9} v, \ \ 
m_e\sim \lambda^{8.4} v, \\
\label{nu123}
m_{\nu_3} &\sim \lambda^{19.1} v, \ \ 
m_{\nu_2}\sim \lambda^{20.3} v, \ \ 
m_{\nu_1} \ll m_{\nu_2}
\end{align}
with $v=v_{SM}/\sqrt{2}$ and $v_{SM}=246$ GeV, where 
we have assumed the neutrino masses 
to be hierarchical and in a normal ordered mass pattern as preferred by recent data,
while $\lambda=0.22$ is the Wolfenstein parameter which parametrises the CKM matrix 
as~\cite{Wolfenstein:1983yz},
\be
V_{us}=\lambda, \ \ V_{cb}\sim \lambda^2, \ \ V_{ub}\sim \lambda^3, 
\label{CKM}
\ee
while the PMNS lepton mixing angles satisfy the approximate relations \cite{King:2012vj},
\be
\tan \theta_{23}\sim 1, \ \ \tan \theta_{12}\sim \frac{1}{\sqrt{2}}, \ \ \theta_{13} \sim \frac{\lambda}{\sqrt{2}}. 
\label{PMNS}
\ee
The above pattern of masses and mixing angles is a complete mystery in the SM, and the origin of the tiny neutrino masses
and large lepton mixing angles requires new physics beyond the Standard Model (BSM).
The flavour puzzle is not just the number of free parameters, it is
the lack of any dynamical understanding of their values, with Yukawa couplings expressed as powers of $\lambda$ above. 
Naively, we might have expected all Yukawa couplings
to be of order unity, like the gauge couplings, but empirically they are not.

The wealth of data of quark and lepton masses and mixing angles can provide some hints concerning possible BSM theories of flavour.
For example, from the above data 
we observe the empirical quark relation discussed by Gatto, Satori and Tonin (GST) \cite{Gatto:1968ss},
\be
V_{us}\sim \sqrt{\frac{m_d}{m_s}},
\label{Vus}
\ee
which hints at the CKM mixing originating from the down type quark mass matrix, with an approximate zero in the first element.
Such a ``texture zero'' was also suggested by Georgi and Jarlskog (GJ)~\cite{Georgi:1979df} to understand the relation between the down quark mass and the electron mass. It is also invoked in the sequential dominance (SD) ~\cite{King:1998jw,King:1999cm,King:1999mb,King:2002nf,King:2002qh}
mechanism for achieving natural hierarchical neutrino masses and mixings arising from the type I seesaw mechanism~\cite{Minkowski:1977sc,Mohapatra:1979ia,Yanagida:1979as,GellMann:1980vs}. 
It seems as though the texture zero is well motivated on phenomenological grounds from the quark, charged lepton and neutrino sectors, and this suggests that the first family is distinguished by some quantum number~\footnote{For example the Froggatt-Nielsen (FN) mechanism~\cite{Froggatt:1978nt}, where 
$U(1)_{FN}$ symmetry, broken by the vacuum expectation value (VEV) of a ``flavon'', distinguishes the families.
Alternatively, modular weights of fermion fields can 
play the role of FN charges, and SM singlet fields with non-zero modular weight called 
``weightons'' can play the role of flavons~\cite{King:2020qaj}. While a simple $Z_2$ symmetry is sufficient to distinguish the first family,
here we shall use a $Z_6$ symmetry which leads to the correct right-handed neutrino hierarchy.}.

Recently new evidence for the experimental anomaly in 
the semi-leptonic $B$ decay ratio $R_{K^{(*)}}$, which violates $\mu - e$ universality in $b\rightarrow s$ decays,
has been presented \cite{Aaij:2021vac}. 
Also the semi-leptonic $B$ decay ratio $R_{D^{(*)}}$ violates 
$\tau$ universality in $b\rightarrow c$ decays. These anomalies motivate new theories of flavour involving leptoquarks,
for example the single vector leptoquark $U^{\mu}_1(3,1,2/3)$ has been shown to address all the B physics anomalies 
\cite{Alonso:2015sja,Calibbi:2015kma,Barbieri:2015yvd,Sahoo:2016pet,Buttazzo:2017ixm,Assad:2017iib,Barbieri:2017tuq,Kumar:2018kmr,Crivellin:2018gyy,Cornella:2019hct,Crivellin:2019szf,Dev:2020qet,Fuentes-Martin:2019ign,Fuentes-Martin:2020luw,Fuentes-Martin:2020hvc,Bhaskar:2021pml,Iguro:2021kdw,Angelescu:2021lln,Cornella:2021sby,Hiller:2021pul} 
with contributions to
the muon $a_{\mu}=(g-2)_{\mu}/2$\cite{Biggio:2016wyy}, while the scalar leptoquarks $S_2(3,{2},7/6)$, $\tilde{S}_2(3,{2},1/6)$, and 
$S_3(\overline{3},3,1/3)$ could also play a role for $R_{K^{(*)}}$ \cite{Hiller:2017bzc}.

Although a vector leptoquark is predicted by Pati-Salam theory  (PS) \cite{Pati:1974yy}, its mass is generally expected to lie above the PeV scale, too heavy to explain the anomalies.
Nevertheless, such a vector leptoquark could arise from a low energy PS gauge group 
\footnote{A low energy PS gauge group has also been considered from a different perspective \cite{Perez:2013osa}.}
as discussed in several works~\cite{DiLuzio:2017vat,Fornal:2018dqn,Baker:2019sli,Calibbi:2017qbu,Bordone:2017bld,Heeck:2018ntp,Matsuzaki:2018jui,Blanke:2018sro}.
However, the ultraviolet completion of such theories remains challenging, and motivates further model building in this direction, in particular models which can simultaneously explain the origin of quark and lepton masses. In this way, the recent anomalies can provide additional 
experimental hints which  can help to shed light on the path towards finding the correct BSM theory of flavour.

\begin{figure}[ht]
\centering
	\includegraphics[scale=0.23]{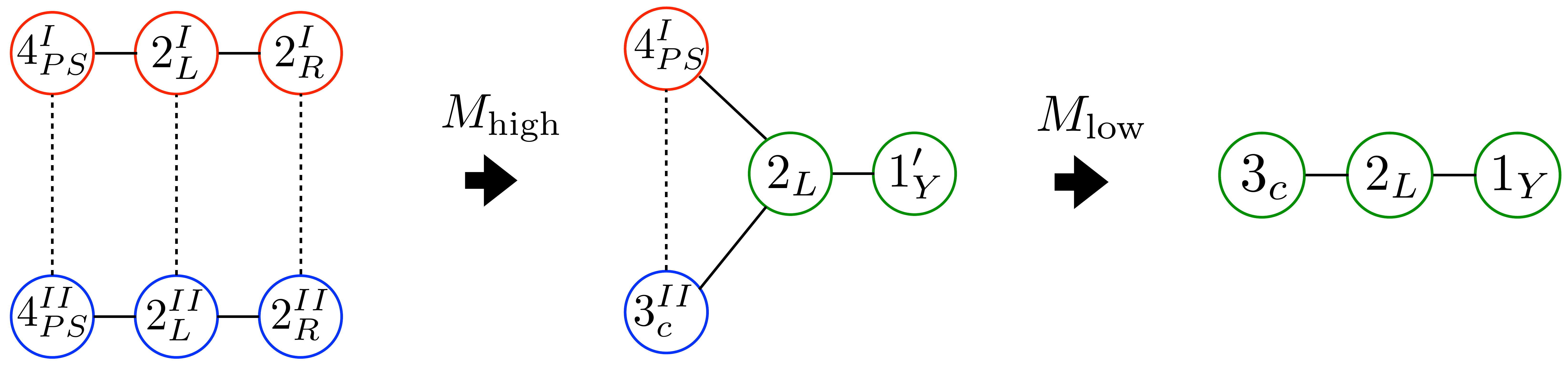}
\caption{The model is based on two copies of the PS gauge group $SU(4)_{PS}\times SU(2)_L\times SU(2)_R$. 
The circles represent the gauge groups with the indicated symmetry breaking as in Eq.\ref{breaking}. }
\label{model}
\end{figure}

In this paper we propose a twin PS theory of flavour capable of explaining some of the anomalies, for natural values of the parameters, as well as providing a theory quark and lepton (including neutrino) masses and mixings. 
At high energies, the theory involves two copies of the PS gauge group, $G_{422}$ \cite{Pati:1974yy},
with the usual three chiral fermion families 
transforming under $G_{422}^{II}$. A fourth vector-like (VL) family, which mediates the second and third family masses,
transforms under $G_{422}^{I}$. The twin PS gauge groups are broken in stages first to 
$G_{4321}$ then to the SM gauge group
$G_{321}$, as in Fig.~\ref{model},
\begin{align}
G_{422}^I \times G_{422}^{II} \stackrel{M_{\mathrm{high}}}{\longrightarrow} G_{4321}
\stackrel{M_{\mathrm{low}}}{\longrightarrow} G_{321}
\label{breaking}
\end{align}
The explanation of the anomalies involves the vector leptoquark $U^{\mu}_1(3,1,2/3)$ from the $SU(4)^{I}_{PS}$,
broken at $M_{\mathrm{low}} \sim 1$ TeV, while the origin of quark and lepton masses depends on the full theory, including the 
high scale PS symmetry, broken at $M_{\mathrm{high}} \gtrsim 1$ PeV, the latter limit being due to the non-observation of $K_L\rightarrow \mu e$
\cite{Valencia:1994cj},
although we later find it to be near the conventional scale of Grand Unified Theories (GUTs).
The first family fermion masses are mediated by a fifth family of VL fermions which transform under $SU(4)^{II}_{PS}$, and
neutrino masses are further suppressed by the type I~\cite{Minkowski:1977sc,Mohapatra:1979ia,Yanagida:1979as,GellMann:1980vs}
seesaw mechanism.
In order to achieve the texture zero in the first element of the mass matrices and hierarchical right-handed neutrino masses
we shall assume a $Z_6$ family symmetry, although we shall not use it to explain the charged fermion mass hierarchies.
Apart from the $Z_6$, no additional symmetries are introduced. 
The model involves ``personal'' Higgs doublets for the second and third family fermion masses, where the origin and nature of these fields is very different from the ``private'' Higgs doublets envisaged in~\cite{Porto:2007ed,Porto:2008hb,BenTov:2012cx,Rodejohann:2019izm}
(although in an Appendix
we show how the model can be recast as a conventional type II two Higgs doublet model (2HDM)).
The twin PS theory of flavour successfully accounts for all quark and lepton (including neutrino) masses and mixings, and predicts a dominant coupling of 
$U^{\mu}_1(3,1,2/3)$ to the third family left-handed doublets.
However the predicted mass matrices, assuming natural values of the parameters,
are not consistent with the single vector leptoquark
solution to the $R_{D^{(*)}}$ anomaly, given its current value.

The layout of the remainder of the paper is as follows.
In section~\ref{high} we define the high energy theory consisting of a 
twin PS gauge group, together with a $Z_6$ family symmetry, and discuss 
the effective operators which will be responsible for the quark and lepton masses and mixings.
In section~\ref{low} we discuss the low energy theory consisting of $G_{4321}$ resulting from the breaking of the 
twin PS theory, and show how the effective Yukawa operators decompose into separate mass matrix structures for quarks and leptons
controlled by personal Higgs fields.
We also discuss the breaking of $G_{4321}$ to the SM gauge group and the EW symmetry breaking via the personal Higgs doublets,
before investigating if some of the leptoquarks predicted by the model could help to explain $R_{D^{(*)}}$.
In section~\ref{masses} we summarise and discuss the predictions for the quark and lepton mass matrices, including the neutrino masses and mixings via the type I seesaw mechanism.
Finally in section~\ref{conclusion} we present our conclusions.
In Appendix~\ref{A} we describe a large mixing angle formalism which may be used to go beyond the mass insertion approximation.
In Appendix~\ref{2HDM} we show how the model may be recast as a 2HDM 
by removing the personal Higgs doublets, introducing additional fields instead.

\section{Twin Pati-Salam Theory of Flavour}
\label{high}
\subsection{The High Energy Model}
\label{2.1}
It is well known that quarks and leptons may be unified into the Pati-Salam (PS) gauge group \cite{Pati:1974yy},
\be
G_{422}=SU(4)_{PS}\times SU(2)_L\times SU(2)_R
\label{PS}
\ee
In traditional PS, the left-handed (LH) chiral quarks and leptons are unified into 
$SU(4)_{PS}$ multiplets with leptons as the fourth colour (red, blue, green, lepton),
\begin{equation}
{\psi_i}(4,2,1)=
\left(\begin{array}{cccc}
u_r & u_b & u_g & \nu \\ d_r & d_b & d_g & e^-
\end{array} \right)_i \equiv (Q_i,L_i)
\label{psi}
\end{equation}
\begin{equation}
\psi^c_j(\bar{4},1,\bar{2})=
\left(\begin{array}{cccc}
u^c_r & u^c_b & u^c_g & \nu^c \\ d^c_r & d^c_b & d^c_g & e^c
\end{array} \right)_j \equiv(u^c_j, d^c_j, \nu^c_j, e^c_j)
\label{psic}
\end{equation}
where $\psi^c_j$ are the CP conjugated RH quarks and leptons (so that they become LH) forming $SU(2)_R$ doublets
and 
$i,j=1\ldots 3$ are family indices. Three right-handed neutrinos (actually their CP conjugates $\nu^c_j$) are 
predicted as part of the gauge multiplets.

The proposed twin PS model in Table~\ref{twinPS} is based on two copies of the PS gauge group, together with a $Z_6$ family symmetry,
\be
G_{422}^I \times G_{422}^{II} \times Z_6
\label{twin}
\ee
which undergoes the breaking in Eq.\ref{breaking}, where $SU(4)^{I}$ is broken at the low scale.
In the proposed model, the usual three chiral fermion families originate from the second PS group $G_{422}^{II}$, broken at the high scale,
and transform under Eq.\ref{twin} as 
\be
\psi_{1,2,3}(1,1,1;4,2,1)_{(\alpha^2,1,1)}, \ \ \psi^c_{1,2,3}(1,1,1;\bar{4},1,\bar{2})_{(\alpha^2,\alpha^5,1)}
\ee
where powers of the $Z_6$ charge $\alpha=e^{i2\pi/6}$ distinguish the families, apart from $\psi_{2,3}$ being indistinguishable leading to large atmospheric mixing.
There are no standard Higgs fields under $G_{422}^{II}$, hence no standard Yukawa couplings involving the chiral fermions. These will be generated effectively via mixing with vector-like (VL) fermions.

We assume high energy Higgs fields which transforms under Eq.\ref{twin} as
\be
H'(1,1,1;4,1,2)_1, \ \ \overline{H}'(1,1,1;\bar{4},1,\bar{2})_1,
\ee
whose VEVs will break the second PS group at a high scale, leaving the first unbroken.
We also assume further Higgs fields $\Phi, \Phi',\overline{\Phi}$, detailed in Table~\ref{twinPS}, which break the two left-right gauge groups into their diagonal subgroup.

The theory also includes a VL fermion family which transforms under Eq.\ref{twin} as
\be
\psi_4(4,2,1;1,1,1)_1,\ \ \overline{\psi_4}(\bar{4},\bar{2},1;1,1,1)_1,  
\ \ \psi^c_4(\bar{4},1,\bar{2};1,1,1)_1,  \ \  \overline{\psi^c_4}({4},1,{2};1,1,1)_1
\ee
carrying quantum numbers under the first PS group,
$G_{422}^{I}$, whose $SU(4)^{I}$ is broken at the low scale.
The theory also involves the scalars $\phi,\overline{\phi},\overline{\phi'},H,\overline{H}$ in 
Table~\ref{twinPS}, with the couplings,
\begin{align}
{\cal L}^{ren}_4 = 
& y^{\psi}_{i4}\overline{H}  \psi_i {\psi^c_4} 
+  y^{\psi}_{43}{H} {\psi_4} \psi^c_3
+x^{\psi}_{i4}{\phi} \psi_i \overline{\psi_4} 
+ x^{\psi^c}_{43} \overline{\psi^c_4} \overline{\phi} \psi^c_3
+ x^{\psi^c}_{42} \overline{\psi^c_4} \overline{\phi'} \psi^c_2
+ M^{\psi}_{4}\psi_4 \overline{\psi_4}
+ M^{\psi^c}_{4}\psi^c_4 \overline{\psi^c_4} \label{L4}
\end{align}
where $i=2,3$ ($i=1$ term forbidden by $Z_6$),
$x,y$ are dimensionless coupling constants and $M_4$ are the VL masses.
These couplings mix the chiral fermions with the VL fermions, and 
will be responsible for generating effective Yukawa couplings of the second and third families.
Since the VL fermions will mix only with the second and third chiral families, they lead to effective couplings to TeV scale
$SU(4)^{I}$ gauge bosons which violate lepton universality.

The theory also includes a fifth VL fermion family $\psi_5,\overline{\psi_5},\psi^c_5,\overline{\psi^c_5}$ split across both PS groups,
and a non-standard Higgs field $h$, as shown in Table~\ref{twinPS},
which couple as,
\begin{align}
{\cal L}^{ren}_{5} = 
& y^{\psi}_{15}{h}  \psi_1 {\psi^c_5} 
+  y^{\psi}_{51}{h} {\psi_5} \psi^c_1+x^{\psi}_{i5}{\Phi} \psi_i \overline{\psi_5} 
+ x^{\psi^c}_{53} \overline{\psi^c_5} {\Phi} \psi^c_3
+ x^{\psi^c}_{52} \overline{\psi^c_5} {\Phi'} \psi^c_2 
+ M^{\psi}_{5}\psi_5 \overline{\psi_5}
+ M^{\psi^c}_{5}\psi^c_5 \overline{\psi^c_5} \label{L5}
\end{align}
where $i=2,3$ ($i=1$ being forbidden by $Z_6$), $x,y$ are dimensionless coupling constants and $M_5$ are the VL masses.
These VL fermions do not couple to the TeV scale
$SU(4)^{I}$ gauge bosons, however they are responsible for effective first family Yukawa couplings.
There are no renormalisable couplings involving a mixture of fourth and fifth VL fermions
to any Higgs fields.

\begin{table}
\centering
\begin{tabular}{| l | c  c c | c c c| c|}
\hline
Field & $SU(4)^I_{PS}$ & $SU(2)^{I}_L$ & $SU(2)^{I}_R$ &  $SU(4)^{II}_{PS}$ & $SU(2)^{II}_L$ & $SU(2)^{II}_R$ & $Z_6$\\ 
\hline \hline
$\psi_{1,2,3}$ 		 & ${\bf 1}$  & ${\bf 1}$  & ${\bf 1}$ & ${\bf 4}$ & ${\bf 2}$ & ${\bf 1}$ & $\alpha^2,1,1$ \\
$\psi^c_{1,2,3}$ 	 & ${\bf 1}$  & ${\bf 1}$ & ${\bf 1}$ & ${\overline{\bf 4}}$ & ${\bf 1}$ & ${\overline{\bf 2}}$ & $\alpha^2,\alpha^5,1$\\
\hline
${H'}$   & ${\bf 1}$ & ${\bf 1}$ & ${\bf 1}$  &${\bf 4}$ &   ${\bf 1}$ & ${\bf 2}$ & $1$ \\
$\overline{H}'$   & ${\bf 1}$ & ${\bf 1}$ & ${\bf 1}$  &  ${\overline{\bf 4}}$   &  ${\bf 1}$ & ${\overline{\bf 2}}$  & $1$\\
\hline
$\Phi$, $\Phi'$ & ${\bf 1}$  &${\bf 2}$ &   ${\bf 1}$  & ${\bf 1}$ &  ${\overline{\bf 2}}$ & ${\bf 1}$  & $1,\alpha$\\
$\overline{\Phi}$   & ${\bf 1}$  &  ${\bf 1}$ & ${\overline{\bf 2}}$ & ${\bf 1}$  &  ${\bf 1}$ &${\bf 2}$  & $1$\\
\hline
 $h$   &  ${\bf 1}$ & ${\overline{\bf 2}}$ &  ${\bf 1}$ &  ${\bf 1}$  &  ${\bf 1}$ &  ${\bf 2}$   & $\alpha^2$    \\
\hline
 $\xi$   &  ${\bf 1}$ & ${\bf 1}$ &  ${\bf 1}$ &  ${\bf 1}$  &  ${\bf 1}$ &  ${\bf 1}$   & $\alpha$    \\
\hline
$\psi_{4}$ 		& ${\bf 4}$ & ${\bf 2}$ & ${\bf 1}$ & ${\bf 1}$ & ${\bf 1}$ & ${\bf 1}$ & $1$ \\
$\overline{\psi_{4}}$ 		& ${\overline{\bf 4}}$   & ${\overline{\bf 2}}$ & ${\bf 1}$ & ${\bf 1}$ & ${\bf 1}$ & ${\bf 1}$ & $1$\\
$\psi^c_{4}$ 		 & ${\overline{\bf 4}}$ & ${\bf 1}$  & ${\overline{\bf 2}}$ & ${\bf 1}$ & ${\bf 1}$ & ${\bf 1}$& $1$\\
$\overline{\psi^c_{4}}$ 		& ${\bf 4}$  & ${\bf 1}$  & ${\bf 2}$ & ${\bf 1}$ & ${\bf 1}$ & ${\bf 1}$ & $1$\\
\hline
$\psi_{5}$ 		& ${\bf 1}$  & ${\bf 2}$  & ${\bf 1}$ & ${\bf 4}$ & ${\bf 1}$ & ${\bf 1}$ & $1$\\
$\overline{\psi_{5}}$ 		& ${\bf 1}$ & ${\overline{\bf 2}}$ & ${\bf 1}$ & ${\overline{\bf 4}}$  & ${\bf 1}$   & ${\bf 1}$ & $1$ \\
$\psi^c_{5}$ 		& ${\bf 1}$  & ${\bf 2}$ & ${\bf 1}$ & ${\overline{\bf 4}}$  & ${\overline{\bf 2}}$ &${\overline{\bf 2}}$ & $1$\\
$\overline{\psi^c_{5}}$ & ${\bf 1}$ & ${\overline{\bf 2}}$ & ${\bf 1}$ & ${\bf 4}$  & ${\bf 2}$   & ${\bf 2}$  & $1$ \\
\hline
$\phi$  &   ${\bf 4}$  & ${\bf 2}$ &  ${\bf 1}$ & ${\overline{\bf 4}}$ & ${\overline{\bf 2}}$ &  ${\bf 1}$  & $1$ \\
$\overline{\phi},\overline{\phi'}$ &   ${\overline{\bf 4}}$ & ${\bf 1}$ & ${\overline{\bf 2}}$ & ${\bf 4}$ &  ${\bf 1}$ & ${\bf 2}$  & $1,\alpha$\\
\hline
\hline
 $H$   & ${\overline{\bf 4}}$ & ${\overline{\bf 2}}$ &  ${\bf 1}$ &  ${\bf 4}$   &  ${\bf 1}$  &  ${\bf 2}$ & $1$ \\
$\overline{H}$   & ${\bf 4}$ &  ${\bf 1}$  &  ${\bf 2}$  & ${\overline{\bf 4}}$  &  ${\overline{\bf 2}}$ & ${\bf 1}$ & $1$ \\
\hline
\hline
\end{tabular}
\caption{The twin PS theory based on 
$G_{422}^I \times G_{422}^{II}$.  The model consists of 
three left-handed chiral fermion families $\psi_{1,2,3},\psi^c_{1,2,3}$ under the second PS group,
plus a VL fourth and fifth fermion family $\psi_{4,5},\psi^c_{4,5}$ and their conjugates. 
The symmetry is broken by the scalars $H',\Phi$, etc.
Two Higgs doublets are contained in $h$.
Additional personal Higgs doublets are contained in $H,\overline{H}$.
A $Z_6$ family symmetry is broken by the Majoron scalar $\xi$.}
\label{twinPS}
\end{table}

\subsection{Effective Yukawa operators}
\label{2.2}

We have already remarked that the usual Yukawa couplings involving purely chiral fermions 
are absent. In this subsection we show how they may be generated effectively once the vector-like fermions are integrated out.

It is instructive to first consider only the fourth VL family, then later consider the fifth one, assuming it to be much heavier than the fourth.
In this case we may write the masses and couplings in Eq.\ref{L4} as a $5\times 5$ matrix in flavour space
\be
	M^{\psi} = \pmatr{
	&\psi^c_1&\psi^c_2&\psi^c_3& \overline{\psi_4} &\psi^c_4\\ 
	\hline
	\psi_1|&0&0&0&0 &0\\
	\psi_2|&0&0&0&0&y^{\psi}_{24}\overline{H} \\
	\psi_3|&0&0&0&x^{\psi}_{34} {\phi}&y^{\psi}_{34}\overline{H} \\ 
	\psi_4|&0& 0 & y^{\psi}_{43} {H}&M^{\psi}_{4}&0\\ 
	\overline{\psi^c_4}|&0&x^{\psi^c}_{42}\overline{\phi'}&x^{\psi^c}_{43}\overline{\phi}&0& M^{\psi^c}_{4}}.	
	\label{M^psi_an_1}
\ee
where the extra zeroes are achieved by $(\psi_2,\psi_3)$ rotations, where such rotations leave the upper $3\times 3$ block of zeroes unchanged, so 
the form of Eq.\ref{M^psi_an_1} is just a choice of basis
\footnote{Note that even without the $Z_6$ symmetry the form of Eq.\ref{M^psi_an_1} 
could be achieved by $3\times 3$ rotations which preserve the zeroes in the upper block. 
Note that a simple $Z_2$ symmetry is sufficient to achieve the texture zero in first entry of the effective mass matrices once the fifth VL family is introduced. However the choice of $Z_6$ symmetry is to enforce the correct hierarchy of right-handed neutrino masses, which would not be possible for $Z_N$ with $N<6$.}.

\begin{figure}[ht]
\centering
	\includegraphics[scale=0.11]{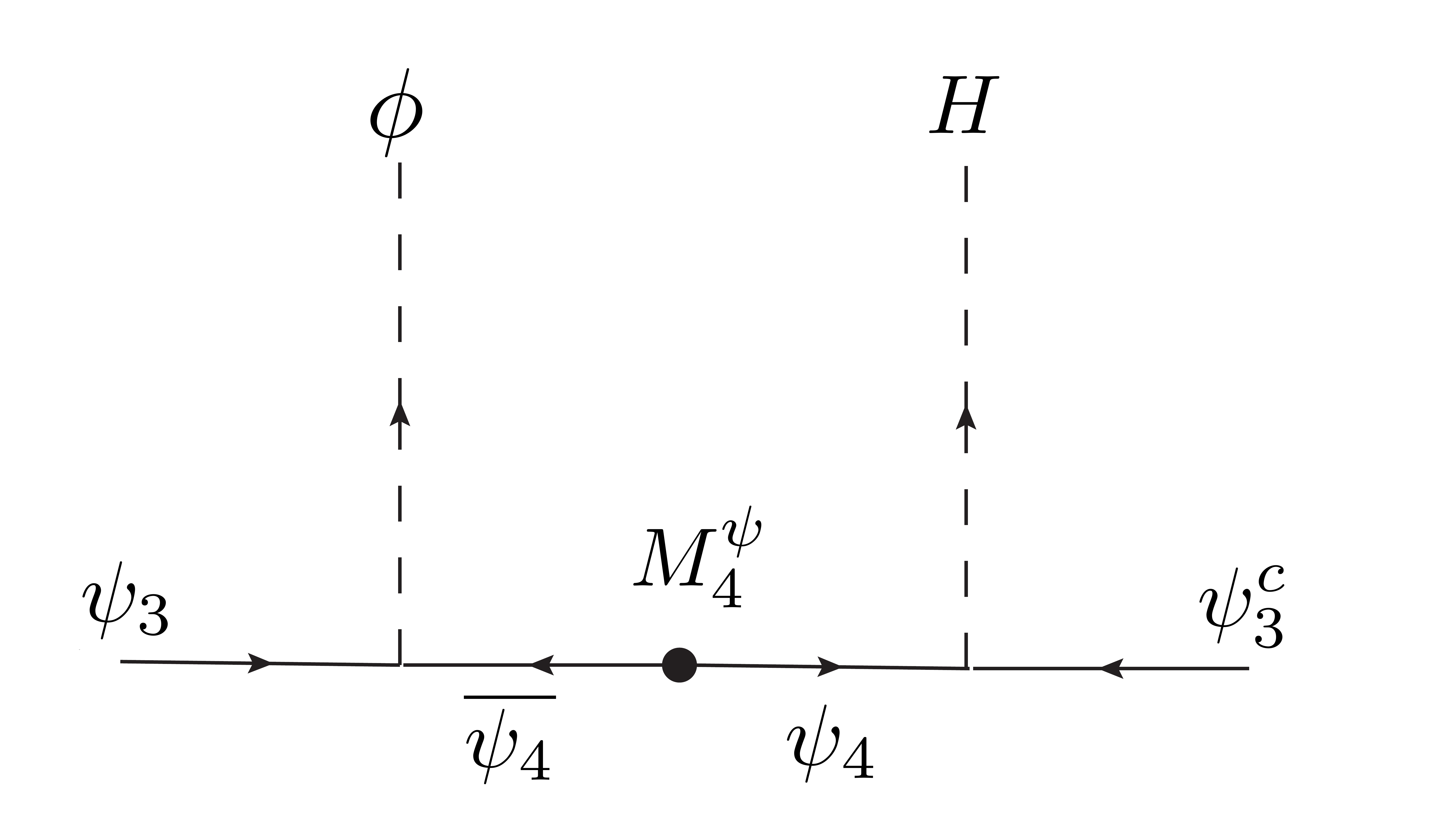}
\hspace*{1ex}
	\includegraphics[scale=0.11]{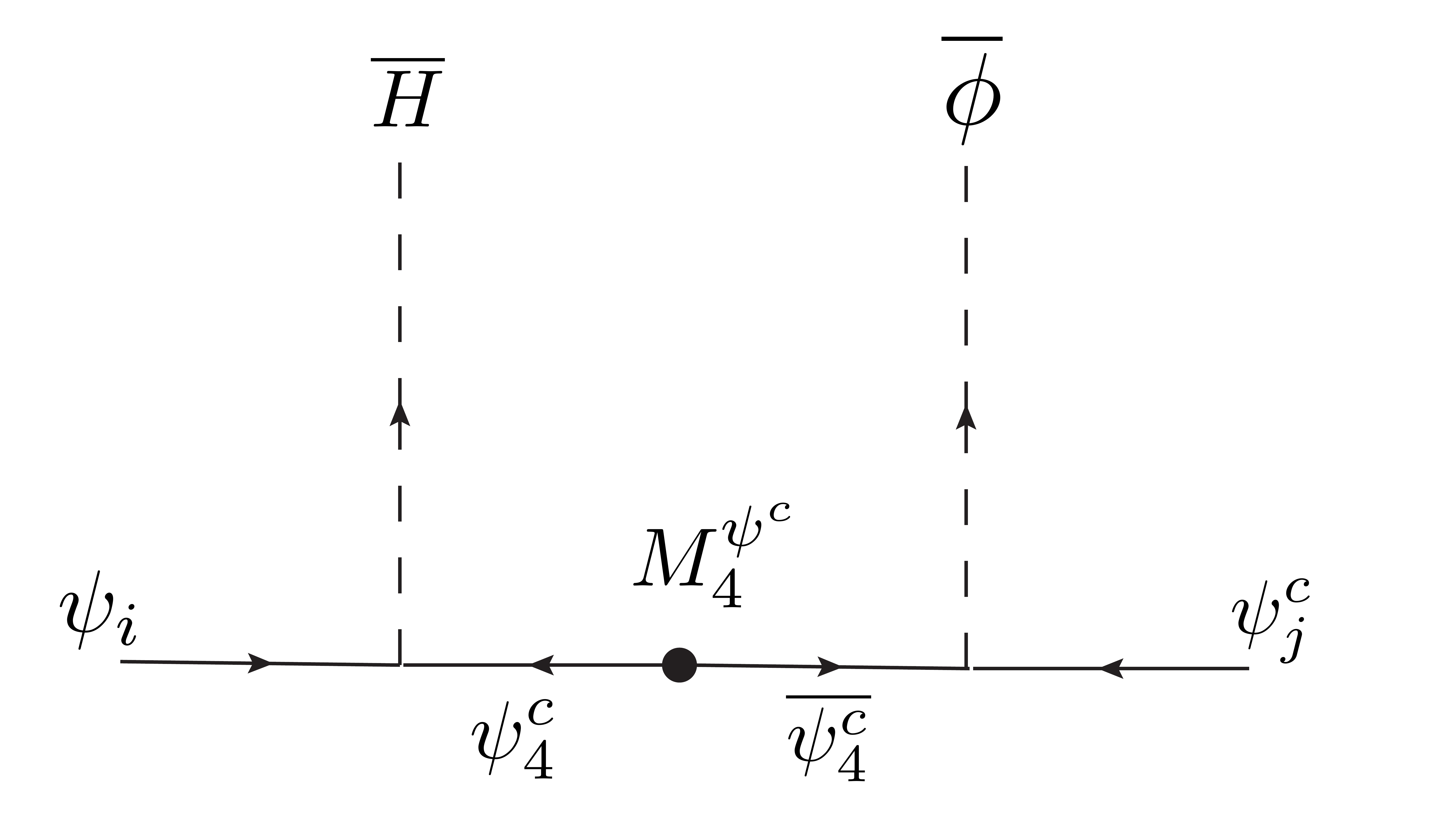}
\caption{Diagrams which lead to the effective Yukawa couplings of the third family (left panel) and second family (right panel) where $i,j=2,3$ are the only non-zero values. In the right panel, we have dropped the distinction between $\overline{\phi}$ and  $\overline{\phi'}$ for simplicity.}
\label{Fig1}
\end{figure}

There are several distinct mass scales in this matrix: the Higgs VEVs $\langle H \rangle$, 
$\langle \overline{H} \rangle$, the Yukon VEVs $\langle \phi \rangle$, $\langle  \overline{\phi} \rangle$
and the VL fourth family masses $M^{\psi}_{4}$, $M^{\psi^c}_{4}$. 
Assuming the latter are heavier than all the VEVs, we may integrate out the fourth family, to generate 
effective Yukawa couplings of the quarks and leptons which originate from the diagrams in 
Fig.~\ref{Fig1}. 

The two diagrams in Fig.\ref{Fig1} lead to effective Yukawa operators (up to an irrelevant minus sign),
after integrating out VL fermions,
\be
{\cal L}^{Yuk}_{4eff}= \frac{x^{\psi}_{34} {\phi}  y^{\psi}_{43} {H} }{M^{\psi}_{4}}  \psi_3 {\psi^c_3}
+
\frac{ y^{\psi}_{i4} \overline{H}  x^{\psi^c}_{42}  \overline{\phi'}  }{M^{\psi^c}_{4}} \psi_i {\psi^c_2} 
+
\frac{ y^{\psi}_{i4} \overline{H}  x^{\psi^c}_{43}  \overline{\phi}  }{M^{\psi^c}_{4}} \psi_i {\psi^c_3}+H.c.
\label{Yuk_mass_insertion4}
\ee
where $i=2,3$.
After Pati-Salam breaking, 
these terms will lead to Yukawa matrices for each of the four charged sectors 
$\psi = u,d,e,\nu$, as we discuss later. 

In the case of neutrinos, Eq.\ref{Yuk_mass_insertion4} leads to the Dirac Yukawa matrix.
There will be a further Majorana
mass matrix for the singlet neutrinos $M^{\nu^c}_{ij}\nu^c_i\nu^c_j$, arising from a symmetric matrix of operators
involving traditional PS fields in the first sector of Table~\ref{twinPS}, in the basis $(\psi^c_1,\psi^c_2,\psi^c_3)$,
\be
\left(
\begin{array}{ccc}
 \tilde{\xi}^2 & \tilde{\xi}^5 & \tilde{\xi}^4 \\
 \tilde{\xi}^5 & \tilde{\xi}^2 &  \tilde{\xi} \\ 
 \tilde{\xi}^4 &  \tilde{\xi} & 1
\end{array}
\right)
\frac{{H}'{H}'}{\Lambda}
\label{M}
\ee
where we have written $ \tilde{\xi}=\xi/\Lambda$, and
dropped the independent dimensionless coefficients $y^{\psi^c}_{ij}$ which multiply each entry of the matrix and will play a role in breaking the degeneracy of the lightest two right-handed neutrinos.
After the $H'$ scalars get their VEVs, the terms in Eq.\ref{M} result in Majorana masses for the right-handed neutrinos, leading to
small physical neutrino masses from the type I seesaw 
mechanism~\cite{Minkowski:1977sc,Mohapatra:1979ia,Yanagida:1979as,GellMann:1980vs}.

The reason we have gone to the basis in Eq.\ref{M^psi_an_1}, 
with more zeros in the $\psi_4,\overline{\psi_4}$ entries, is that the effective Yukawa operators in Eq.\ref{Yuk_mass_insertion4}
have the suggestive matrix form,
\be
{\cal L}^{Yuk}_{4eff}= 
\pmatr{
	&\psi^c_1&\psi^c_2&\psi^c_3\\ 
	\hline
	\psi_1|&0&0&0\\
	\psi_2|&0&0&0\\
	\psi_3|&0&0&x^{\psi}_{34} y^{\psi}_{43}}
	\frac{{\phi} }{M^{\psi}_{4}}  {H} 
	+
\pmatr{
	&\psi^c_1&\psi^c_2&\psi^c_3\\ 
	\hline
	\psi_1|&0&0&0\\
	\psi_2|&0 & y^{\psi}_{24} x^{\psi^c}_{42} & y^{\psi}_{24} x^{\psi^c}_{43}\\
	\psi_3|&0 & y^{\psi}_{34} x^{\psi^c}_{42} & y^{\psi}_{34} x^{\psi^c}_{43}}
	\frac{\overline{\phi} }{M^{\psi^c}_{4}}\overline{H}  
\label{Yuk_mass_insertion_1}
\ee
where the dimensionless couplings $x,y$ in the matrices are expected to be of order unity,
and we have dropped the distinction between $\overline{\phi}$ and  $\overline{\phi'}$ for simplicity.
If we assume that $\phi,\overline{\phi}$ fields develop vacuum expectation values (VEVs)
with a hierarchy of scales,
\be
\frac{\langle \overline{\phi} \rangle}{M^{\psi^c}_{4}}\ll \frac{\langle  {\phi} \rangle}{M^{\psi}_{4}}  \lesssim 1
\label{hierarchy}
\ee
then the first matrix in Eq.\ref{Yuk_mass_insertion_1} 
generates larger effective third family Yukawa couplings, while the second matrix generates suppressed second
family Yukawa couplings and mixings.
Since the sum of the two matrices has rank 1, the first family will be massless, assuming only the fourth VL family.
Indeed the first family masses are protected by an approximate $U(1)$ family symmetry which emerges accidentally 
as a result of the special rank 1 nature of the effective Yukawa matrices and the fact that so far only a fourth VL family has been 
considered. The second mild inequality in Eq.\ref{hierarchy} means that the mass insertion approximation breaks down for the third family Yukawa couplings, so strictly speaking we should use a large angle mixing formalism as discussed in Appendix~\ref{A}.
However in the interests of clarity, we shall continue to use the mass insertion approximation even for the third family.

The first family masses depend on the fifth VL family and related fields in Table~\ref{twinPS}.
Including both fourth and fifth VL families, the masses and couplings in Eqs.\ref{L4} and \ref{L5} can be written
as a $7\times 7$ matrix in flavour space, in the basis of Eq.\ref{M^psi_an_1},
\be
	M^{\psi} = \pmatr{
	&\psi^c_1&\psi^c_2&\psi^c_3&\overline{\psi_4}  &\psi^c_4&\overline{\psi_5} &\psi^c_5\\ 
	\hline
	\psi_1|&0&0&0&0 &0& 0 &y^{\psi}_{15}{h}  \\
	\psi_2|&0&0&0 &0&y^{\psi}_{24}\overline{H}&x^{\psi}_{25} {\Phi}&0 \\
	\psi_3|&0&0&0&x^{\psi}_{34} {\phi} &y^{\psi}_{34}\overline{H} & x^{\psi}_{35} {\Phi}& 0 \\ 
	\psi_4|& 0&0& y^{\psi}_{43} {H}&M^{\psi}_{4} &0&0 & 0 \\ 
	\overline{\psi^c_4}|&0&x^{\psi^c}_{42}\overline{\phi'}&x^{\psi^c}_{43}\overline{\phi}&0& M^{\psi^c}_{4}& 0 & 0 \\
	\psi_5|& y^{\psi}_{51} {h}& 0 & 0 &0 & 0&M^{\psi}_{5}  &0 \\ 
	\overline{\psi^c_5}|&0 &x^{\psi^c}_{52}{\Phi'}&x^{\psi^c}_{53}{\Phi}&  0  &0 &0& M^{\psi^c}_{5}
	 }.	
	\label{M^psi_an_2}
\ee

\begin{figure}[ht]
\centering
	\includegraphics[scale=0.11]{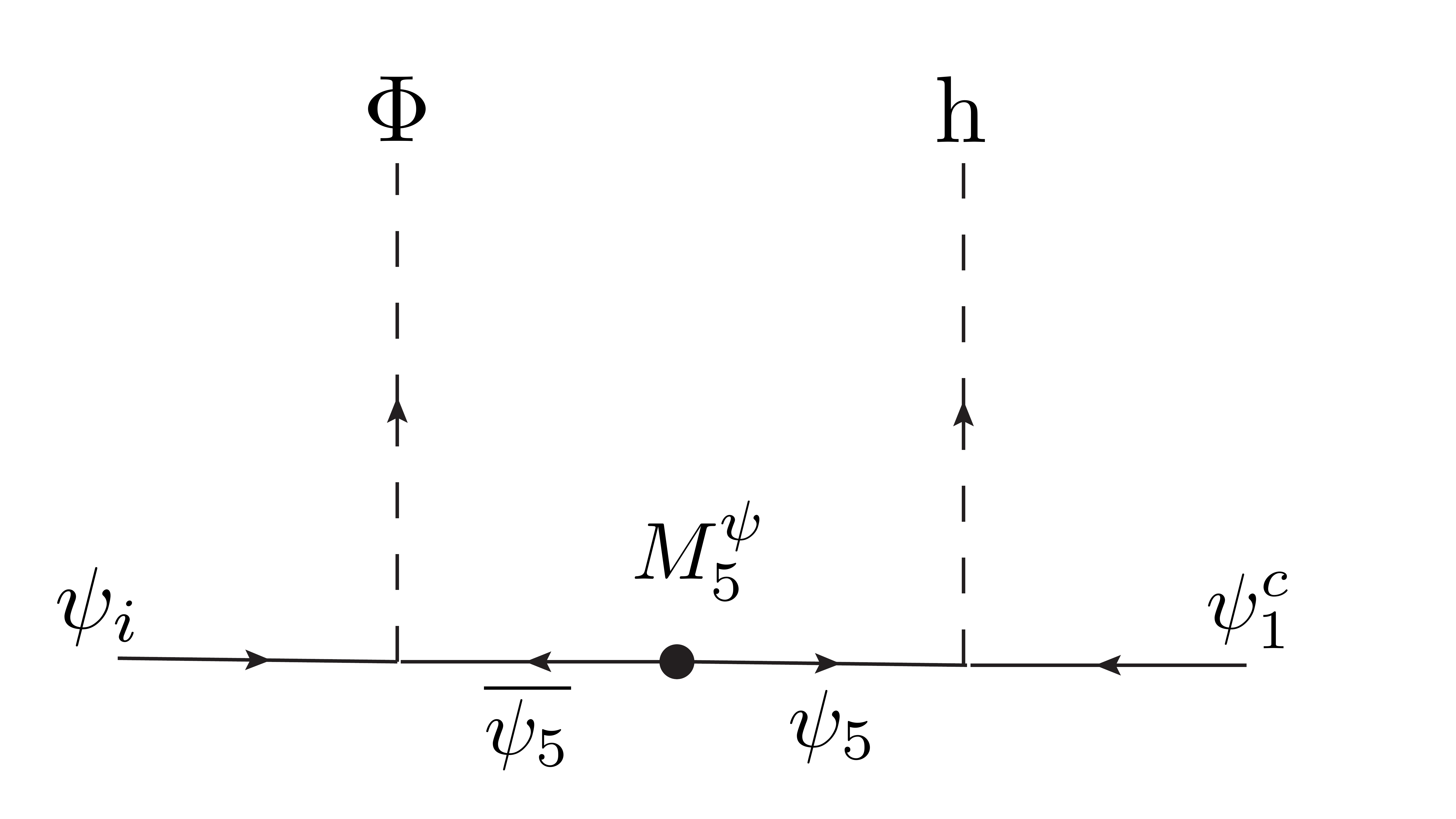}
\hspace*{1ex}
	\includegraphics[scale=0.11]{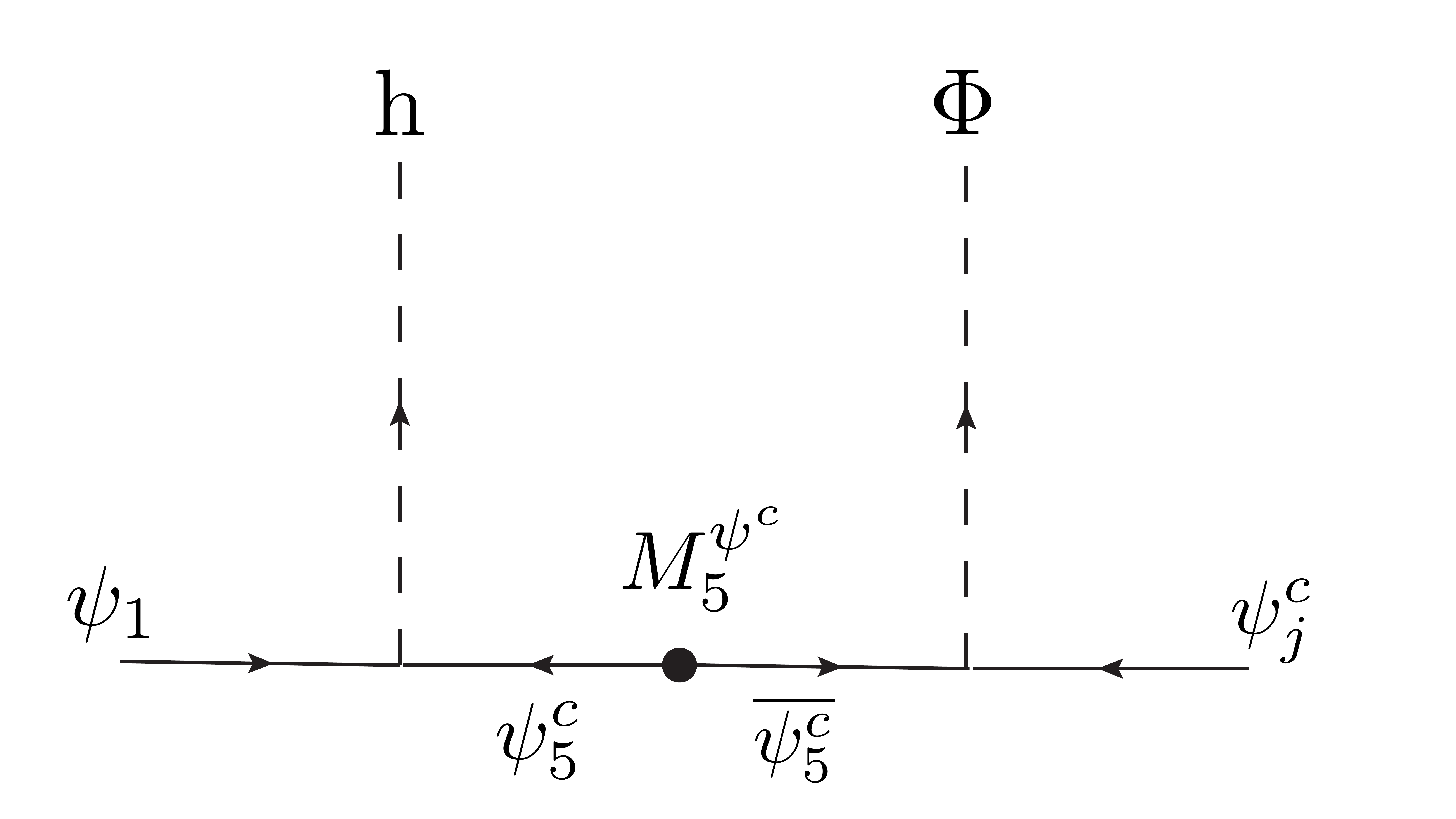}
\caption{Diagrams which contribute effectively to the first column (left panel) and first row (right panel) of the Yukawa matrices,
where $i,j=2,3$ are the only non-zero values.
 In the right panel, we have dropped the distinction between $\Phi$ and $\Phi'$ for simplicity.}
\label{Fig2}
\end{figure}

The first family couplings involving the fifth VL family will yield non-zero masses involving the first family
from the diagrams in Fig.~\ref{Fig2}.
The diagrams in Fig.\ref{Fig2} lead to effective Yukawa operators (up to an irrelevant minus sign),
after integrating out VL fermions,
\be
{\cal L}^{Yuk}_{5eff}= \frac{x^{\psi}_{i5} {\Phi}  y^{\psi}_{51} {h} }{M^{\psi}_{5}}  \psi_i {\psi^c_1}
+
\frac{ y^{\psi}_{15} {h}  x^{\psi^c}_{52}  {\Phi'}  }{M^{\psi^c}_{5}} \psi_1 {\psi^c_2} 
+
\frac{ y^{\psi}_{15} {h}  x^{\psi^c}_{53}  {\Phi}  }{M^{\psi^c}_{5}} \psi_1 {\psi^c_3} + H.c.
\label{Yuk_mass_insertion5}
\ee
which contribute to the first column and row of the Yukawa matrices, with the $(1,1)$ texture zero enforced.
The effective Yukawa operators in Eq.\ref{Yuk_mass_insertion5}
have the matrix form,
\be
{\cal L}^{Yuk}_{5eff}= 
\pmatr{
	&\psi^c_1&\psi^c_2&\psi^c_3\\ 
	\hline
	\psi_1|&0&0&0\\
	\psi_2|&x^{\psi}_{25} y^{\psi}_{51}&0&0\\
	\psi_3|&x^{\psi}_{35} y^{\psi}_{51}&0&0}
	\frac{{\Phi} }{M^{\psi}_{5}}  {h} 
	+
\pmatr{
	&\psi^c_1&\psi^c_2&\psi^c_3\\ 
	\hline
	\psi_1|&0&y^{\psi}_{15} x^{\psi^c}_{52}&y^{\psi}_{15} x^{\psi^c}_{53}\\
	\psi_2|&0 & 0 & 0\\
	\psi_3|&0 & 0 & 0}
	\frac{{\Phi} }{M^{\psi^c}_{5}}{h}  
\label{Yuk_mass_insertion_2}
\ee
where the dimensionless couplings $x,y$ in the matrices are expected to be of order unity,
and in the right term we have dropped the distinction between $\Phi$ and $\Phi'$ for simplicity.

Including both the fourth and fifth VL families, the matrix of operators responsible for the effective Yukawa matrix can be written 
using Eqs.\ref{Yuk_mass_insertion_1} and \ref{Yuk_mass_insertion_2} as,
\be
M^{\psi}_{ij}=A_{ij}\frac{ {\phi}  {H} }{M^{\psi}_{4}}
+B_{ij}\frac{  \overline{H} \overline{\phi} }{M^{\psi^c}_{4}}
+C_{ij}\frac{ {\Phi}  {h} }{M^{\psi}_{5}}
+D_{ij}\frac{{h} {\Phi} }{M^{\psi^c}_{5}}
\label{Mpsi}
\ee
where 
\begin{align}
A_{ij}&=x^{\psi}_{i4} y^{\psi}_{4j}\delta_{i3}\delta_{j3}, \ \ B_{ij}= y^{\psi}_{i4} x^{\psi^c}_{4j}(1-\delta_{i1})(1-\delta_{1j}), \\
C_{ij}&=x^{\psi}_{i5} y^{\psi}_{5j}(1-\delta_{i1})\delta_{j1}, \ \ D_{ij}= y^{\psi}_{i5} x^{\psi^c}_{5j}\delta_{i1}(1-\delta_{1j})
\label{ABCD}
\end{align}
If we assume a hierarchy of scales, extending Eq.\ref{hierarchy} to the fifth family,
\be
 \frac{\langle  {\Phi} \rangle}{M^{\psi}_{5}} , \frac{\langle {\Phi} \rangle}{M^{\psi^c}_{5}}\ll
\frac{\langle \overline{\phi} \rangle}{M^{\psi^c}_{4}}\ll \frac{\langle  {\phi} \rangle}{M^{\psi}_{4}}  \lesssim 1
\label{hierarchy2}
\ee
then the term proportional to $A_{ij}$ dominates the third family masses, the term proportional to $B_{ij}$ dominates the second family masses, while the terms proportional to $C_{ij}$ and $D_{ij}$ 
contribute to the first column and row, respectively, with both terms maintaining the texture zero in the first entry of the mass matrix.
The hierarchy of quark and lepton masses in the SM Yukawa couplings is re-expressed as the hierarchy of scales in Eq.\ref{hierarchy2}.
This is not just a reparameterisation of the hierarchy, since it involves extra dynamics and testable experimental predictions, such as the VL fermion spectrum with $M^{\psi}_{4}\sim 1$~TeV. It also leads to connections with $B$ physics as we shall see.

\begin{table}
\centering
\begin{tabular}{| l | c  c c c | c |}
\hline
Field & $SU(4)^I_{PS}$ & $SU(4)^{II}_{PS}$ & $SU(2)^{I+II}_L$ & $SU(2)^{I+II}_R$ &  $Z_6$\\ 
\hline \hline
$\psi_{1,2,3}$ 		 & ${\bf 1}$ & ${\bf 4}$ & ${\bf 2}$ & ${\bf 1}$ & $\alpha^2,1,1$\\
$\psi^c_{1,2,3}$ 	& ${\bf 1}$ & ${\overline{\bf 4}}$ & ${\bf 1}$ & ${\overline{\bf 2}}$ & $\alpha^2,\alpha^5,1$\\
\hline
${H'}$   & ${\bf 1}$  &${\bf 4}$ &   ${\bf 1}$ & ${\bf 2}$ & $1$ \\
$\overline{H}'$   & ${\bf 1}$  &  ${\overline{\bf 4}}$   &  ${\bf 1}$ & ${\overline{\bf 2}}$  & $1$\\
\hline
$\Phi$, $\Phi'$  &   ${\bf 1}$  & ${\bf 1}$  & ${\bf 1+3}$ &  ${\bf 1}$ & $1,\alpha$  \\
$\overline{\Phi}$ &   ${\bf 1}$  & ${\bf 1}$   &  ${\bf 1}$ &  ${\bf 1+3}$& $1$ \\
\hline
$h$   &   ${\bf 1}$  & ${\bf 1}$ & ${\overline{\bf 2}}$   &  ${\bf 2}$ & $\alpha^2$ \\
\hline
 $\xi$   &  ${\bf 1}$ & ${\bf 1}$ &  ${\bf 1}$ &  ${\bf 1}$   & $\alpha$    \\
\hline
$\psi_{4}$ 		& ${\bf 4}$ & ${\bf 1}$ & ${\bf 2}$ & ${\bf 1}$& $1$ \\
$\overline{\psi_{4}}$ 		& ${\overline{\bf 4}}$ & ${\bf 1}$   & ${\overline{\bf 2}}$ & ${\bf 1}$& $1$\\
$\psi^c_{4}$ 		 & ${\overline{\bf 4}}$ & ${\bf 1}$ & ${\bf 1}$ & ${\overline{\bf 2}}$& $1$\\
$\overline{\psi^c_{4}}$ 		& ${\bf 4}$  & ${\bf 1}$ & ${\bf 1}$ & ${\bf 2}$ & $1$ \\
\hline
$\psi_{5}$ 		& ${\bf 1}$ & ${\bf 4}$  & ${\bf 2}$ & ${\bf 1}$& $1$ \\
$\overline{\psi_{5}}$ 		& ${\bf 1}$ & ${\overline{\bf 4}}$     & ${\overline{\bf 2}}$ & ${\bf 1}$& $1$\\
$\psi^c_{5}$ 		& ${\bf 1}$ & ${\overline{\bf 4}}$  & ${\bf 1+3}$ & ${\overline{\bf 2}}$& $1$\\
$\overline{\psi^c_{5}}$ 		& ${\bf 1}$ & ${\bf 4}$  & ${\bf 1+3}$ & ${\bf 2}$ & $1$ \\
\hline
$\phi$  &   ${\bf 4}$  & ${\overline{\bf 4}}$ & ${\bf 1+3}$ &  ${\bf 1}$  & $1$ \\
$\overline{\phi}$, $\overline{\phi'}$ &   ${\overline{\bf 4}}$ & ${\bf 4}$ &  ${\bf 1}$ &  ${\bf 1+3}$& $1,\alpha$ \\
\hline
\hline
 $H$   & ${\overline{\bf 4}}$  &  ${\bf 4}$ & ${\overline{\bf 2}}$   &  ${\bf 2}$ & $1$ \\
$\overline{H}$   & ${\bf 4}$ & ${\overline{\bf 4}}$  &  ${\overline{\bf 2}}$ & ${\bf 2}$  & $1$\\
\hline
\hline
\end{tabular}
\caption{Under the subgroup $G_{4422}$ the fields in Table~\ref{twinPS} transform as shown.
}
\label{tab:funfields1}
\end{table}

\section{The Low Energy $G_{4321}$ Theory}
\label{low}

\subsection{High scale symmetry breaking to $G_{4321}$}

In this subsection we shall discuss the high scale symmetry breaking
\begin{align}
G_{422}^I \times G_{422}^{II} \stackrel{M_{high}}{\longrightarrow} G_{4321}
\label{breaking1}
\end{align}

We can think of this as a two part symmetry breaking:
(i) the two pairs of left-right groups break down to a diagonal left-right subgroup, (ii) the second PS group is broken.
However the scales of these two parts of breaking, and their order, is not yet specified.

(i) To achieve the two pairs of left-right groups breaking to their diagonal left-right subgroup we shall assume the VEVs 
\begin{equation}
\langle \Phi \rangle \sim  v_{\Phi}, \ \ \ \   \langle \overline{\Phi}\rangle  \sim v_{\overline{\Phi}}
\label{PhiVEV}
\end{equation}
which lead to the symmetry breakings, respectively,
\be
SU(2)^I_L\times SU(2)^{II}_L\rightarrow SU(2)^{I+II}_L, \ \ \ \ 
SU(2)^I_R\times SU(2)^{II}_R \rightarrow SU(2)^{I+II}_R
\ee
Since the two $SU(4)_{PS}$ groups remain intact, the above symmetry breaking corresponds to
\be
G_{422}^I \times G_{422}^{II} \rightarrow G_{4422}
\label{4422}
\ee
where 
\be
G_{4422}\equiv SU(4)^{I}_{PS}\times SU(4)^{II}_{PS}\times SU(2)^{I+II}_L\times SU(2)^{I+II}_R
\label{4422def}
\ee
We summarise the transformation of the fields under $G_{4422}$ in Table~\ref{tab:funfields1}.

(ii) Then we assume the high scale PS group is broken via 
the Higgs $H',\overline{H'}$ in Table~\ref{tab:funfields1},
\begin{equation}
{H'}(1,4,1,2)=
\left(\begin{array}{cccc}
u_{H'}^r & u_{H'}^b & u_{H'}^g & \nu_{H'} \\ d_{H'}^r & d_{H'}^b & d_{H'}^g & e_{H'}^-
\end{array} \right) \label{H'}
\end{equation}
and
\begin{equation}
{\bar{H'}}(1,\bar{4},1,\bar{2})=
\left(\begin{array}{cccc}
\bar{d}_{H'}^r & \bar{d}_{H'}^b & \bar{d}_{H'}^g & e_{H'}^+ \\
\bar{u}_{H'}^r & \bar{u}_{H'}^b & \bar{u}_{H'}^g & \bar{\nu}_{H'}
\end{array} \right). \label{barH'}
\end{equation}
which develop VEVs in their right-handed neutrino components, 
\begin{equation}
\langle \nu_{H'}\rangle \sim \langle \bar{\nu}_{H'}\rangle \gtrsim 1 \ {\rm PeV}
\label{HpVEV}
\end{equation}
leading to the further symmetry breaking of the gauge group,
\be
G_{4422}\rightarrow G_{4321}\equiv SU(4)_{PS}^{I}\times SU(3)_c^{II}\times SU(2)^{I+II}_L\times U(1)_{Y'}
\label{4321}
\ee
where $SU(4)_{PS}^{II}$ is broken to $SU(3)_c^{II}\times U(1)^{II}_{B-L}$ ($4\rightarrow 3_{1/6}+1_{-1/2}$), while $SU(2)_R$ is broken to $U(1)_{T_{3R}}$
and the Abelian generators are broken to $U(1)_{Y'}$ where
\be
Y'=T^{II}_{B-L}+T^{I+II}_{3R}
\ee
The broken $SU(4)_{PS}^{II}$ generators are associated with gauge bosons which will 
mediate various processes at acceptable rates. The non-observance of $K_L\rightarrow \mu e$ 
is responsible for the limit in Eq.~\ref{HpVEV}, which is why we refer to this as high scale symmetry breaking.

The combined symmetry breakings (i) and (ii) in Eqs.~\ref{4422} and \ref{4321} are equivalent to that in Eq.\ref{high}, with 
the fields transforming under $G_{4321}$ as shown in Table~\ref{tab:funfields2}.
In particular, the Higgs scalars $H,\overline{H}$ decompose under $G_{4422}\rightarrow G_{4321}$ as,
\begin{align}
\label{H3}
{H}(\bar{4},{4},\bar{2},{2})& \rightarrow {H}_{t}(\bar{4},{3},\bar{2},2/3), \ \ {H}_b(\bar{4},{3},\bar{2},-1/3), \ \ {H}_{\tau}(\bar{4},1,\bar{2},-1), \ \  {H}_{\nu_{\tau}}(\bar{4},1,\bar{2},0) \\
\overline{H}(4,\bar{4},\bar{2},2)& \rightarrow  {H}_{c}(4,\bar{3},\bar{2},1/3), \ \ {H}_s(4,\bar{3},\bar{2},-2/3), \ \ {H}_{\mu}(4,1,\bar{2},0), \ \ {H}_{\nu_{\mu}}(4,1,\bar{2},1) 
\label{H2}
\end{align}
where the notation anticipates that a separate personal Higgs field contributes to each of the second and third family quark and lepton masses as shown below. 

\begin{table}
\vspace{-0.5in}
\centering
\begin{tabular}{| l | c c c c | c |}
\hline
Field & $SU(4)_{PS}^I$ & $SU(3)_{c}^{II}$ & $SU(2)^{I+II}_L$ & $U(1)_{Y'}$ & $Z_6$ \\ 
\hline
\hline
$Q_{1,2,3}$ 		 & ${\bf 1}$ & ${\bf 3}$ & ${\bf 2}$ & $1/6$ & $\alpha^2,1,1$ \\
$u^c_{1,2,3}$ 		 & ${\bf 1}$ & ${\overline{\bf 3}}$ & ${\bf 1}$ & $-2/3$ & $\alpha^2,\alpha^5,1$\\
$d^c_{1,2,3}$ 		 & ${\bf 1}$ & ${\overline{\bf 3}}$ & ${\bf 1}$ & $1/3$ & $\alpha^2,\alpha^5,1$\\
$L_{1,2,3}$ 		 & ${\bf 1}$ & ${\bf 1}$ & ${\bf 2}$ & $-1/2$ & $\alpha^2,1,1$\\
$e^c_{1,2,3}$ 		 & ${\bf 1}$ & ${\bf 1}$ & ${\bf 1}$ & $1$ & $\alpha^2,\alpha^5,1$\\
$\nu^c_{1,2,3}$         & ${\bf 1}$ & ${\bf 1}$ & ${\bf 1}$ & $0$ & $\alpha^2,\alpha^5,1$\\
\hline
$\Phi$, $\Phi'$  &   ${\bf 1}$  & ${\bf 1}$  & ${\bf 1+3}$ &  $0$  & $1,\alpha$ \\
$\overline{\Phi}$ &   ${\bf 1}$  & ${\bf 1}$   &  ${\bf 1}$ &  $0$ & $1$\\
\hline
$h_u$   &   ${\bf 1}$  & ${\bf 1}$ & ${\overline{\bf 2}}$   &  $1/2$ & $\alpha^2$ \\
 $h_d$   &   ${\bf 1}$  & ${\bf 1}$ & ${\overline{\bf 2}}$   &  $-1/2$ & $\alpha^2$ \\
\hline
 $\xi$   &  ${\bf 1}$ & ${\bf 1}$ &  ${\bf 1}$ &  $0$   & $\alpha$    \\
\hline
$\psi_{4}$ 		& ${\bf 4}$ & ${\bf 1}$ & ${\bf 2}$ & $0$& $1$ \\
$\overline{\psi}_{4}$ 		& ${\overline{\bf 4}}$ & ${\bf 1}$   & ${\overline{\bf 2}}$ & $0$& $1$\\
$\psi^c_{4u\nu}$ 		 & ${\overline{\bf 4}}$ & ${\bf 1}$ & ${\bf 1}$ & $-1/2$& $1$\\
$\psi^c_{4de}$ 		 & ${\overline{\bf 4}}$ & ${\bf 1}$ & ${\bf 1}$ & $1/2$& $1$\\
$\overline{\psi^c}_{4u\nu}$ 		& ${\bf 4}$  & ${\bf 1}$ & ${\bf 1}$ & $1/2$  & $1$\\
$\overline{\psi^c}_{4ed}$ 		& ${\bf 4}$  & ${\bf 1}$ & ${\bf 1}$ & $-1/2$  & $1$\\
\hline
$Q_{5}$ 		 & ${\bf 1}$ & ${\bf 3}$ & ${\bf 2}$ & $1/6$ & $1$ \\
$\overline{Q_{5}}$ 		 & ${\bf 1}$ & ${\overline{\bf 3}}$ & ${\overline{\bf 2}}$ & $-1/6$ & $1$ \\
$u^c_{5}$ 		 & ${\bf 1}$ & ${\overline{\bf 3}}$ & ${\bf 1+3}$ & $-2/3$ & $1$\\
$\overline{u^c_{5}}$ 		 & ${\bf 1}$ & ${{\bf 3}}$ & ${\bf 1+3}$ & $2/3$ & $1$\\
$d^c_{5}$ 		 & ${\bf 1}$ & ${\overline{\bf 3}}$ & ${\bf 1+3}$ & $1/3$ & $1$\\
$\overline{d^c_{5}}$ 		 & ${\bf 1}$ & ${{\bf 3}}$ & ${\bf 1+3}$ & $-1/3$& $1$ \\
$L_{5}$ 		 & ${\bf 1}$ & ${\bf 1}$ & ${\bf 2}$ & $-1/2$ & $1$\\
$\overline{L_{5}}$ 		 & ${\bf 1}$ & ${\bf 1}$ & ${\overline{\bf 2}}$ & $1/2$ & $1$\\
$e^c_{5}$ 		 & ${\bf 1}$ & ${\bf 1}$ & ${\bf 1+3}$ & $1$  & $1$\\
$\overline{e^c_{5}}$ 		 & ${\bf 1}$ & ${\bf 1}$ & ${\bf 1+3}$ & $-1$ & $1$\\
$\nu^c_{5}$         & ${\bf 1}$ & ${\bf 1}$ & ${\bf 1+3}$ & $0$ & $1$\\
$\overline{\nu^c_{5}}$         & ${\bf 1}$ & ${\bf 1}$ & ${\bf 1+3}$ & $0$& $1$ \\
\hline
${\phi}_3$ & ${\bf 4}$   &  ${\overline{\bf 3}}$ &  ${\bf 1+3}$ & $-1/6$ & $1$ \\
${\phi}_1$ &   ${\bf 4}$ & ${\bf 1}$ &  ${\bf 1+3}$ & $1/2$ & $1$ \\
$\overline{\phi}_3$, $\overline{\phi}_3'$ &   ${\overline{\bf 4}}$ & ${\bf 3}$ &  ${\bf 1}$ & $1/6$ & $1,\alpha$ \\
$\overline{\phi}_1$, $\overline{\phi}_1'$ &   ${\overline{\bf 4}}$ & ${\bf 1}$ &  ${\bf 1}$ & $-1/2$  & $1,\alpha$\\
\hline
\hline
${H}_t$    & $\overline{\bf 4}$ & ${{\bf 3}}$  & ${\overline{\bf 2}}$ & ${2/3}$ & $1$ \\
${H}_b$    & $\overline{\bf 4}$ & ${\bf 3}$  &  ${\overline{\bf 2}}$ & $-1/3$  & $1$\\
${H}_{\tau}$    & $\overline{\bf 4}$ & ${\bf 1}$  & ${\overline{\bf 2}}$ & $-1$  & $1$\\
${H}_{\nu_{\tau}}$    & $\overline{\bf 4}$ & ${\bf 1}$  &  ${\overline{\bf 2}}$ & ${0}$ & $1$ \\
${H}_c$    & ${\bf 4}$ & ${\overline{\bf 3}}$  &  ${\overline{\bf 2}}$ & ${1/3}$  & $1$\\
${H}_s$    & ${\bf 4}$ & ${\overline{\bf 3}}$  &  ${\overline{\bf 2}}$ & $-2/3$ & $1$ \\
${H}_{\mu}$    & ${\bf 4}$ & ${\bf 1}$  &  ${\overline{\bf 2}}$ & $0$  & $1$\\
${H}_{\nu_{\mu}}$    & ${\bf 4}$ & ${\bf 1}$  &  ${\overline{\bf 2}}$ & ${1}$  & $1$\\
\hline
\end{tabular}
\caption{Under the subgroup $G_{4321}$, the fields of Tables~\ref{twinPS},\ref{tab:funfields1} decompose as shown.
We have dropped the Higgs $H', \overline{H'}$, and the $\overline{\phi_3}$ states with $Y'=(7/6,1/6,-5/6)$,
$\overline{\phi_1}$ states with $Y'=(1/2,-1/2,-3/2)$, and $\overline{\Phi}$ states with $Y'=(1,0,-1)$,
coming from the $SU(2)_R$ triplets.
We have labelled the Higgs fields arising from the decomposition of $H$ and $\overline{H}$ in a suggestive personal Higgs notation,
$H_{t,b,\tau,\nu_{\tau}}$ and $H_{c,s,\mu,\nu_{\mu}}$,
according to the fermion to which it gives mass.
}
\label{tab:funfields2}
\end{table}

\begin{figure}[ht]
\centering
	\includegraphics[scale=0.056]{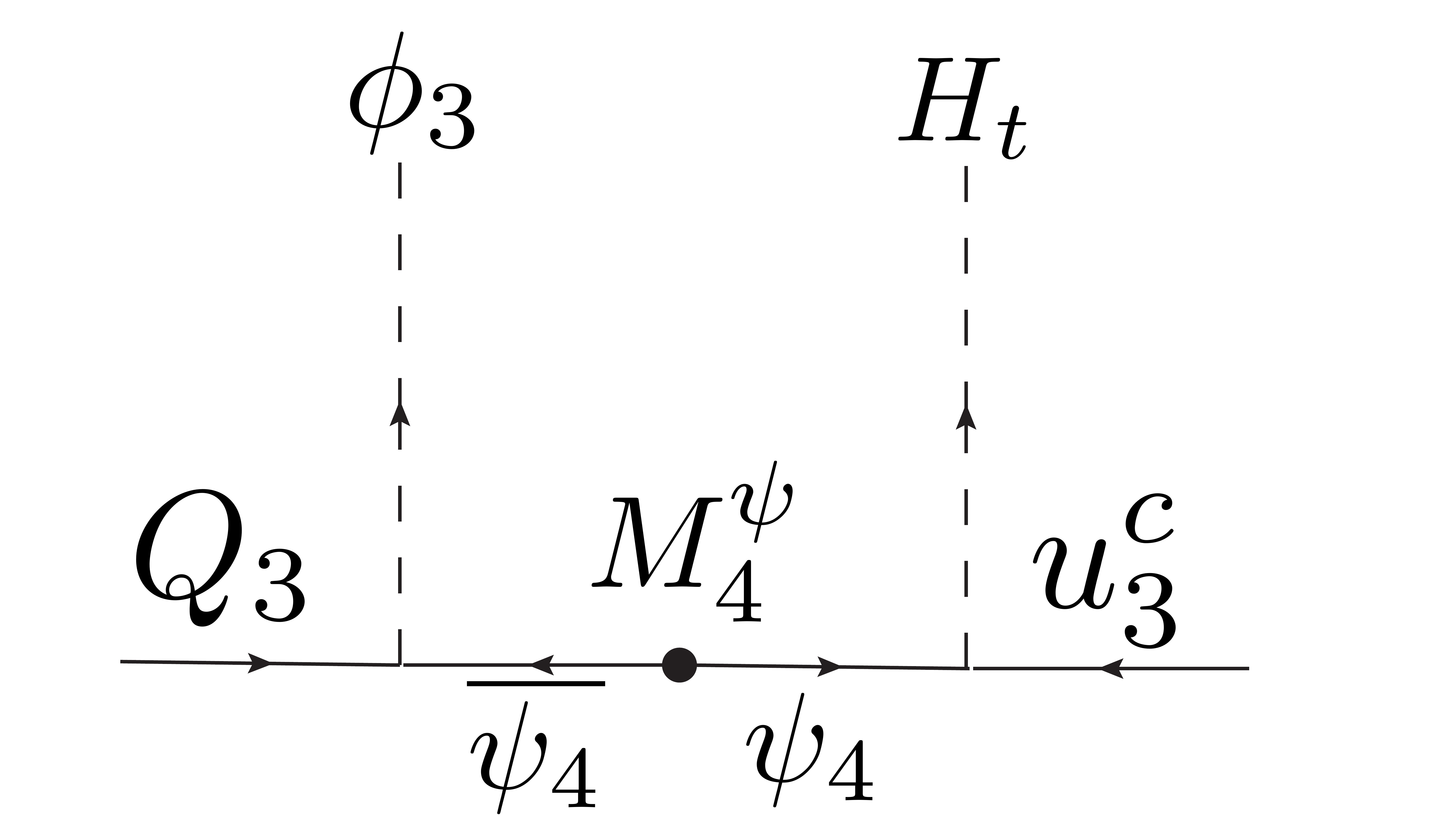} 
	\includegraphics[scale=0.056]{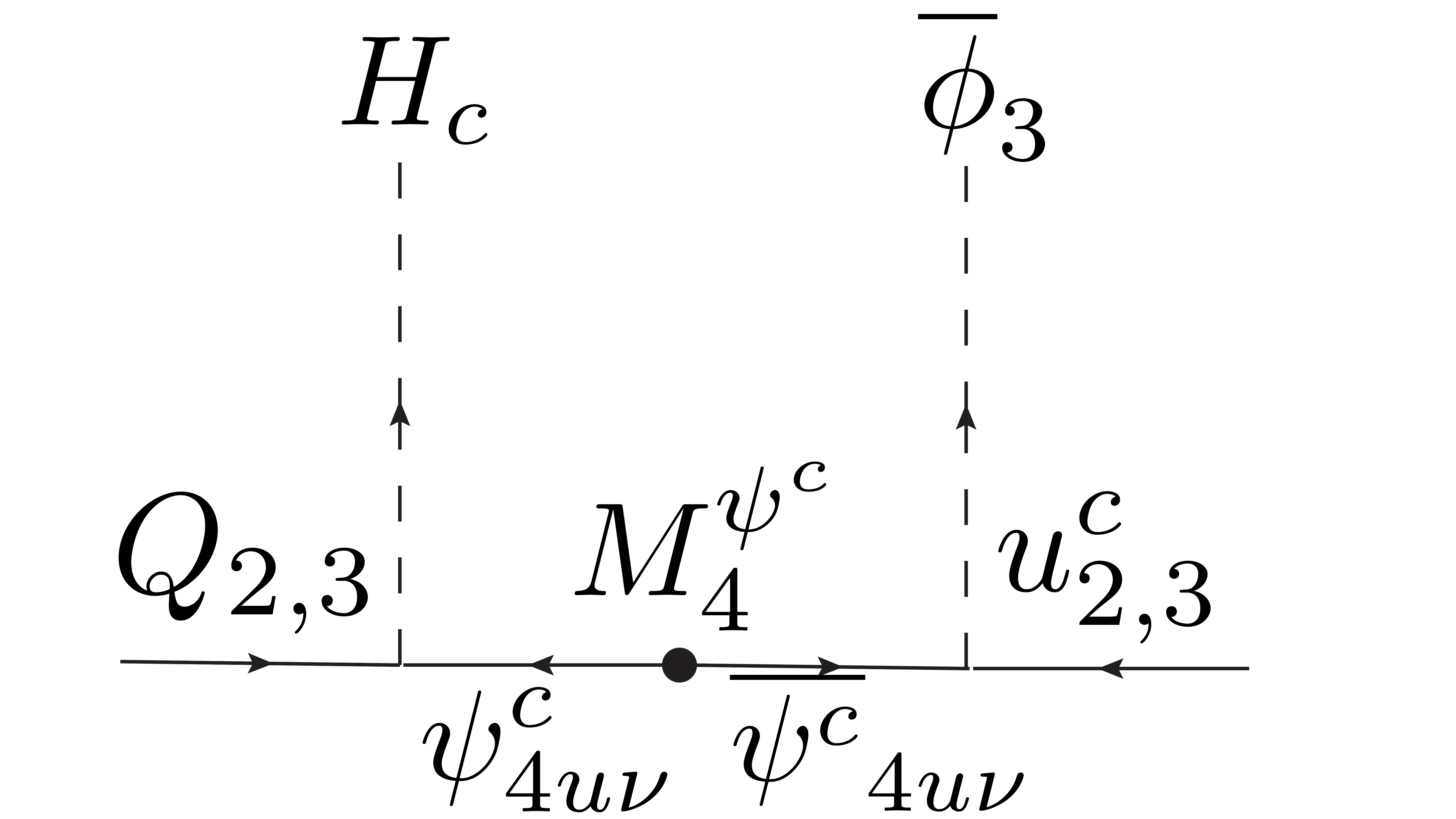}
	\includegraphics[scale=0.056]{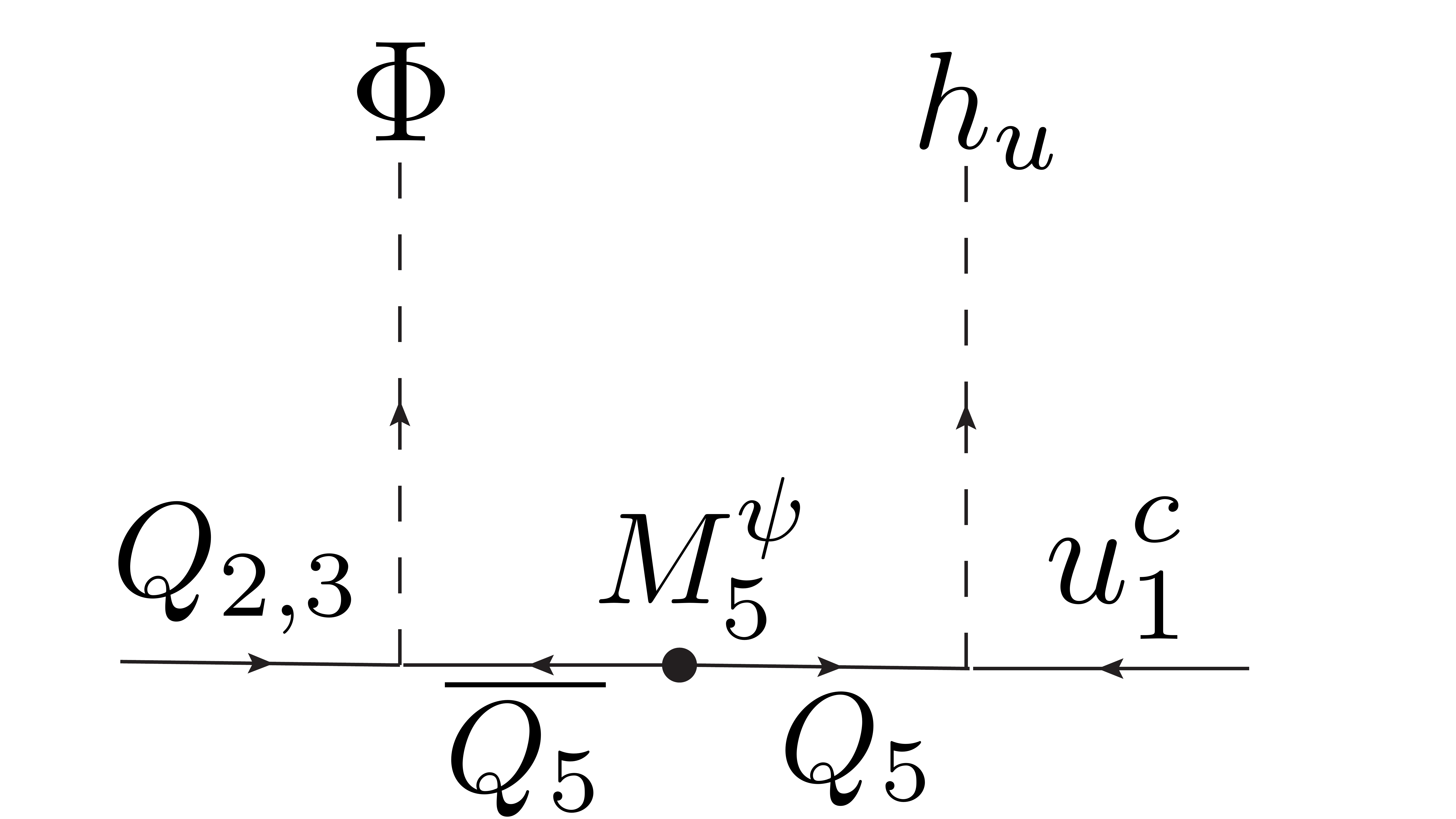}
	\includegraphics[scale=0.056]{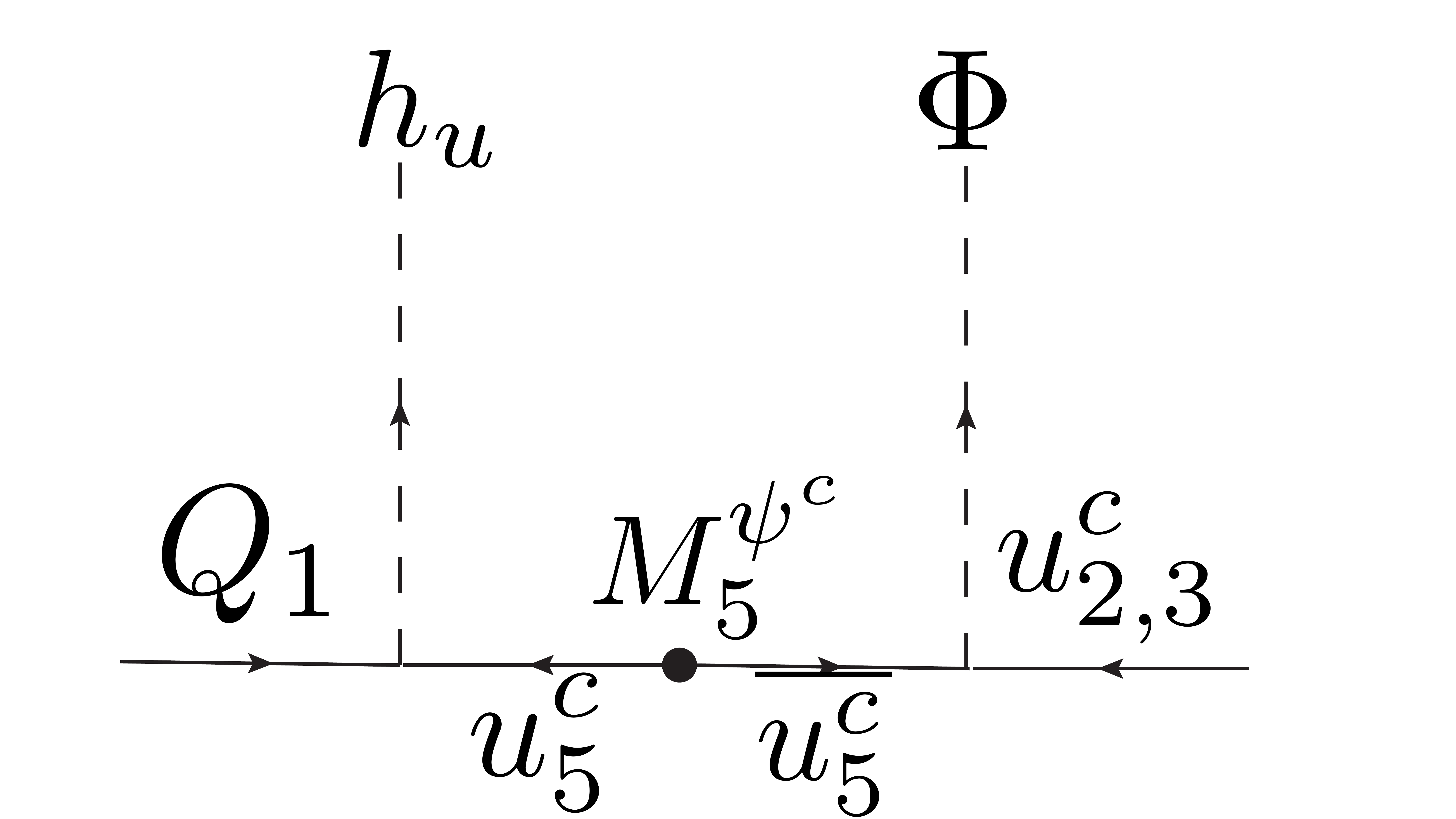}\vspace{0.15in}
	\includegraphics[scale=0.056]{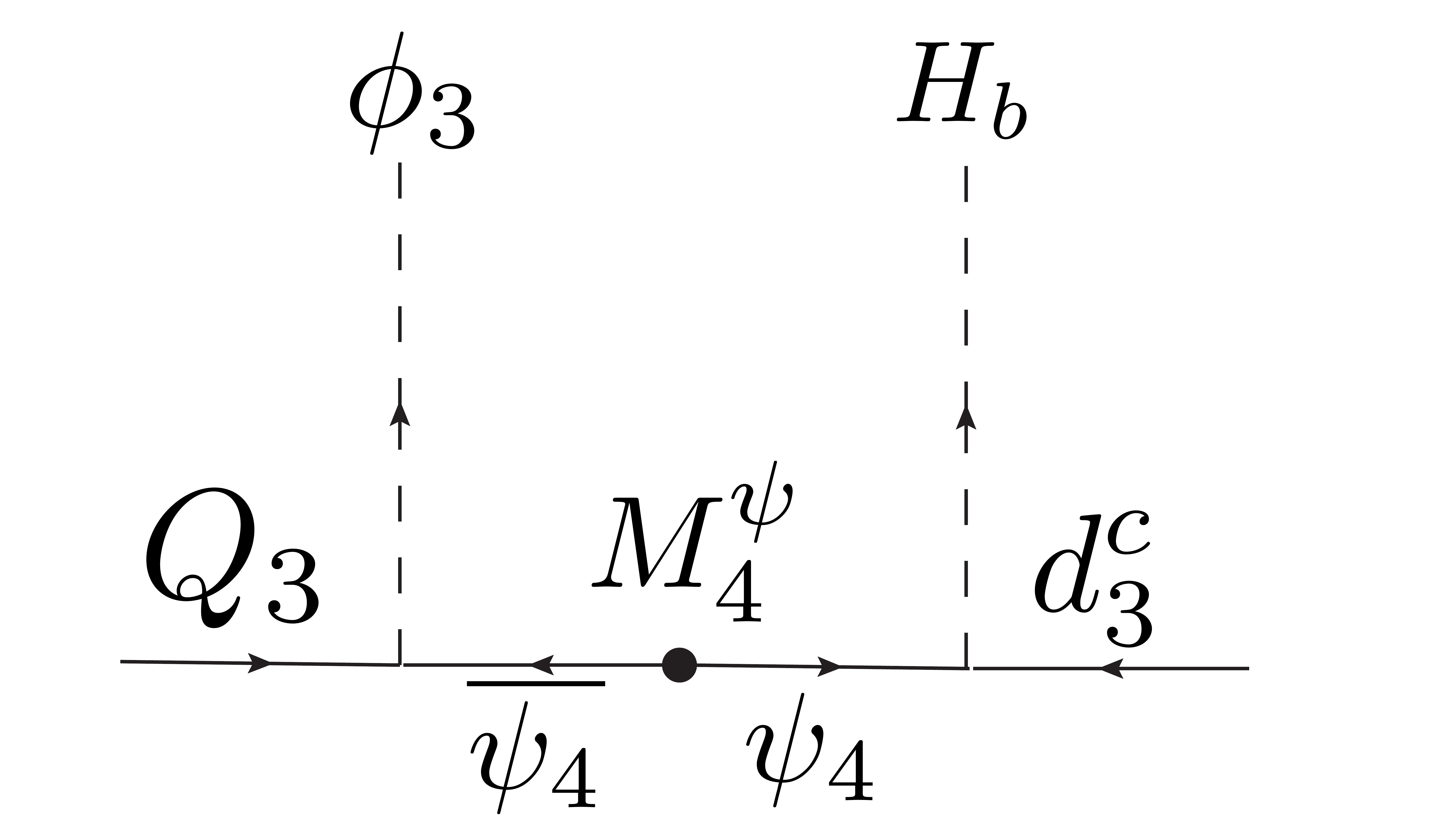}
	\includegraphics[scale=0.056]{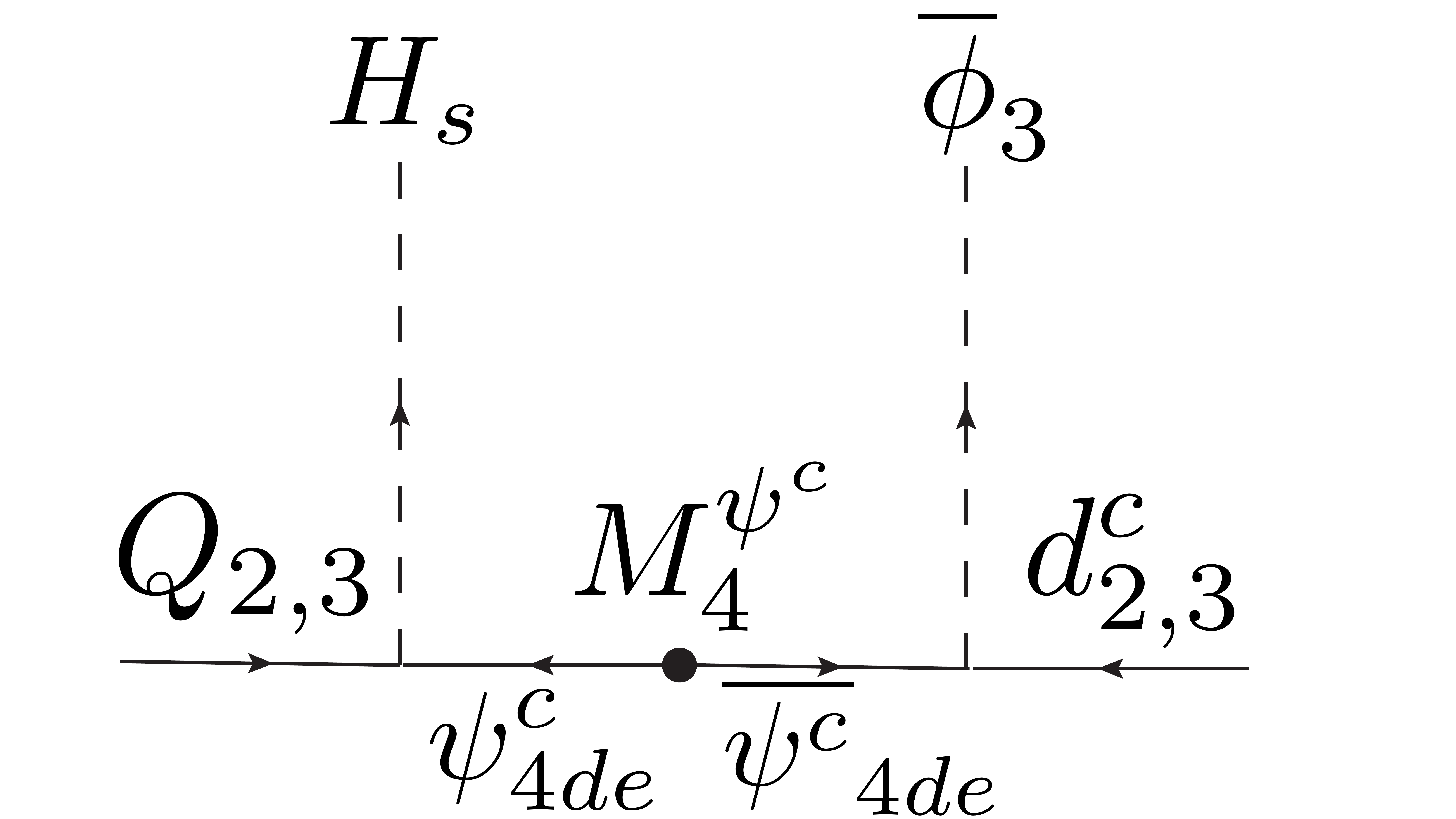}
	\includegraphics[scale=0.056]{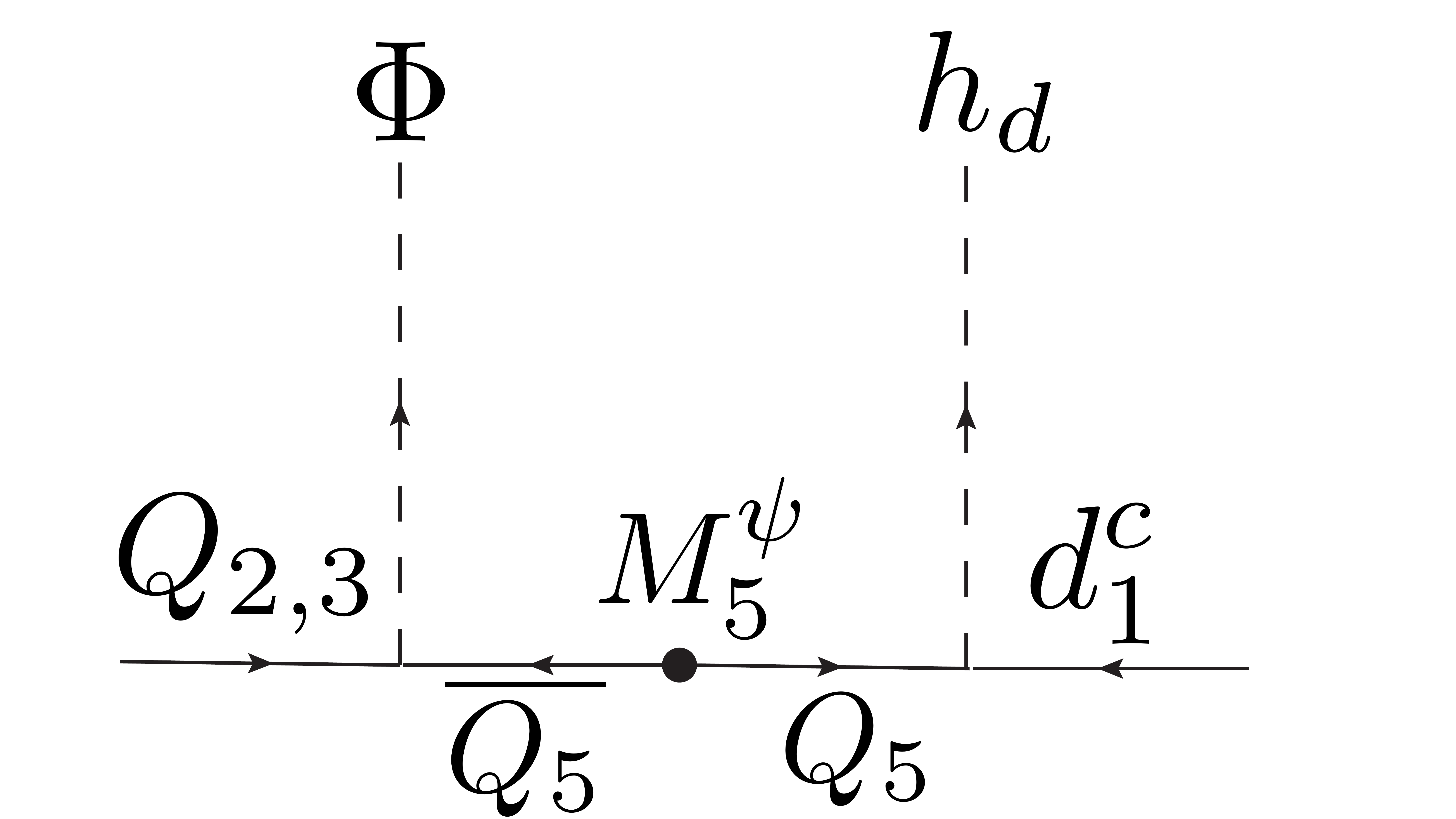}
	\includegraphics[scale=0.056]{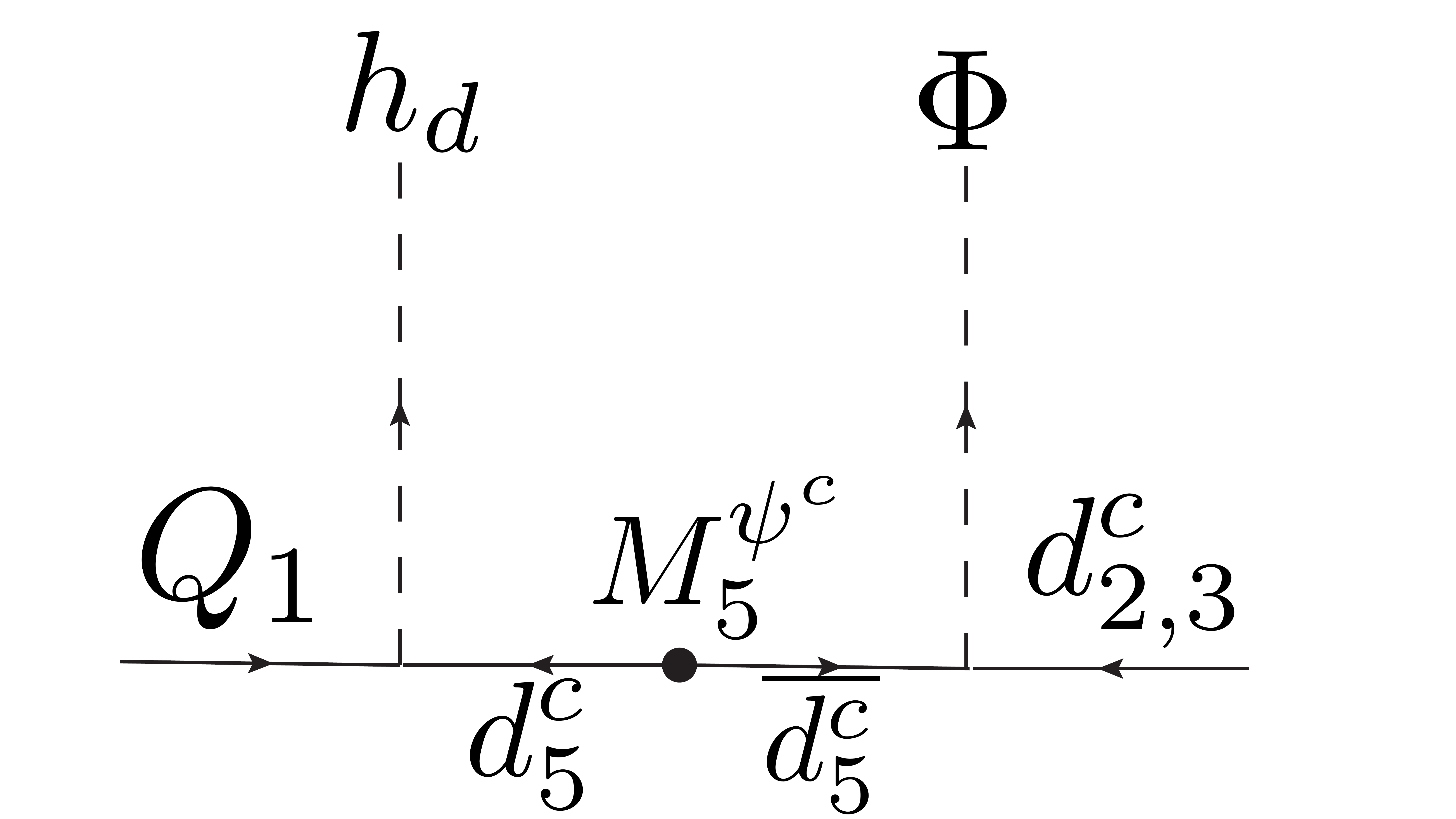}\vspace{0.15in}
	\includegraphics[scale=0.056]{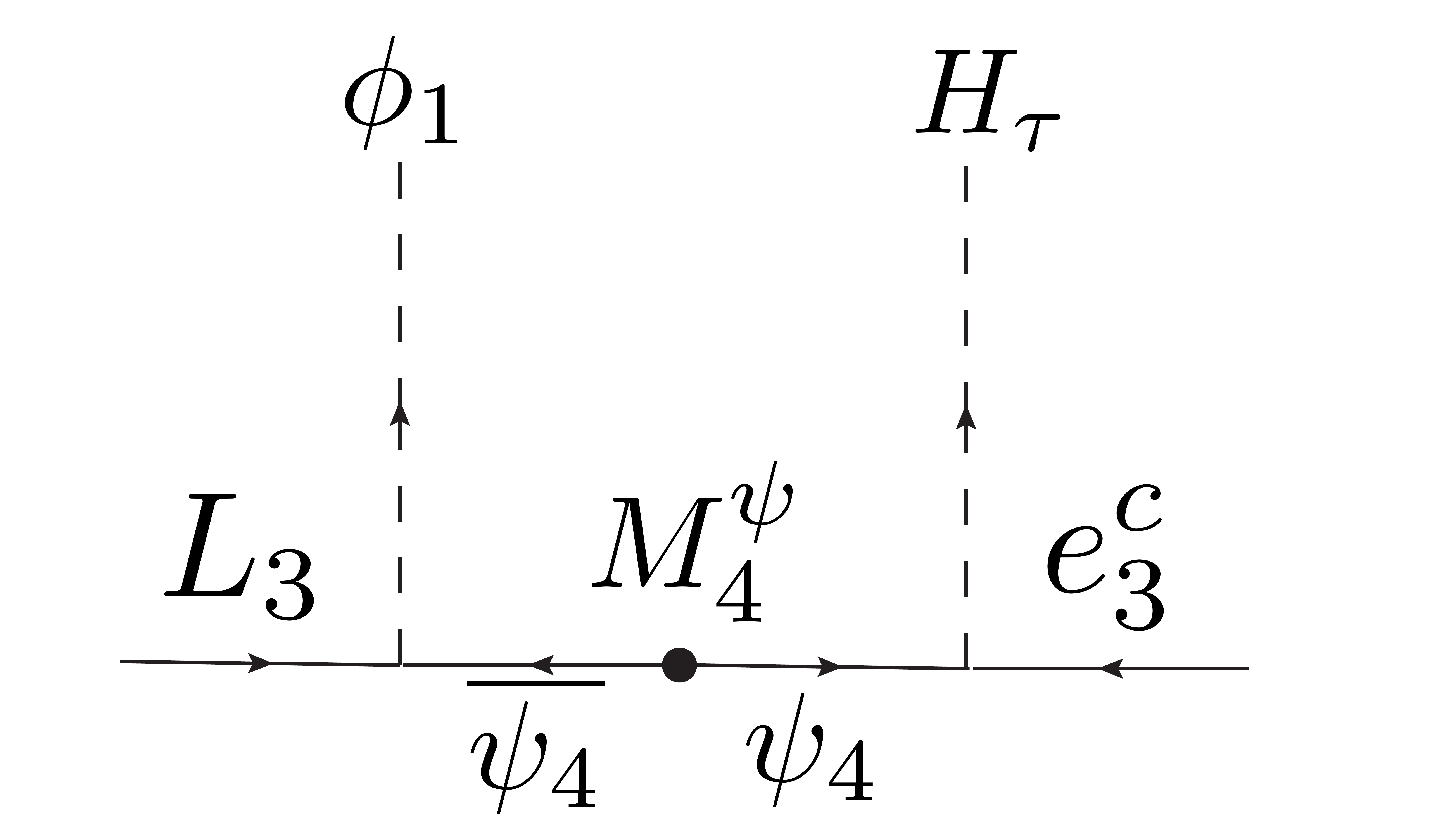}
	\includegraphics[scale=0.056]{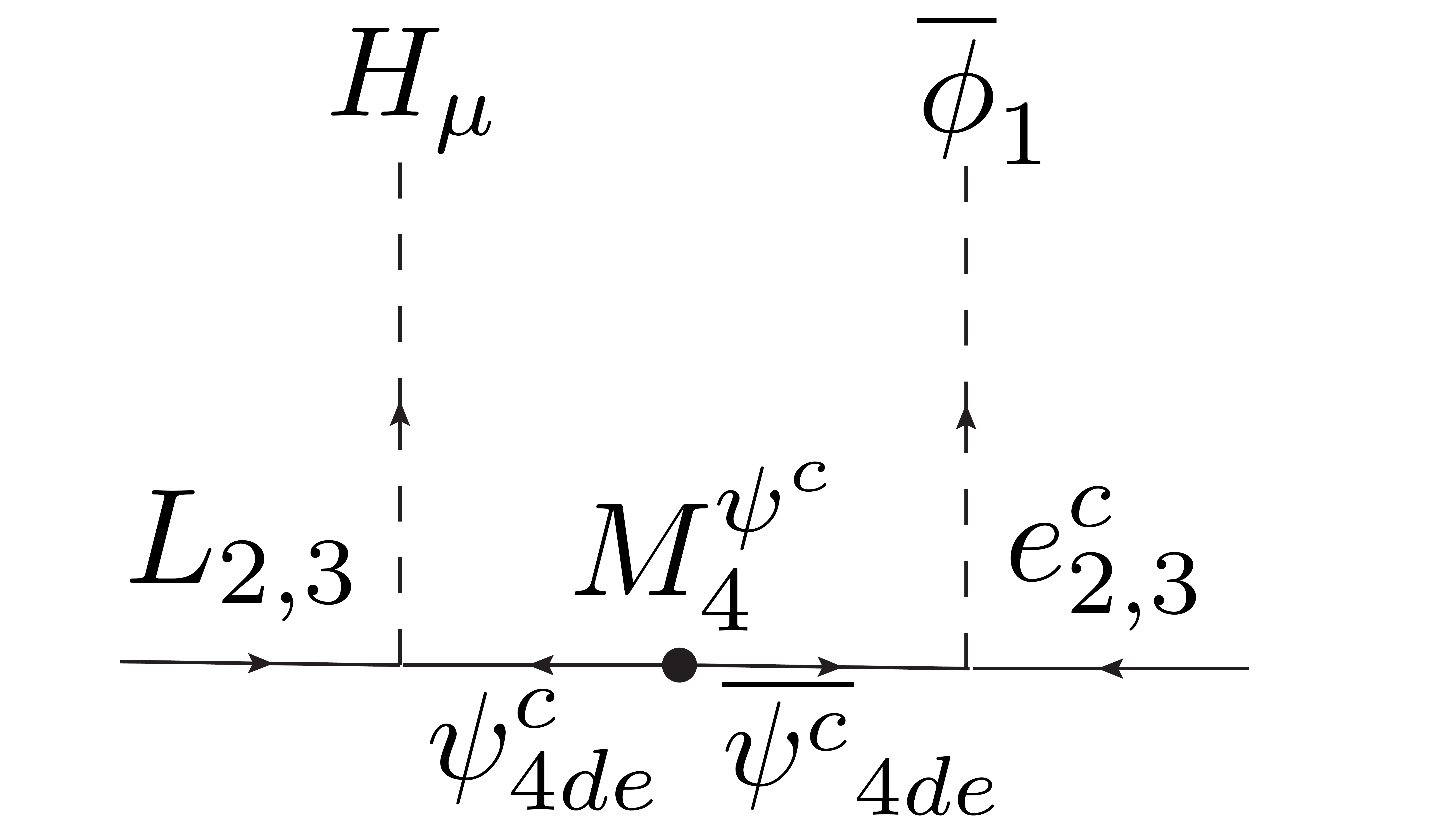}
	\includegraphics[scale=0.056]{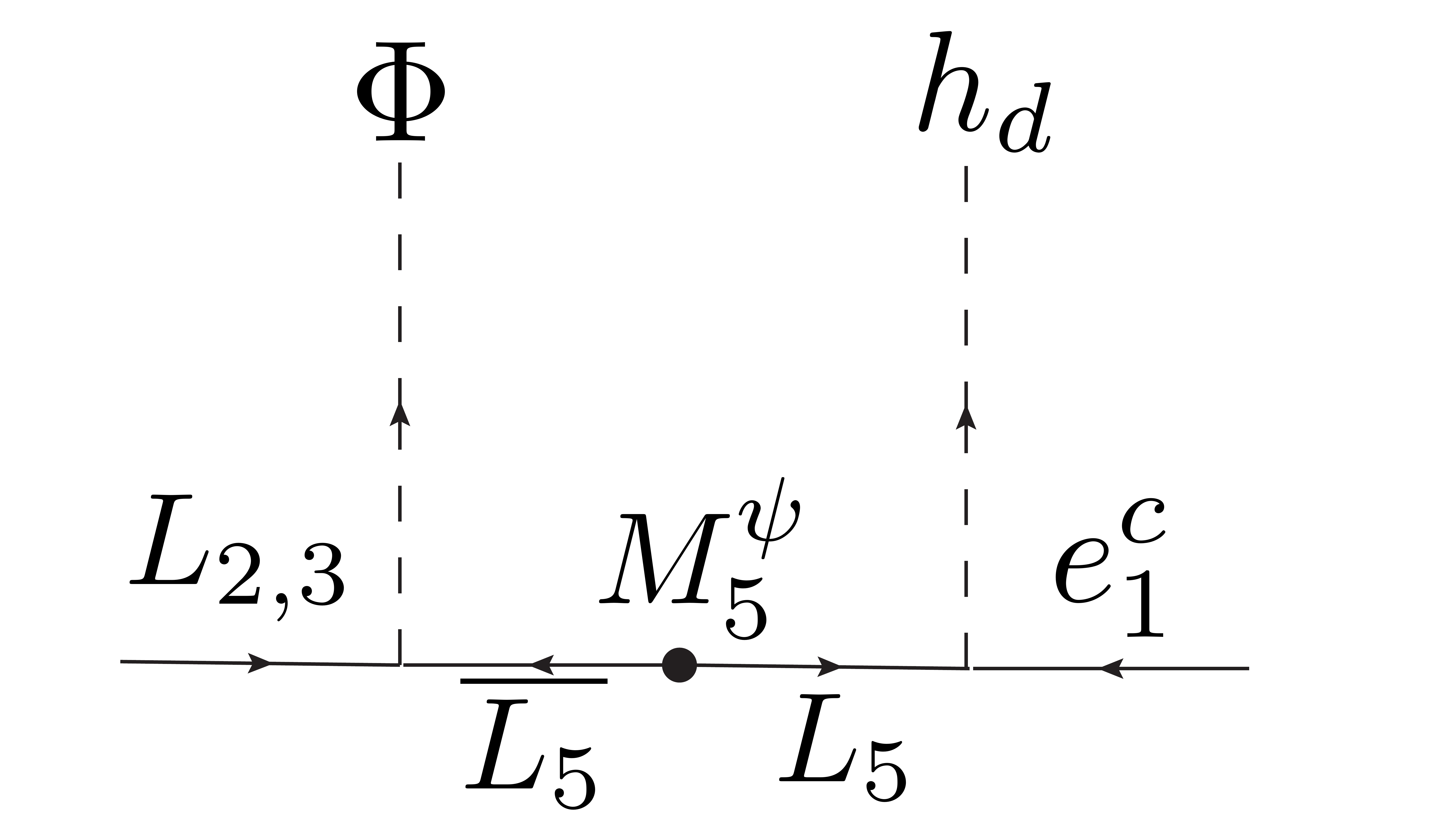}
	\includegraphics[scale=0.056]{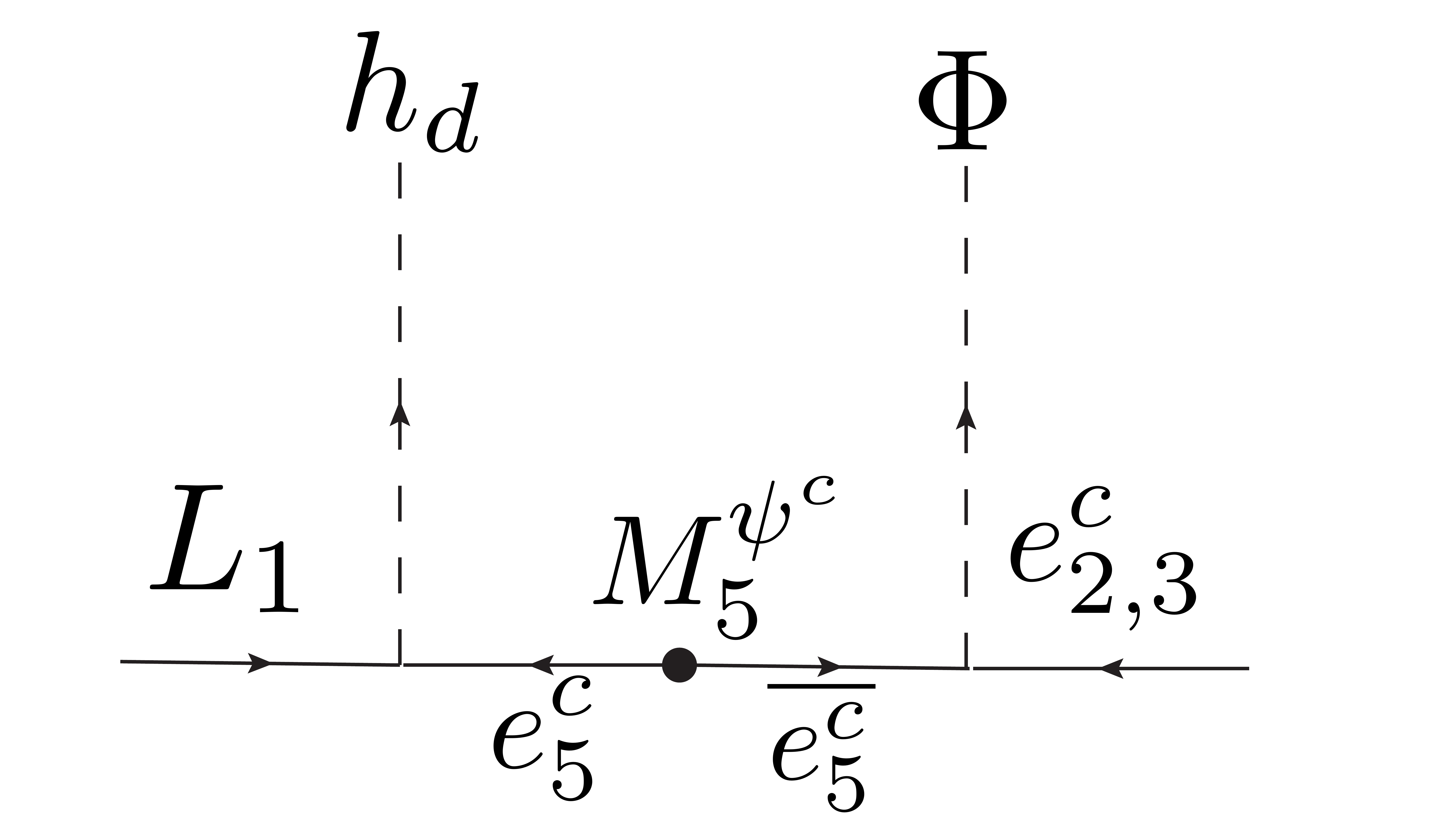}\vspace{0.15in}
	\includegraphics[scale=0.056]{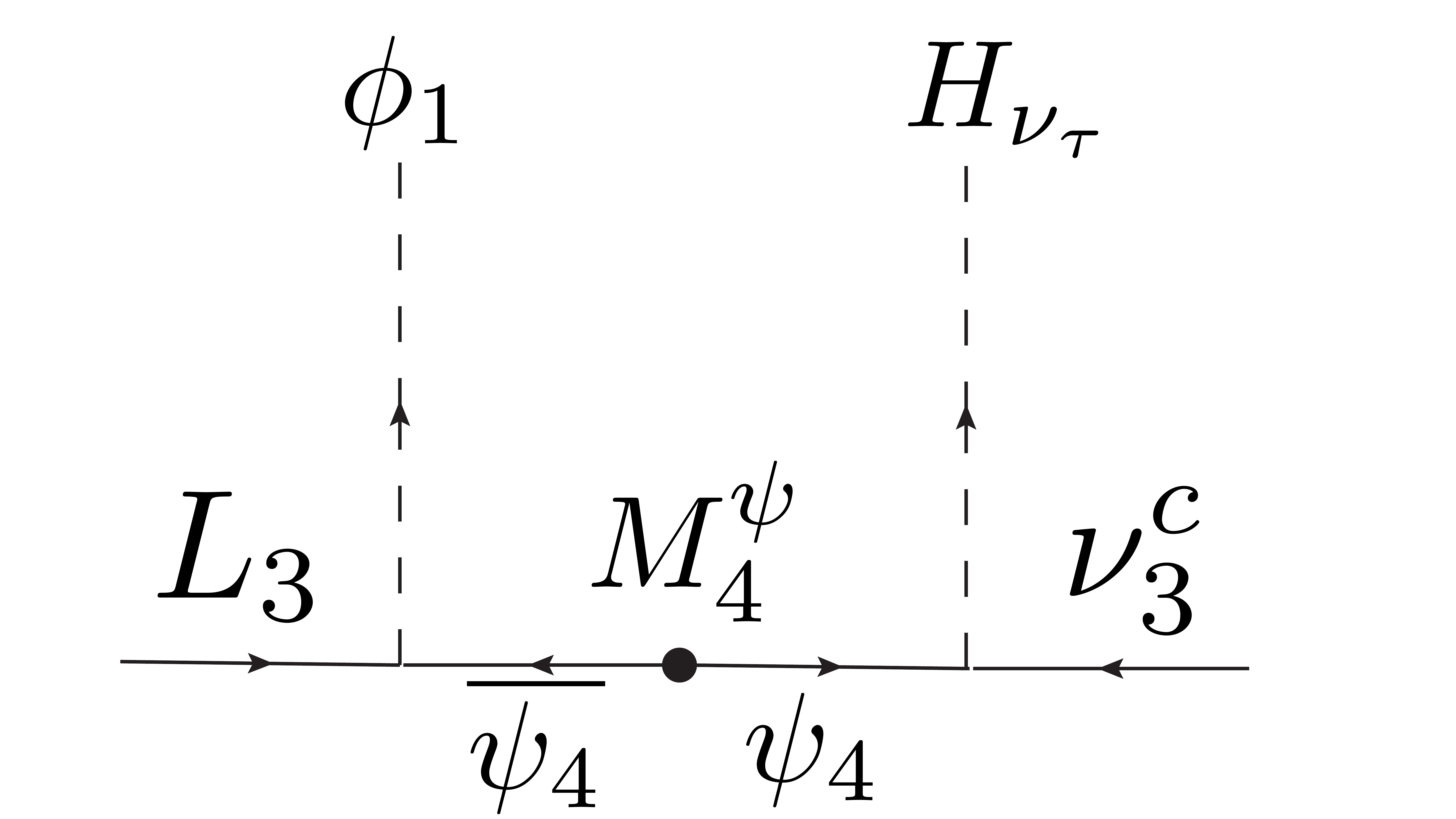}
	\includegraphics[scale=0.056]{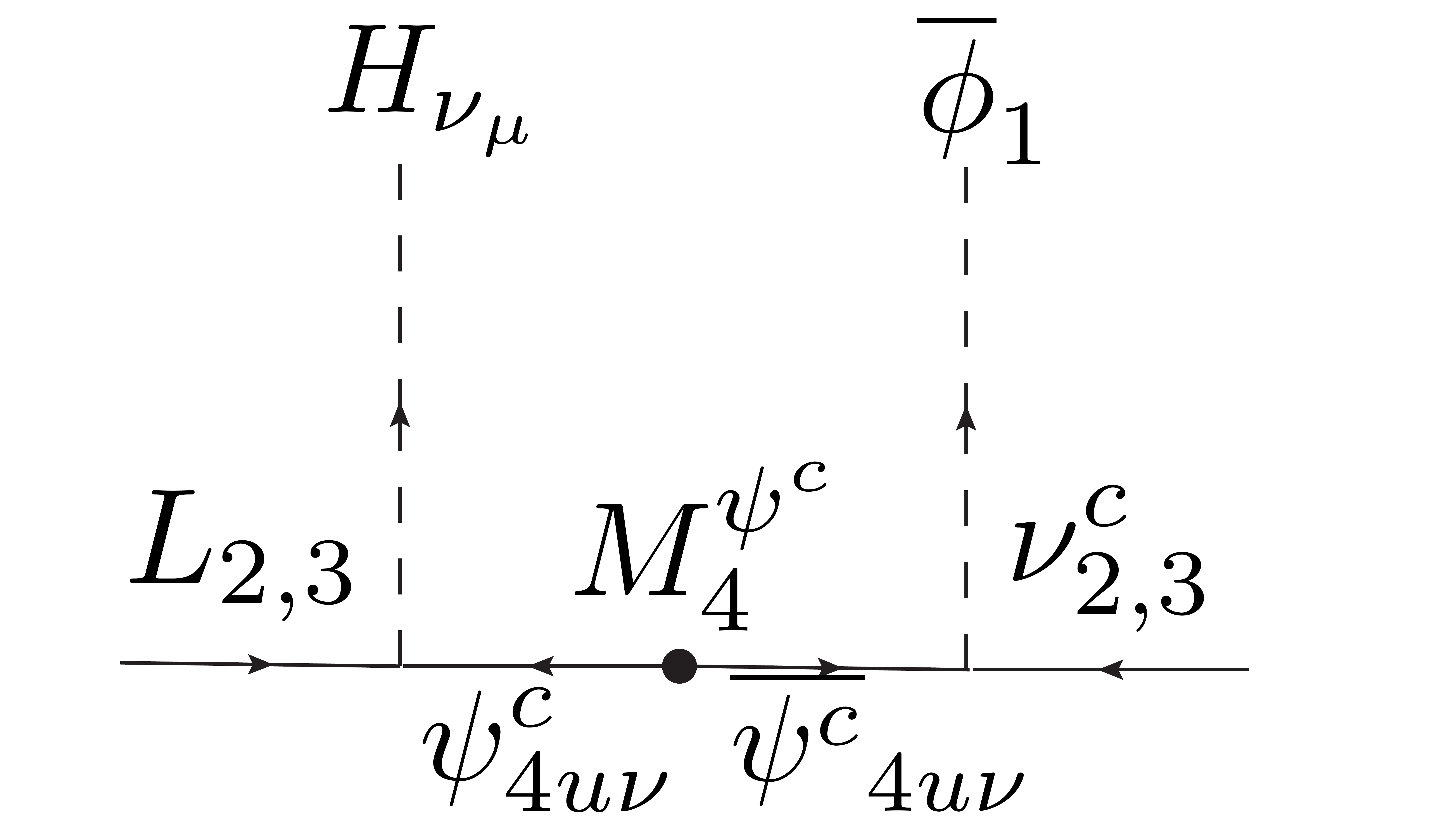}
	\includegraphics[scale=0.056]{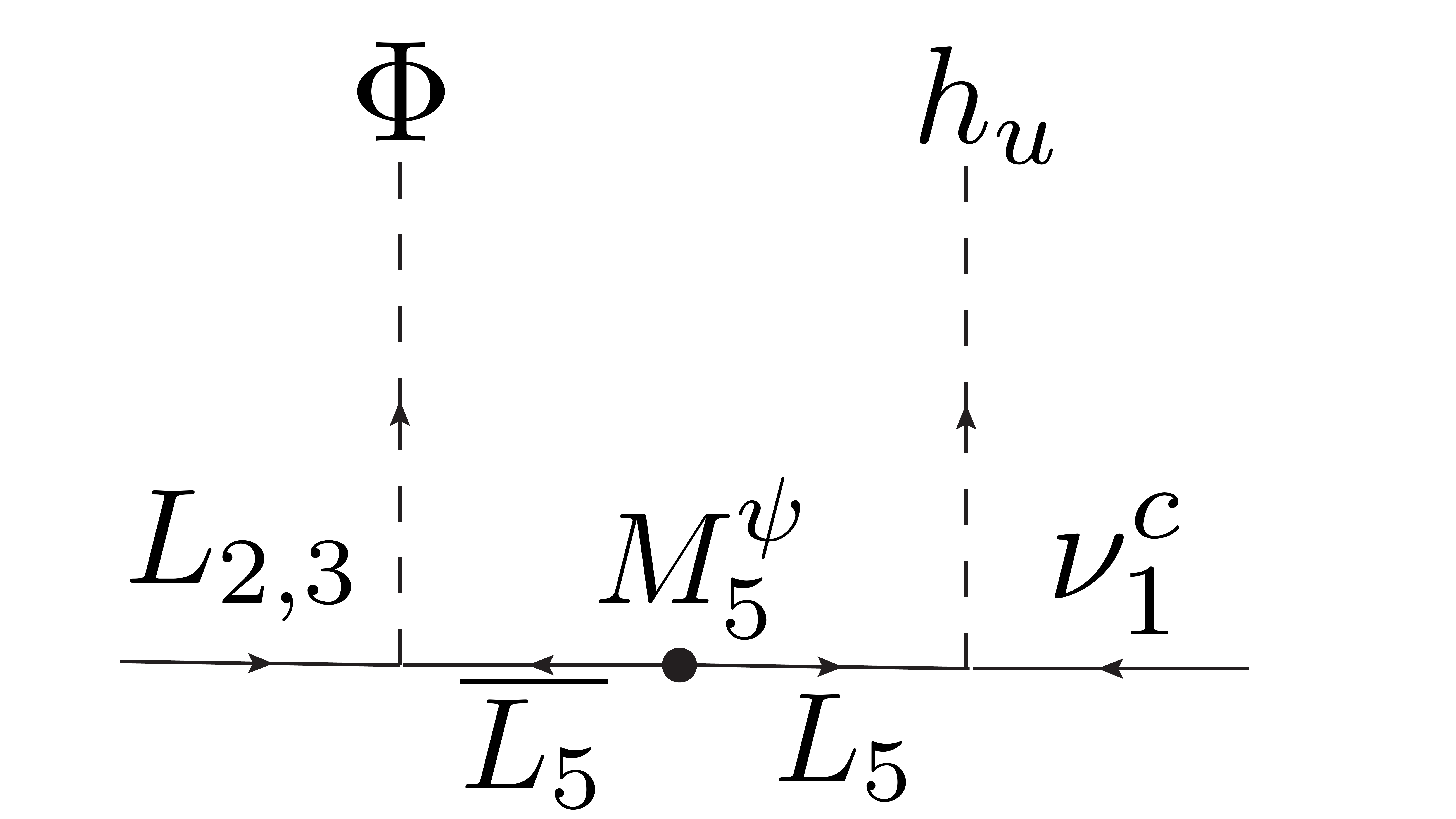}
	\includegraphics[scale=0.056]{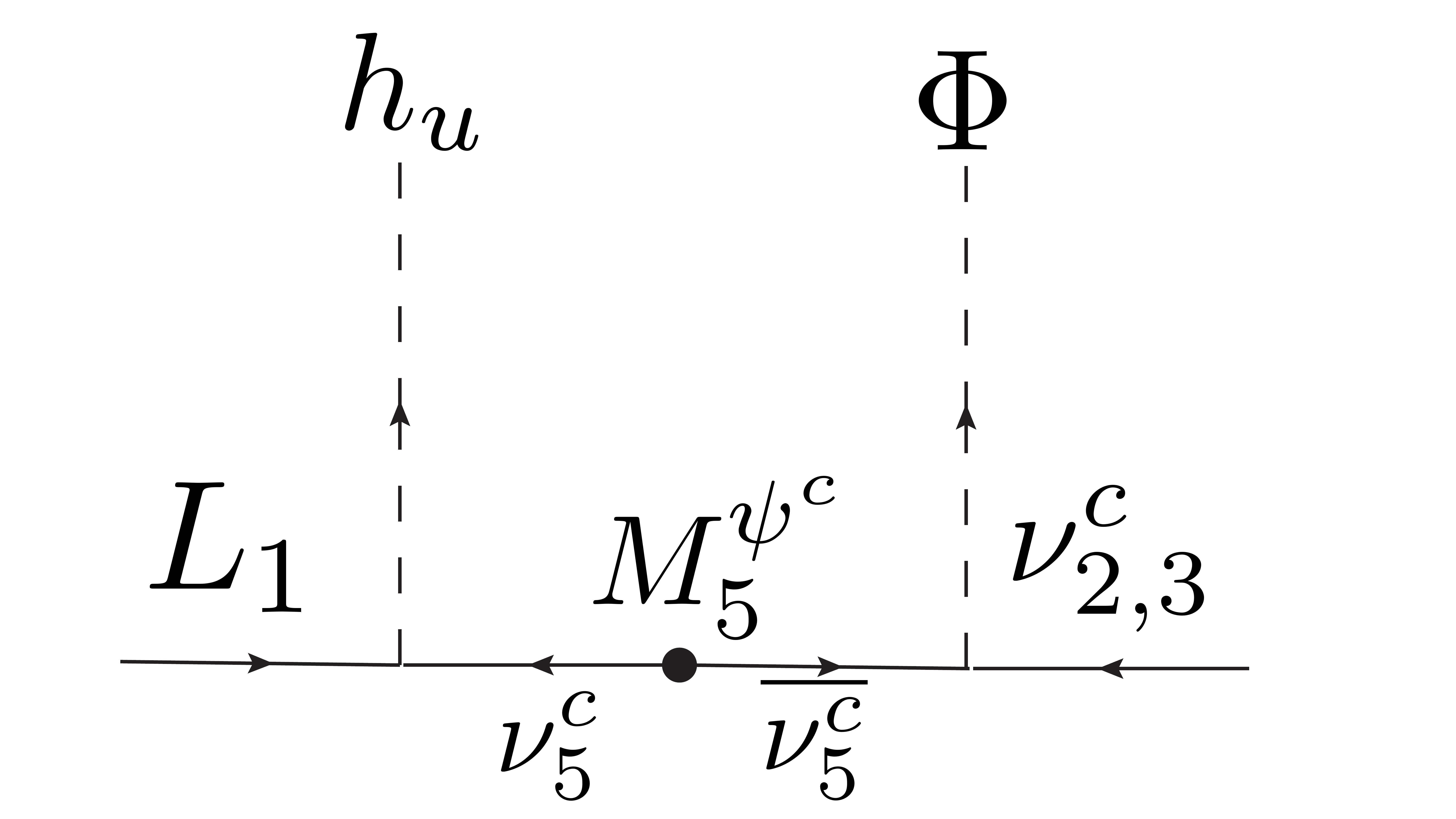}
			\caption{Diagrams which yield the mass matrices for all quarks and leptons in the low energy $G_{4321}$ theory,
			arising from the decomposition of Figs.~\ref{Fig1},\ref{Fig2}. The couplings respect $G_{4321}$ according to the assignments in Table~\ref{tab:funfields2}.
			Each row of diagrams represents a particular charged fermion, $u,d,e,\nu$, generating the effective mass matrices in Eqs.\ref{Mu},\ref{Md},\ref{Me},\ref{Mnu}, respectively, 
			 The columns of diagrams generate the entries in the mass matrix proportional to $A,B,C,D$, respectively.
			Note the 8 independent personal Higgs fields $H_t,\ldots, H_{\nu_{\mu}} $
			in the first two columns of diagrams, associated with the 3rd and 2nd familes.}
\label{Fig3}
\end{figure}

The effective operator matrix in Eqs.\ref{Mpsi} decomposes under the gauge group $G_{4321}$ into separate quark and lepton operator matrices (which yield mass matrices after the scalars get their VEVs),
\begin{align}
M_{u}&=A\frac{  {\phi}_3  }{M^{\psi}_{4}} {H}_{t}
+ B\frac{\overline{\phi}_3  }{M^{\psi^c}_{4}} {H}_{c} 
+C\frac{ {\Phi}  }{M^{\psi}_{5}} {h_u}
+D\frac{  {\Phi} }{M^{\psi^c}_{5}} {h_u}
\label{Mu}\\
M_{d}&=A\frac{  {\phi}_3  }{M^{\psi}_{4}} {H}_{b}
+ B\frac{\overline{\phi}_3  }{M^{\psi^c}_{4}} {H}_{s} 
+C\frac{ {\Phi} }{M^{\psi}_{5}} {h_d} 
+D\frac{  {\Phi} }{M^{\psi^c}_{5}}{h_d} 
\label{Md}\\
M_{e}&=A\frac{  {\phi}_1  }{M^{\psi}_{4}} {H}_{\tau}
+ B\frac{\overline{\phi}_1  }{M^{\psi^c}_{4}} {H}_{\mu} 
+C\frac{ {\Phi} }{M^{\psi}_{5}} {h_d} 
+D\frac{  {\Phi} }{M^{\psi^c}_{5}} {h_d}
\label{Me}\\
M_{\nu}^D&=A\frac{  {\phi}_1  }{M^{\psi}_{4}} {H}_{\nu_{\tau}}
+ B\frac{\overline{\phi}_1  }{M^{\psi^c}_{4}} {H}_{\nu_{\mu}} 
+C\frac{ {\Phi} }{M^{\psi}_{5}} {h_u} 
+D\frac{ {\Phi} }{M^{\psi^c}_{5}}{h_u} 
\label{Mnu}
\end{align}
where $A, B,C, D$ are the universal matrices in Eq.\ref{ABCD} with dimensionless elements of order unity
which preserve the unified structure. The diagrams responsible for these operators given in Fig.~\ref{Fig3},
show that $A$ only has a non-zero (3,3) element, $B$ only has non-zero elements in the (2,3) block, while
$C$ and $D$ only contribute to the first column and row, whilst maintaining a zero in the (1,1) element of the mass matrices.
Consequently, as discussed further below Eq.\ref{hierarchy2},
the $A$ and $B$ terms are mainly responsible for the third and second family quark and lepton masses, respectively, while the $C,D$ terms are responsible for the first family masses, which vanish without them. 
Hence we observe from Eqs.\ref{Mu}-\ref{Mnu} that 
there are eight personal Higgs scalar doublets associated with each of the fermion masses of the third and second families:
\begin{align}
m_t\leftrightarrow H_t,\ \ \ \  m_b\leftrightarrow H_b,\ \ \ \  m_{\tau}\leftrightarrow H_{\tau},\ \ \ \ m^D_{\nu_{\tau}}\leftrightarrow H_{\nu_{\tau}},\\
m_c\leftrightarrow {H}_c,\ \ \ \  m_s\leftrightarrow {H}_s,\ \ \ \  m_{\mu}\leftrightarrow {H}_{\mu},\ \ \ \  
m^D_{\nu_{\mu}}\leftrightarrow {H}_{\nu_{\tau}}.
\end{align}
where the neutrino masses shown above are the Dirac (D) masses, not the physical neutrino masses ($M_{\nu}^D$ in Eq.\ref{Mnu}
being the Dirac mass matrix).
As regards the first family masses, the situation 
 is similar to a 2HDM of type II, with $h_u$ contributing to the up quark mass and first family neutrino Dirac mass, while
$h_d$ contributes to the down quark mass and the electron mass.

The high energy gauge group only allows two different VL masses 
$M^{\psi}_{4}$, $M^{\psi^c}_{4}$, 
according to Eq.\ref{L4}, even after the Higgs $H,\overline{H}$ each split into 4 multiplets, and 
the VL fermions $\psi^c_4,\overline{\psi^c_4}$ each split into two multiplets.
In principle, $SU(2)_R$ symmetry breaking effects could split the $M^{\psi^c}_{4}$ masses into separate masses for 
$M^{\psi^c}_{4u\nu}$ and $M^{\psi^c}_{4ed}$, however such splitting does not happen due to the fact that $H',\overline{H'}$
do not couple to them, since the fourth family transforms under the first PS group, while $H',\overline{H'}$ transform under 
the second PS group. Furthermore the low energy PS group $SU(4)_{PS}^I$ strictly ensures that the quark and lepton components 
of the fourth family masses remain equal.

However, the fifth family masses $M^{\psi}_{5}$, $M^{\psi^c}_{5}$ can be split below the high energy PS breaking scale
by couplings to $H',\overline{H'}$, which share the second $SU(4)_{PS}^{II}$ gauge group, as seen in Table~\ref{twinPS}
\footnote{Note that there is no $SU(2)_R$ splitting since the fifth family transforms under $SU(2)_R^I$ while $H',\overline{H'}$ transform under
 $SU(2)_R^{II}$.}. 
Thus we may consider the following operators, with $SU(4)^{II}_{PS}$ indices shown explicitly,
\be
\frac{({H'}\overline{H'})_{\alpha}^{\beta}}{\Lambda}(\psi_5)^{\alpha}(\overline{\psi_5})_{\beta}+
\frac{({H'}\overline{H'})_{\alpha}^{\beta}}{\Lambda}(\psi^c_5)^{\alpha}(\overline{\psi^c_5})_{\beta}
\ee
where $\Lambda$ is a high scale cut-off above the high scale PS breaking scale. If we assume that 
two Higgs fields combine into an adjoint of $SU(4)^{II}_{PS}$, as discussed in \cite{King:1994he},
 \be
 ({H'}\overline{H'})_{\alpha}^{\beta}\sim {H'}_{\alpha}\overline{H'}^{\beta}-\frac{1}{4}{H'}^{\gamma}\overline{H'}_{\gamma}\delta_{\alpha}^{\beta}
 \ee
then after the Higgs fields receive the VEVs in Eq.\ref{HpVEV}, this results in quark-lepton mass splittings proportional to the generator $T^{II}_{B-L}=(1/\sqrt{6}){\rm diag}(1,1,1,-3)$
leading to different contributions to the fifth family quarks and leptons \cite{King:1994he},
 \be
\frac{\langle \nu_{H'}\rangle  \langle \bar{\nu}_{H'}\rangle}{\Lambda}(Q_5\overline{Q_5}-3L_5\overline{L_5})+
\frac{\langle \nu_{H'}\rangle  \langle \bar{\nu}_{H'}\rangle}{\Lambda}(u^c_5\overline{u^c_5}+d^c_5\overline{d^c_5}
-3\nu^c_5\overline{\nu^c_5}-3e^c_5\overline{e^c_5})
\label{fifthmasssplitting}
\ee
If the mass terms in Eq.\ref{fifthmasssplitting} dominate over the original mass terms, then they 
can be responsible for the smallness of the electron mass compared to the down quark mass,
which otherwise would be degenerate since their $C,D$ terms in Eqs.\ref{Md},\ref{Me} share the same Higgs fields and are otherwise identical.
As discussed earlier, we also have a texture zero in the first element of the fermion mass matrices, 
as motivated by the observed quark mixing relation in Eq.\ref{Vus}.

\subsection{Low scale symmetry breaking of $G_{4321}$ to the SM}
\label{2.4}

In this subsection we shall discuss the low scale symmetry breaking
\begin{align}
 G_{4321} \stackrel{M_{low}}{\longrightarrow} G_{321}
 \label{4321to321}
\end{align}
To achieve this symmetry breaking we shall use 
the scalar field $\phi (4,\bar{4},1,1)$ in Table~\ref{tab:funfields1} 
which decomposes to 
$\phi_3 ({4},\bar{3},1,-1/6)$ and $\phi_1 ({4},1,1,1/2)$ in Table~\ref{tab:funfields2}, and similarly for $\overline{\phi}$, $\overline{\phi'}$.
These fields are assumed to develop low scale VEVs,
\beq  
\label{vevconf}
\vev{\phi_3} = 
\left(
\begin{array}{ccc}
\tfrac{v_3}{\sqrt{2}} & 0 & 0 \\
0 & \tfrac{v_3}{\sqrt{2}} & 0 \\ 
0 & 0 & \tfrac{v_3}{\sqrt{2}} \\
0 & 0 & 0
\end{array}
\right) \, , \ \ 
\vev{\phi_1} = 
\left(
\begin{array}{c}
0 \\ 
0 \\ 
0 \\
\tfrac{v_1}{\sqrt{2}}
\end{array}
\right) \, ,
\eeq
and analogously for $\overline{\phi}_{1,3}$, $\overline{\phi'}_{1,3}$  where we assume relatively low scale VEVs,
\begin{equation}
v_1 \sim v_3  \lesssim 1 \ {\rm TeV}, \ \ 
\label{HVEV}
\end{equation}
and similarly for $\bar{v}_1,\bar{v}_3$, $\bar{v}'_1,\bar{v}'_3$,
leading to the symmetry breaking of $G_{4321}$ down to the SM gauge group $G_{321}$,
\be
SU(4)_{PS}^{I}\times SU(3)_c^{II}\times SU(2)^{I+II}_L\times U(1)_{Y'} \rightarrow SU(3)_c\times SU(2)_L\times U(1)_{Y}
\ee
The $SU(4)_{PS}^{I}$ is broken to $SU(3)_{c}^{I}\times U(1)^{I}_{B-L}$ ($4\rightarrow 3_{1/6}+1_{-1/2}$),
with $SU(3)_{c}^{I}\times SU(3)_c^{II}$ broken to the diagonal subgroup $SU(3)_c^{I+II}$ identified as SM QCD $SU(3)_c$.
We identify $SU(2)^{I+II}_L$ as the SM EW group $SU(2)_L$.
The Abelian generators are broken to SM hypercharge $U(1)_{Y}$ where
\be
Y=T^{I}_{B-L}+Y' = T^{I}_{B-L} +T^{II}_{B-L}+T_{3R}
\ee

The physical massive scalar spectrum includes a real color octet, three SM singlets and 
a complex scalar transforming as $(3,1,2/3)$. 
The heavy gauge bosons include a vector leptoquark $U_1^{\mu} = (3,1,2/3)$ from $SU(4)_{I}\rightarrow SU(3)_{I}\times U(1)^{I}_{B-L}$,
a heavy gluon $g'_{\mu} = (8,1,0)$ from $SU(3)_{I}\times SU(3)_{II}\rightarrow SU(3)_{I+II}$
and a $Z'_{\mu} = (1,1,0)$ from $U(1)^{I}_{B-L}\times U(1)_{Y'} \rightarrow U(1)_{Y}$.  

The heavy gauge boson masses resulting from the symmetry breaking in Eq.\ref{4321to321}
are generalisations of the results in \cite{Diaz:2017lit},
\begin{align}
\label{MV}
M_{U_1} &= \tfrac{1}{2} g_4 \sqrt{v_1^2 +  \bar{v}_1^2 +  \bar{v'}_1^2+ v_3^2+ \bar{v}_3^2+ {\bar{v'}}_3^2} \, , \\
\label{Mgp}
M_{g'} &= \tfrac{1}{\sqrt{2}} \sqrt{g_4^2 + g_3^2}  \sqrt{v_3^2 + \bar{v}_3^2 + {\bar{v'}}_3^2}\, , \\
\label{MZp}
M_{Z'} &= \tfrac{1}{2} \sqrt{\tfrac{3}{2}} \sqrt{g_4^2 + \tfrac{2}{3} g_1^2}\sqrt{v_1^2 + \bar{v}_1^2 + {\bar{v'}}_1^2
+ \tfrac{1}{3} v_3^2 + \tfrac{1}{3} \bar{v}_3^2+ \tfrac{1}{3} {\bar{v'}}_3^2} 
\, .
\end{align}

Under the breaking in Eq.\ref{4321to321}
to the SM gauge group, the fourth VL family in Table~\ref{tab:funfields2}
decomposes into fermions with the usual SM quantum numbers of the 
chiral quarks and leptons in Eq.\ref{psi},\ref{psic},
but including partners in conjugate representations,
\begin{align}
\psi_4 &\rightarrow (Q_4,L_4)\equiv (Q_{L4},L_{L4}), \ \ \ \ 
\overline{\psi_4}\rightarrow (\overline{Q_4},\overline{L_4}) \stackrel{CP}{\rightarrow} (Q_{R4},L_{R4}) ,\\
\psi^c_{4u\nu}&\rightarrow (u^c_4,\nu^c_4) \stackrel{CP}{\rightarrow} (u_{R4},\nu_{R4}), \ \ \ \ 
\overline{\psi^c}_{4u\nu}\rightarrow (\overline{u^c_4},\overline{\nu^c_4})\equiv  (u_{L4},\nu_{L4}),\\
\psi^c_{4ed}&\rightarrow (d^c_4,e^c_4) \stackrel{CP}{\rightarrow}  (d_{R4},e_{R4}), \ \ \ \ 
\overline{\psi^c}_{4ed}\rightarrow (\overline{d^c_4},\overline{e^c_4})\equiv  (d_{L4},e_{L4}),
\end{align}
where we have converted to left (L) and right (R) convention in the last step, either by a simple equivalence, or using a $CP$ transformation where applicable.
Similarly we shall write the three chiral familes of quarks and leptons in L, R convention as,
\be
Q_i, L_i \equiv  Q_{Li}, ,L_{Li}, \ \ \ \  u^c_j, d^c_j,  e^c_j, \nu^c_j  \stackrel{CP}{\rightarrow} u_{Rj},d_{Rj},e_{Rj},\nu_{Rj}, \ \ (i,j=1,2,3)
\ee
The heavy gauge bosons $U_1,g',Z'$ couple to the chiral fermions 
and VL fourth family fermions with left-handed interactions~\cite{DiLuzio:2017vat},
\begin{align}
& \frac{g_4}{\sqrt{2}}  \left( \bar{Q}_{L4} \gamma^\mu L_{L4} + \textrm{H.c.} \right) U_{1\mu} \nonumber \\
 + & \frac{g_4 g_s}{g_3} \left( \bar{Q}_{L4} \gamma^\mu T^a Q_{L4} - \frac{g_3^2}{g_4^2} \,
\bar{Q}_{Li} \gamma^\mu T^a Q_{Li} \right) g'^a_\mu \label{L} \\
+\frac{\sqrt{3} \,g_4 g_Y}{\sqrt{2} \,g_1} & \left( \frac{1}{6} \bar{Q}_{L4} \gamma^\mu Q_{L4} -\frac{1}{2} \bar{L}_{L4} \gamma^\mu L_{L4} 
- \frac{g_1^2}{9 g_4^2} \,\bar{Q}_{Li} \gamma^\mu Q_{Li}
+ \frac{ g_1^2}{3 g_4^2} \,\bar{L}_{Li} \gamma^\mu 
L_{Li} \right) Z'_\mu \nonumber
\end{align}
and right-handed interactions,
\begin{align}
& \frac{g_4}{\sqrt{2}} \left(\bar{Q}_{R4} \gamma^\mu L_{R4} + \textrm{H.c.}\right)U_{1\mu} \nonumber \\
 + & \frac{g_4 g_s}{g_3} \left( \bar{Q}_{R4} \gamma^\mu T^a Q_{R4} - \frac{g_3^2}{g_4^2} \,\left(\bar{u}_{Rj} \gamma^\mu T^a u_{Rj} 
+ \bar{d}_{Rj} \gamma^\mu T^a d_{Rj} \right) \right) g'^a_\mu \ \   + \label{R} \\
\frac{\sqrt{3} g_4 g_Y}{\sqrt{2} g_1} &
\left( \frac{1}{6} \bar{Q}_{R4} \gamma^\mu Q_{R4} -\frac{1}{2}  \bar{L}_{R4} \gamma^\mu L_{R4} 
- \frac{2 g_1^2}{9 g_4^2} \left ( 2 \bar{u}_{Rj} \gamma^\mu u_{Rj} - \bar{d}_{Rj} \gamma^\mu d_{Rj} -3\bar{e}_{Rj} \gamma^\mu e_{Rj} \right)
 \right) Z'_\mu \nonumber
\end{align}
where the SM gauge couplings of $SU(3)_c$ and $U(1)_Y$ are given by~\cite{DiLuzio:2017vat},
\beq 
\label{matchinggsgY}
g_s = \frac{g_4 g_3}{\sqrt{g_4^2 + g_3^2}} \, , \qquad g_Y = \frac{g_4 g_1}{\sqrt{g_4^2 + \frac{2}{3} g_1^2}} \, , 
\eeq
where $g_{4,3,2,1}$ are the gauge couplings of $G_{4321}$.
In the above expressions we have ignored the mixing between the chiral fermions and the VL fermions,
and dropped the right-handed neutrino couplings. We have also dropped the EW singlet VL fourth family couplings, 
assuming them to be much heavier that the EW doublets, $M^{\psi^c}_4\gg M^{\psi}_4$.

In the present model, all flavour changing is generated from Eq.\ref{L} after making the rotation as in Eq.\ref{34matrix},
\be
\pmatr{
Q_{L3}\\
Q_{L4}
}
\rightarrow
\pmatr{
c^{Q}_{34}&s^{Q}_{34}\\
-s^{Q}_{34}&c^{Q}_{34}
}
\pmatr{
Q_{L3}\\
Q_{L4}
}, \ \ \ \ 
\pmatr{
L_{L3}\\
L_{L4}
}
\rightarrow
\pmatr{
c^{L}_{34}&s^{L}_{34}\\
-s^{L}_{34}&c^{L}_{34}
}
\pmatr{
L_{L3}\\
L_{L4}
}
\label{34}
\ee
where the large mixing angles
$s^Q_{34}, s^L_{34} $ were introduced in Eq.\ref{largeangle}, beyond the mass insertion approximation, and 
all other mixing with the fourth VL family is suppressed by small mixing angles in this model.
The transformation in Eq.\ref{34} leads to non-universal third family terms in Eq.\ref{L}.
The further CKM type transformations required to diagonalise the quark and lepton mass matrices predicted by the model (see later),
then lead to flavour changing operators originating from the non-universal third family terms.

\begin{figure}[ht]
\centering
	\includegraphics[scale=0.10]{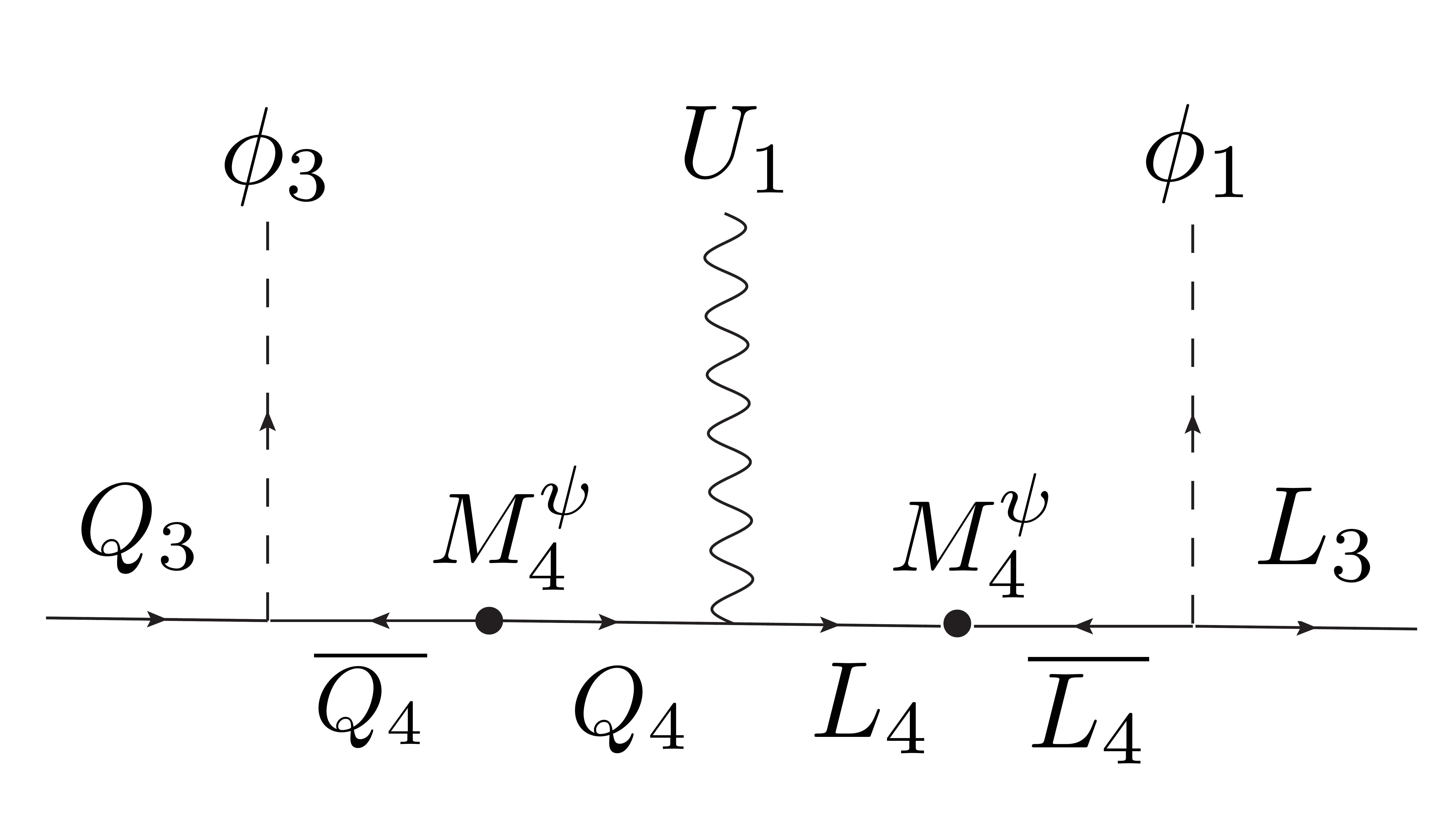}\ \ 
	\includegraphics[scale=0.10]{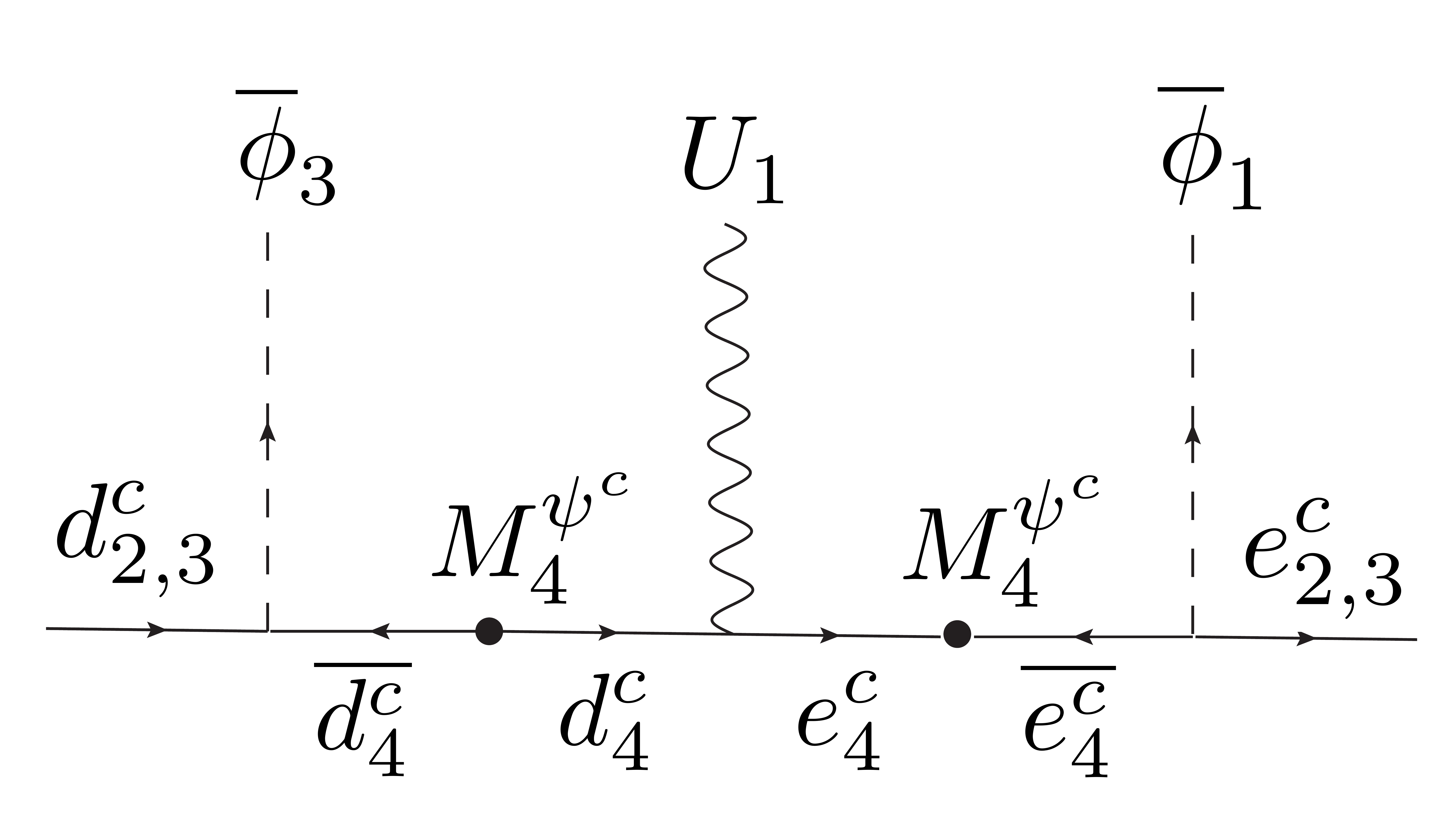}
\caption{Diagrams in the model which leads to the effective $U_1$ vector leptoquark couplings 
to quarks to leptons in the mass insertion approximation. The left (right) panels show the left (right) handed couplings.
The diagram in the left-panel will dominate since $M^{\psi}_4\ll M^{\psi^c}_4$, leading to the 
approximate effective operator in Eq.\ref{U1op}. }
\label{Fig4}
\end{figure}

The key feature of the heavy gauge boson couplings is that, while the heavy gluon $g'_{\mu}$ and the $Z'_{\mu}$ couple to all 
chiral and VL quarks and leptons, the heavy vector leptoquark $U_1^{\mu}$ only couples to the fourth family VL fermions in the original 
basis of Eqs.\ref{L} and \ref{R}.
The reason is that $U_1^{\mu}$ originates entirely from $SU(4)_{I}$, which remains unbroken to low scales, and 
under which the chiral quarks and leptons are singlets.
In the present model, effective $U_1^{\mu}$ vector leptoquark couplings to chiral quarks and leptons can be generated from 
Eq.\ref{L},
after the rotations in 
Eq.\ref{34}, leading to the effective operator,
\begin{align}
 \frac{g_4}{\sqrt{2}}  \bar{Q}_{L4} \gamma^\mu L_{L4} \rightarrow 
\frac{g_4}{\sqrt{2}} s^Q_{34} s^L_{34} 
\bar{Q}_{L3} \gamma^\mu L_{L3} \, U_{1\mu} 
\approx \frac{g_4}{\sqrt{2}} 
 \frac{x^{\psi}_{34} {\langle \phi_1 \rangle } }{M^{\psi}_{4}} \frac{x^{\psi}_{34} {\langle \phi_3 \rangle } }{M^{\psi}_{4}}
\bar{Q}_{L3} \gamma^\mu L_{L3} \, U_{1\mu} 
\label{U1op}
\end{align}
plus H.c., where we have also shown the result in mass insertion approximation 
from the diagrams in Fig.~\ref{Fig4}, with the left-hand diagram dominating due to $M^{\psi}_4\ll M^{\psi^c}_4$ from Eq.\ref{hierarchy2}.
Equivalently, the dominance of this operator follows from the large third family quark and lepton masses which imply large mixing angles
$s^Q_{34}, s^L_{34}$. Similar operators involving 
right-handed couplings to the second family, arising from the right-hand diagram in Fig.~\ref{Fig4},
will be suppressed.
Since the first family quarks and leptons only couple to fifth family VL fermions, which do not interact at all with $U_1^{\mu}$, similar operators involving the first family will be absent. 

The operator in Eq.\ref{U1op} has the right structure of vector leptoquark $U_1^{\mu}$ couplings to account for the $B$-physics anomalies
in  $R_{K^{(*)}}$ and $R_{D^{(*)}}$ as discussed in many papers mentioned in the Introduction.
For example, according to the analysis in \cite{Buttazzo:2017ixm}, a single operator as in Eq.\ref{U1op}, involving only the third family doublets,
can account for both the anomalies simultaneously, once the further transformations required to diagonalise the quark and lepton mass matrices are taken into account, leading to, in the notation of \cite{Buttazzo:2017ixm},
\begin{align}
\frac{g_4}{\sqrt{2}} s^Q_{34} s^L_{34} 
\bar{Q}_{L3} \gamma^\mu L_{L3} \, U_{1\mu} \equiv
g_U
\bar{Q}_{L3} \gamma^\mu L_{L3} \, U_{1\mu}  \rightarrow 
g_U \beta_{i\alpha}
\bar{Q}_{Li} \gamma^\mu L_{L\alpha} \, U_{1\mu} 
\label{U1op2}
\end{align}
In the effective field theory analysis of \cite{Buttazzo:2017ixm} these further transformations were regarded as 
relatively free parameters with good global fits obtained for 
$\beta_{s\tau}\approx 4|V_{cb}|$, with $\beta_{b\mu} < 0.5$ and $\beta_{s\mu}<  5|V_{cb}|$
constrained to lie on narrow contours \cite{Buttazzo:2017ixm}. However in the present model 
the quark and lepton mass matrices are predicted, and the natural expectation is that these mixing 
parameters are of order $|V_{cb}|$, as we shall discuss later.
The values of $g_U$ and $M_{U_1}$ are also constrained by the global fit to the B physics anomalies \cite{Buttazzo:2017ixm}, 
for example  $g_U\approx1.1$ and $M_{U_1}\approx 1.6$ TeV provides a good fit
consistent with LHC searches, and corresponds to the benchmark point discussed in the next subsection for $s^Q_{34} \approx s^L_{34}\approx 1/\sqrt{2} $.

\subsection{Flavour Changing Neutral Currents (FCNCs)}
\label{FCNC}
The heavy gauge bosons $g',Z'$ will generate FCNCs from the couplings in Eq.\ref{L}, after the rotation in Eq.\ref{34} followed by the rotations required to diagonalise the quark and lepton mass matrices.  A detailed analysis of FCNCs in the $G_{4321}$ model has been recently performed in \cite{Cornella:2021sby}, but here in this subsection we summarise the key issues which are relevant for the twin PS model.

The first observation is that in Eqs.\ref{L}, \ref{R} the first three families of quarks and leptons all couple equally to $g',Z'$ for the three families of a given charge. This means that the rotations used to go to the basis in Eq.\ref{M^psi_an_2} will not induce any flavour violation. However, the couplings of the fourth vector-like fermions to $g',Z'$ are non-universal, so any mixing of the three families with the fourth family will induce non-universality in the light states. In the twin PS model, there is only significant mixing of the third left-handed chiral family with the fourth family, 
and it is a good approximation to only consider the rotation in Eq.\ref{34}. 

After the rotations in Eq.\ref{34}, the Lagrangian in Eq.\ref{L} will generate non-universal $g',Z'$ couplings to the third family quark and lepton doublets, while the first two families continue to have universal couplings to good approximation. This is equivalent to an approximate $U(2)^5$ global symmetry which will protect against the most dangerous FCNCs involving the first two families. However the non-universal third family doublet couplings will lead to FCNCs once the quark and lepton mass matrices (considered in detail in the next section) are diagonalised. Fortunately these matrices turn out to have small off-diagonal elements, so FCNCs are suppressed. For example, there will be tree-level FCNCs arising in $B_s$ mixing suppressed by $V_{ts}\sim 0.04$.

Consider the example of benchmark parameters~\cite{DiLuzio:2017vat}:
$v_1\approx \bar{v}_1\approx \bar{v}'_1\approx 312$ GeV, $v_3\approx \bar{v}_3\approx \bar{v}'_3\approx 488$ GeV, 
$g_4 \approx 3$, $g_3 \approx g_s \approx 1$,
$g_1 \approx g_Y \approx 0.36$, which leads to 
$M_{Z'} \approx  1.4$~TeV, $M_{U_1} \approx 1.6$~TeV, and $M_{g'} \approx  2.0$~TeV.
This set of parameters has the typical feature that $g_4\gg g_3,g_1$ so that the heavy gauge bosons $g',Z'$ have suppressed couplings to 
light quarks and leptons, according to Eqs.\ref{L},\ref{R}, which will inhibit the direct production of these states at the LHC. 

As discussed above, 
the fourth family doublets with large couplings to $g',Z'$ will generate non-universal third family couplings to these gauge bosons,
after the replacements in Eq.\ref{34}. These non-universal third family couplings to $g',Z'$ will subsequently 
lead to tree-level FCNCs following the transformations required to diagonalise the $3\times 3$ quark and lepton mass matrices. 
The typical constraint from $B_s$ mixing \cite{DiLuzio:2019jyq} has the parametric form
\be
\frac{(s^{Q}_{34})^2V_{ts}}{M}\lesssim \frac{1}{220\ {\rm TeV}}
\ee
where $M$ represents the $g',Z'$ masses, whose benchmark values are $M\sim 1$ TeV. With $V_{ts}\sim 0.04$, this constraint is satisfied  
providing that $(s^{Q}_{34})^2\lesssim 0.1$, ignoring the other dimensionless couplings which are of order unity.

In addition to the couplings in Eq.\ref{U1op},
the rotations in Eq.\ref{34} will generate vector leptoquark $U_1$ interactions which couple the third and fourth family doublets,
\begin{align}
 \frac{g_4}{\sqrt{2}}  \bar{Q}_{L4} \gamma^\mu L_{L4} \rightarrow 
\frac{g_4}{\sqrt{2}} s^Q_{34} 
\bar{Q}_{L3} \gamma^\mu L_{L4} \, U_{1\mu} 
\label{U1op3}
\end{align}
Such couplings allow a one loop box diagram contribution to $B_s$ mixing proportional to the internal vector-like lepton mass squared \cite{DiLuzio:2018zxy}.
The mass of the vector-like lepton can be lowered by including an additional scalar $\Omega_{15}$ which transforms under 
$SU(4)^I_{PS}$ in the adjoint representation, whose VEV contributes to the fourth family masses \cite{DiLuzio:2018zxy}.

\subsection{Electroweak symmetry breaking }
\label{EWSB}

In this subsection we discuss electroweak (EW) symmetry breaking in this model.
The low energy Higgs 
fields originate from multiplets which transforms under $G_{4422}$ as
\begin{equation}
{H}(\bar{4},{4},{2},\bar{2}), \ \ \overline{H}(4,\bar{4},\bar{2},2)
\end{equation}
Under $G_{4422}\rightarrow G_{4321}$ these decompose into the personal Higgs scalar doublets as in Eq.\ref{H3},\ref{H2} and Table~\ref{tab:funfields2}.
Under the breaking to the SM gauge group $G_{4321}\rightarrow G_{321}$, the personal Higgs scalar doublets
further decompose into personal Higgs EW doublets, plus other colour and charge exotic doublets,
\begin{align}
\label{Ht}
{H}_{t}(\bar{4},{3},\bar{2},2/3) &\rightarrow  
{H}_{t}(1,\bar{2},1/2)+{H}_{t}(8,\bar{2},1/2)+{H}_{t}({3},\bar{2},7/6)
\\
\label{Hb}
{H}_b(\bar{4},{3},\bar{2},-1/3) &\rightarrow  
{H}_b(1,\bar{2},-1/2)+{H}_b(8,\bar{2},-1/2)+{H}_b({3},\bar{2},1/6)
\\
\label{Htau}
{H}_{\tau}(\bar{4},1,\bar{2},-1) &\rightarrow 
{H}_{\tau}(1,\bar{2},-1/2) +{H}_{\tau}(\bar{3},\bar{2},-7/6) 
 \\
\label{Hnutau}
{H}_{\nu_{\tau}}(\bar{4},1,\bar{2},0) &\rightarrow 
{H}_{\nu_{\tau}}(1,\bar{2},1/2)+{H}_{\nu_{\tau}}(\bar{3},\bar{2},-1/6)
\\
\label{Hc}
 {H}_{c}(4,\bar{3},\bar{2},1/3) &\rightarrow  
{H}_{c}(1,\bar{2},1/2)+{H}_{c}(8,\bar{2},1/2)+{H}_{c}(\bar{3},\bar{2},-1/6)
\\
\label{Hs}
{H}_s(4,\bar{3},\bar{2},-2/3) &\rightarrow  
{H}_s(1,\bar{2},-1/2)+{H}_s(8,\bar{2},-1/2)+{H}_s(\bar{3},\bar{2},-7/6)
\\
\label{Hmu}
{H}_{\mu}(4,1,\bar{2},0) &\rightarrow 
{H}_{\mu}(1,\bar{2},-1/2) +{H}_{\mu}(3,\bar{2},1/6) 
 \\
\label{Hnumu}
 {H}_{\nu_{\mu}}(4,1,\bar{2},1) &\rightarrow 
{H}_{\nu_{\mu}}(1,\bar{2},1/2)+{H}_{\nu_{\mu}}(3,\bar{2},7/6)
\end{align}
where the SM gauge group $G_{321}$ reps are shown.
Higgs VEVs may appear in the first eight EW doublets in Eqs.\ref{Ht}-\ref{Hnumu} which are both colour singlets and have electrically neutral components, together with the two EW doublets $h_u,h_d$ in Table~\ref{tab:funfields2},
leading to the familiar electroweak symmetry breaking $G_{321}\rightarrow G_{31}$,
\be
SU(3)_c\times SU(2)_L\times U(1)_{Y} \rightarrow SU(3)_c\times U(1)_Q
\ee
where the electric charge generator is given by the familiar result
\be
Q=T_{3L}+Y
\ee
Eqs.\ref{Ht}-\ref{Hnumu} predict 8 Higgs EW doublets, 4 colour octet scalar EW doublets, and 8 scalar EW doublets identified as leptoquarks,
the physical implications of the latter being briefly discussed in the next subsection.

Models with multiple light Higgs doublets face the phenomenological challenge of FCNCs arising from tree-level exchange of the EW scalar doublets in the Higgs basis. 
Therefore we need to assume that only one pair of Higgs doublets $H_u$ and $H_d$ are light, given by linear combinations of the 
EW doublets,
\begin{align}
H_u &=\tilde{\alpha}_u H_t +\tilde{\beta}_u H_c +\tilde{\gamma}_u H_{\nu_{\tau}}+\tilde{\delta}_u H_{\nu_{\mu}} 
+ \tilde{\varepsilon}_u h_u \nonumber \\
H_d &=\tilde{\alpha}_d H_b +\tilde{\beta}_d H_s +\tilde{\gamma}_d H_{\tau}+\tilde{\delta}_d H_{\mu} + \tilde{\varepsilon}_d h_d
\label{Higgs}
\end{align}
where $\tilde{\alpha}_{u,d},\tilde{\beta}_{u,d},\dots$ are complex elements of two unitary Higgs mixing matrices, 
$U_{H_u}$ and $U_{H_d}$.
The orthogonal linear combinations are assumed to be very heavy, well above the TeV scale in order to sufficiently suppress the FCNCs.
The situation is familiar from $SO(10)$ models \cite{Fukuyama:2012rw} 
where there are 6 Higgs doublets arising from the $10$, $120$ and $\overline{126}$ representations, 
denoted as $H_{10}$, $H_{120}$, $H_{\overline{126}}$, two from each,
but below the $SO(10)$ breaking scale only two Higgs doublets are assumed to be light,
similar to $H_u$ and $H_d$ above.

We further assume that only the light Higgs doublet states get VEVs,
\be
\langle H_u \rangle = v_u, \ \  \langle H_d \rangle= v_d,
\label{vuvd}
\ee
while the heavy linear combinations do not, i.e. we assume that in the Higgs basis the linear combinations which do not get VEVs are very heavy.
We shall not discuss the Higgs potential which is responsible for this, however we note that, as in $SO(10)$,
the general requirement is that the Higgs mass squared matrix of doublets must have an approximately zero determinant 
analogous to the case of
$SO(10)$ with $H_{10},H_{120},H_{\overline{126}}$~\cite{Fukuyama:2012rw}. Of course the requirement is not so severe as in $SO(10)$ due to the smaller hierarchy of mass scales required for acceptable FCNCs. Although the discussion of the Higgs potential is beyond the scope of this paper, we note that this would probably involve the following three features: additional Higgs fields; a discussion of renormalisation group (RG) effects; and fine-tuning. We also note that a similar assumption was made in the three-site PS model \cite{Bordone:2017bld} which was also proposed to explain the anomalies via a low scale PS breaking, where the Higgs doublets from $(15,2,2)$ and $(1,2,2)$ were assumed to give rise to one set of light Higgs doublets.

Assuming that the above conditions are met, one may 
invert the unitary transformations in Eq.\ref{Higgs}, and hence express each of the personal Higgs doublets in terms of the light EW doublets $H_u$, $H_d$, 
\begin{align}
\label{Ht30}
{H}_{t}(1,\bar{2},1/2)&\equiv \alpha_u H_u(1,\bar{2},1/2)+\ldots
\\
\label{Hb30}
{H}_b(1,\bar{2},-1/2)&\equiv   \alpha_d H_d(1,\bar{2},-1/2)+\ldots
\\
\label{Htau30}
{H}_{\tau}(1,\bar{2},-1/2) &\equiv  \gamma_d H_d(1,\bar{2},-1/2)+\ldots
 \\
\label{Hnutau30}
{H}_{\nu_{\tau}}(1,\bar{2},1/2)&\equiv  \gamma_u H_u(1,\bar{2},1/2)+\ldots
\\
\label{Hc30}
{H}_{c}(1,\bar{2},1/2)&\equiv \beta_u H_u(1,\bar{2},1/2)+\ldots
\\
\label{Hs30}
{H}_s(1,\bar{2},-1/2)&\equiv \beta_d H_d(1,\bar{2},-1/2)+\ldots
\\
\label{Hmu30}
{H}_{\mu}(1,\bar{2},-1/2) &\equiv \delta_d H_d(1,\bar{2},-1/2)+\ldots
\\
{H}_{\nu_{\mu}}(1,\bar{2},1/2)& \equiv \delta_u H_u(1,\bar{2},1/2)+\ldots
\label{Hnumu30}
\\
{h}_{u}(1,\bar{2},1/2)& \equiv \varepsilon_u H_u(1,\bar{2},1/2)+\ldots
\label{Hu30}
\\
{h}_{d}(1,\bar{2},-1/2) &\equiv \varepsilon_d H_d(1,\bar{2},-1/2)+\ldots
\label{Hd30}
\end{align}
ignoring the heavy states indicated by dots.

When the light Higgs $H_u$, $H_d$ gain their VEVs in Eq.\ref{vuvd}, the personal Higgs in the original basis can be thought of as gaining VEVs
$\langle H_t\rangle =\alpha_u v_u$, etc..
This approach will be used in the next section, when constructing the low energy quark and lepton mass matrices.

\subsection{Scalar Leptoquarks and $R_{D^{(*)}}$}
\label{scalarleptoquarks}

We have seen that the scalar fields ${H}$ and $\overline{H}$ of the twin PS model decompose into various states 
as shown in Eqs.\ref{Ht}-\ref{Hnumu}. Amongst these scalars are 
scalar EW doublet leptoquarks, identified as the well studied types 
$S_2(3,{2},7/6)$ and $\tilde{S}_2(3,{2},1/6)$ (4 copies of each), which could play a role in $R_{K^{(*)}}$ \cite{Hiller:2017bzc}. 
In this subsection we briefly investigate if $S_2$ could accommodate the $R_{D^{(*)}}$ anomaly as suggested in \cite{Sakaki:2013bfa}. 

We also note that the 4321 model~\cite{DiLuzio:2017vat} predicts the scalar leptoquark $S_1(3,{1},2/3)$,
arising from the symmetry breaking scalars $\Omega_3$ and $\Omega_1$ (which correspond to $\phi_3$ and $\phi_1$ in the twin PS model). 
The $S_1$ could also accommodate the $R_{D^{(*)}}$ anomaly if sufficiently large couplings to the relevant quarks and leptons can be generated \cite{Sakaki:2013bfa}. However this does not appear to be possible in the 4321 model \cite{DiLuzio:2018zxy}.
Note that the 4321 model does not involve the scalar fields ${H}$, $\overline{H}$ so leptoquarks
$S_2$, $\tilde{S}_2$ are not predicted.

\begin{figure}[ht]
\centering
	\includegraphics[scale=0.11]{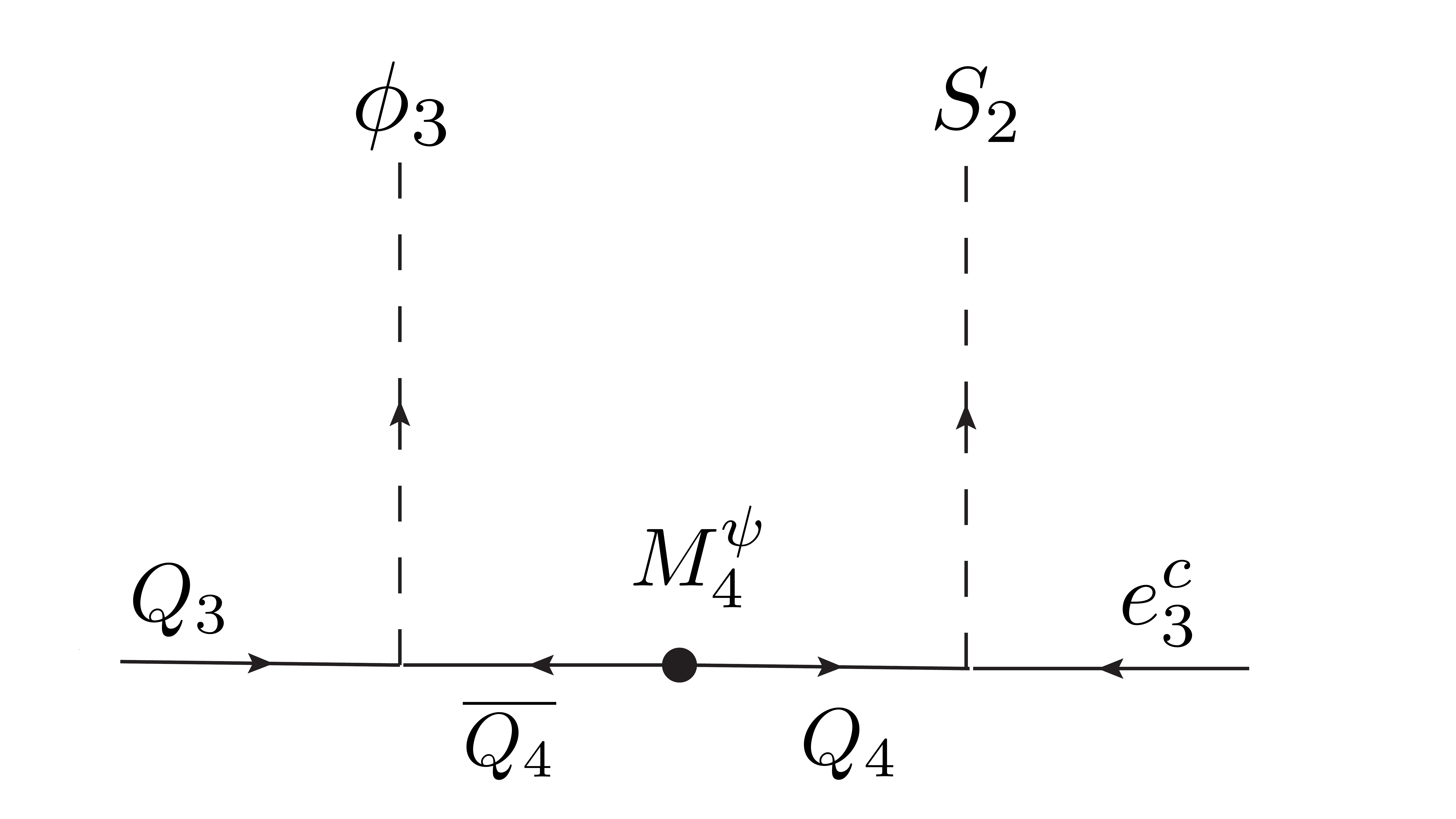}
\hspace*{1ex}
	\includegraphics[scale=0.11]{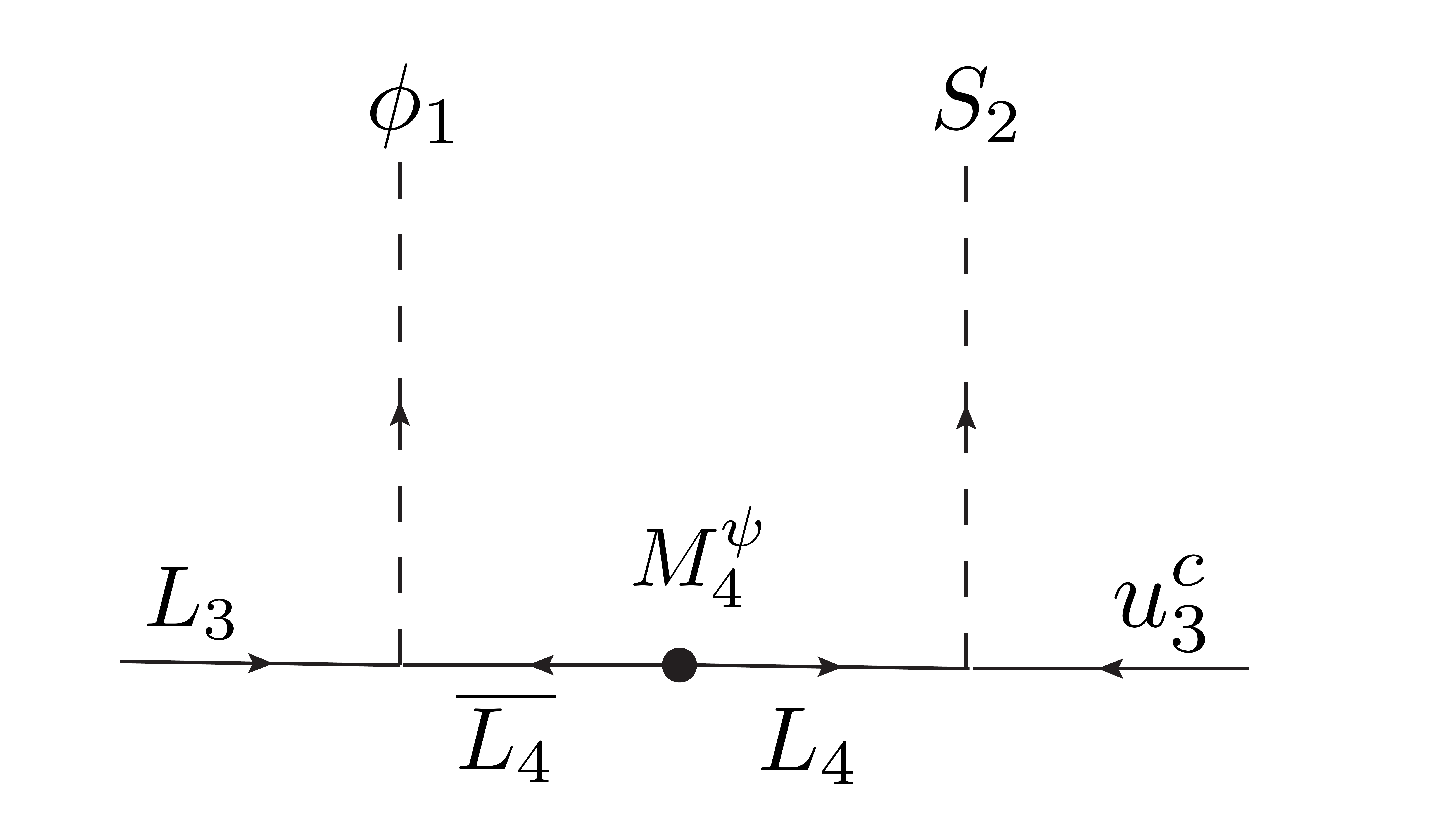}
\caption{Diagrams which lead to the $S_2$ leptoquark effective couplings relevant for $R_{D^{(*)}}$.}
\label{Fig6}
\end{figure}

Although a detailed discussion of the phenomenology of the scalar leptoquarks predicted by the twin PS model are beyond the scope of the present paper, here we briefly discuss if the scalar leptoquark $S_2$ predicted by the twin PS model could contribute 
significantly to the $R_{D^{(*)}}$ anomaly.
The dominant couplings of the leptoquark $S_2$ are due to the diagrams in Fig.\ref{Fig6} which arise from decomposition of the left panel of Fig.\ref{Fig1} which is also responsible for the third family Yukawa couplings.

The two diagrams in Fig.\ref{Fig6} lead to effective operators (up to an irrelevant minus sign),
after integrating out VL fermions, of similar structure to the third family Yukawa couplings,
\be
{\cal L}^{S_2}_{4eff}= \frac{x^{\psi}_{34} y^{\psi}_{43} {\langle \phi_3 \rangle }}{M^{\psi}_{4}} S_2 Q_3 {e^c_3} + 
\frac{x^{\psi}_{34} y^{\psi}_{43} {\langle \phi_1 \rangle }}{M^{\psi}_{4}} S_2 L_3 {u^c_3}
 \approx y_t S_2 Q_3 {e^c_3} + y_{\tau}S_2 L_3 {u^c_3}
\label{Yuk_mass_insertion6}
\ee
plus H.c., where the effective couplings are approximately equal to those of the top quark and tau lepton, both of which are expected to be of order unity
\footnote{The mixing angle formalism in Appendix~\ref{A} is required for an accurate treatment, however the result that the leptoquark couplings are approximately those of the top quark and tau lepton remains valid.}.
Converting to more familiar LR notation, the effective couplings in Eq.\ref{Yuk_mass_insertion6} may be written as,
\be
{\cal L}^{S_2}_{4eff}
 \approx S_2(y_t \overline{Q}_3 i\sigma_2 {\tau_R} + y_{\tau}\theta^{u_R}_{23}\overline{c}_R L_{3L}) + {\rm H.c.}
\label{S2}
\ee
where we have inserted a mixing angle of $\theta^{u_R}_{23}$ arising from the up type quark mass matrix
in going from $u^c_3$ to $c_R$.
According to \cite{Sakaki:2013bfa}, $S_2$ could accommodate the $R_{D^{(*)}}$ anomaly if the product of the couplings in Eq.\ref{S2} is imaginary and somewhat larger than unity in magnitude, neither of which seems possible in this model, where we expect the mixing angle to be very small,
$\theta^{u_R}_{23}\sim m_c/m_t$, according to the up type quark mass matrix in the next section.
We conclude that it seems unlikely that $S_2$ could contribute significantly to the $R_{D^{(*)}}$ anomaly in the twin PS model.

\section{Quark and Lepton Masses and Mixings}
In this section we summarise and discuss the quark and lepton masses and mixings, including the neutrino sector and the seesaw mechanism.
\label{masses}
\subsection{Dirac mass matrices}
The fermion mass matrices from Eqs.\ref{Mu},\ref{Md},\ref{Me},\ref{Mnu} may be written as,
\begin{align}
M_{u}&=A m^0_t+ B m^0_c+C m^0_u+D \bar{m}^0_u
\label{Mu2}\\
M_{d}&=A m^0_b + B m^0_s +C m^0_d+D \bar{m}^0_d
\label{Md2}\\
M_{e}&=A m^0_{\tau}+ B m^0_{\mu} + C m^0_{e}+D \bar{m}^0_e
\label{Me2}\\
M_{\nu}^D&=A m^D_{\nu_{\tau}} + B m^D_{\nu_{\mu}} +C m^D_{\nu_e} +D \bar{m}^D_{\nu_e}
\label{Mnu2}
\end{align}
where $A,B,C,D$ are universal (same for $u,d,e,\nu$) dimensionless matrices,
\begin{align}
\label{ABC}
A  &=  
\left(
\begin{array}{ccc}
0 &0 &0 \\
0 & 0 & 0 \\ 
0 & 0 & x^{\psi}_{34} y^{\psi}_{43}
\end{array}
\right) , \ \ 
B  =  
\left(
\begin{array}{ccc}
0 &0 &0 \\
0 & y^{\psi}_{24} x^{\psi^c}_{42} & y^{\psi}_{24} x^{\psi^c}_{43} \\ 
0 & y^{\psi}_{34} x^{\psi^c}_{42} & y^{\psi}_{34} x^{\psi^c}_{43}
\end{array}
\right) ,  \\
C & =  
\left(
\begin{array}{ccc}
0 &0 &0 \\
x^{\psi}_{25} y^{\psi}_{51} & 0 & 0 \\ 
x^{\psi}_{35} y^{\psi}_{51} & 0 & 0
\end{array}
\right) , \ \ 
D  =  
\left(
\begin{array}{ccc}
0 & y^{\psi}_{15} x^{\psi^c}_{52}&y^{\psi}_{15} x^{\psi^c}_{53} \\
0 & 0 & 0 \\ 
0 & 0 & 0
\end{array}
\right) ,
\end{align}
while $x,y$ are complex dimensionless order unity coefficients, and we have defined,
\begin{align}
m^0_t & =\frac{\langle  {\phi}_3 \rangle }{M^{\psi}_{4}} \alpha_u v_u,\ \ 
m^0_c = \frac{\langle \overline{\phi}_3 \rangle }{M^{\psi^c}_{4}}  \beta_u v_u, \ \ 
m^0_u=  \frac{\langle  {\Phi} \rangle }{M^{\psi}_{5}}  \varepsilon_u v_u, \ \ 
 \bar{m}^0_u=  \frac{\langle  {{\Phi}} \rangle }{M^{\psi^c}_{5}}  \varepsilon_u v_u,
\label{tcu2}  \\
 m^0_b&= \frac{\langle  {\phi}_3 \rangle }{M^{\psi}_{4}} \alpha_d v_d,\ \ 
m^0_s = \frac{\langle \overline{\phi}_3 \rangle }{M^{\psi^c}_{4}}  \beta_d v_d,\ \ 
m^0_d = \frac{\langle  {\Phi} \rangle }{M^{\psi}_{5}}  \varepsilon_d v_d, \ \ 
 \bar{m}^0_d=  \frac{\langle  {{\Phi}} \rangle }{M^{\psi^c}_{5}}  \varepsilon_d v_d,
 \label{bsd2} \\
m^0_{\tau} &= \frac{\langle  {\phi}_1 \rangle }{M^{\psi}_{4}}  \gamma_d v_d, \ \ 
m^0_{\mu}= \frac{\langle \overline{\phi}_1 \rangle }{M^{\psi^c}_{4}}  \delta_d v_d, \ \ 
m^0_e = \frac{\langle  {\Phi} \rangle }{3M^{\psi}_{5}}  \varepsilon_d v_d, \ \ 
 \bar{m}^0_e=  \frac{\langle  {{\Phi}} \rangle }{3M^{\psi^c}_{5}} \varepsilon_d v_d,
\label{taumue2} \\
m^D_{\nu_{\tau}}& = \frac{\langle  {\phi}_1 \rangle }{M^{\psi}_{4}}  \gamma_u v_u, \ \ 
m^D_{\nu_{\mu}}= \frac{\langle \overline{\phi}_1 \rangle }{M^{\psi^c}_{4}} \delta_u v_u, \ \ 
m^D_{\nu_e} = \frac{\langle  {\Phi} \rangle }{3M^{\psi}_{5}}   \varepsilon_u v_u, \ \ 
 \bar{m}^D_{\nu_e}=  \frac{\langle  {{\Phi}} \rangle }{3M^{\psi^c}_{5}} \varepsilon_u v_u,
\label{nu2}
\end{align}
where we have expressed the personal Higgs fields in terms of the light Higgs doublets
using Eqs.\ref{Ht30}-\ref{Hd30}, with VEVs in Eq.\ref{vuvd}, 
and taken the fifth family lepton masses to be three times larger than the fifth family quark masses, according to Eq.\ref{fifthmasssplitting}. 
Since we have assumed the hierarchy in Eq.\ref{hierarchy2}, it is natural to assume that each term in Eqs.\ref{tcu2},\ref{bsd2},\ref{taumue2} roughly corresponds to a particular charged fermion mass of the second and third family, as the notation suggests (the neutrinos will be discussed separately),
with each fermion mass controlled by its own personal Higgs as discussed below Eqs.\ref{Mu}-\ref{Mnu}. 
However, unlike private Higgs models~\cite{Porto:2007ed,Porto:2008hb,BenTov:2012cx,Rodejohann:2019izm}, the fermion mass hierarchies are controlled by the heavy fourth and fifth family messenger masses, rather than requiring a hierarchy of Higgs VEVs, which do not need to be very small, as discussed below. Eq.\ref{nu2} refers to the Dirac neutrino masses, where the Dirac neutrino mass matrix in Eq.\ref{Mnu2}
enters the type I seesaw mechanism and will be discussed in the following subsection.

By comparing Eqs.\ref{tcu2},\ref{bsd2},\ref{taumue2} to Eqs.\ref{tcu},\ref{bsd},\ref{taumue},
 a number of requirements emerge to achieve a correct description of the charged fermion masses of the second and third families:
\begin{itemize}
\item The dominant VEV is $\langle {H}_t \rangle =  \alpha_u v_u \sim v = 175$ GeV for the correct top mass
\item Also the large top mass requires $ \langle  {\phi}_3 \rangle \sim M^{\psi}_{4} $ 
\item $m_b/m_t\sim  \langle  {H}_b\rangle /\langle  {H}_t \rangle 
\sim  \lambda^{2.5}$ implies $\langle {H}_b \rangle = \alpha_d v_d \sim  \lambda^{2.5} v\sim  5$ 
GeV 
\item $m_s/m_c\sim  \langle  {H}_s\rangle /\langle  {H}_c \rangle =  ( \beta_d v_d)/ (\beta_u v_u) \sim  \lambda^{1.7}\sim 1/13$ 
\item $m_s/m_{\mu}\sim \frac{ \langle \overline{\phi}_3 \rangle \langle  {H}_s\rangle }{ \langle \overline{\phi}_1 \rangle \langle  {H}_{\mu} \rangle} 
\sim  1$ 
\end{itemize}
We conclude that all second and third family masses can be accommodated with the above conditions satisfied.
As claimed, the personal Higgs VEVs here are not very small and could be around 1-10 GeV, apart from that associated with the top quark whose VEV is approximately that of the SM Higgs doublet, recalling that we have absorbed the factor of 
$\sqrt{2}$ into the VEVs according to $v=v_{SM}/\sqrt{2}$ and $v_{SM}=246$ GeV.

Approximate forms of Eqs.\ref{Mu2},\ref{Md2},\ref{Me2},\ref{Mnu2} can also be useful for analytic estimates as follows,
\begin{align}
\label{Mud}
M_u & \sim  
\left(
\begin{array}{ccc}
0 &  \bar{m}^0_u &  \bar{m}^0_u \\
m^0_u & m_c & m_c \\ 
m^0_u & m_c & m_t
\end{array}
\right) , \ \ \ \ 
M_d \sim 
\left(
\begin{array}{ccc}
0 &  \bar{m}^0_d &  \bar{m}^0_d \\
m^0_d & m_s & m_s \\ 
m^0_d & m_s & m_b
\end{array}
\right) \\
\label{Menu}
M_{\nu}^{D} & \sim 
\left(
\begin{array}{ccc}
0 &  \bar{m}^D_{\nu_e} &  \bar{m}^D_{\nu_e} \\
m^D_{\nu_e} & m^D_{\nu_{\mu}} & m^D_{\nu_{\mu}} \\ 
m^D_{\nu_e} & m^D_{\nu_{\mu}} & m^D_{\nu_{\tau}} \end{array}
\right), \ \ \ \ 
M_e  \sim 
\left(
\begin{array}{ccc}
0 &  \bar{m}^0_e &  \bar{m}^0_e \\
m^0_e & m_{\mu} & m_{\mu} \\ 
m^0_e & m_{\mu} & m_{\tau}
\end{array}
\right)
\end{align}
assuming $m_f^0\sim m_f$ for the second and third family charged fermions
and dropping the dimensionless coefficients.
If $M^{\psi}_{5}\sim M^{\psi^c}_{5}$, then 
the matrices are approximately symmetric, up to order unity dimensionless coefficients $x,y$ which we have dropped here, hence,
\be
m^0_u\sim \bar{m}^0_u \sim \sqrt{m_u m_c},\ \ 
m^0_d\sim \bar{m}^0_d \sim \sqrt{m_d m_s} ,\ \ 
m^0_e\sim \bar{m}^0_e \sim \sqrt{m_e m_{\mu}}  \ \ 
\label{ude2}
\ee
which follows from the perturbative diagonalisations $m_u\approx \bar{m}^0_u m^0_u/ m_c$, etc..
The crude approximations made in writing Eqs.\ref{Mud}, \ref{Menu} are useful in giving insight into 
the Higgs VEVs $\langle {h}_u \rangle,\langle {h}_d\rangle$ in Eq.\ref{ude2}
associated with the up and down quarks, which are related to the first family mass parameters via 
Eqs.\ref{tcu2},\ref{bsd2},\ref{taumue2}.
These Higgs VEVs
need not be very small, partly because they are associated with the geometric mean of the first and second family masses, and partly because there is additional suppression coming from the ratio of scales ${\langle  {\Phi} \rangle }/{M^{\psi}_{5}}$
and ${\langle  {\Phi} \rangle }/{M^{\psi^c}_{5}}$.
From Eq.\ref{ude2} and Eqs.\ref{tcu2},\ref{bsd2},\ref{taumue2}, we find 
\be
\sqrt{\frac{m_u m_c}{m_d m_s}}\sim \frac{\langle {h}_u \rangle}{\langle {h}_d \rangle} =  \frac{\varepsilon_u v_u}{ \varepsilon_d v_d}
\sim 2.5, \ \ 
\sqrt{\frac{m_e m_{\mu}}{m_d m_s}}\sim \sqrt{\frac{m_e}{m_d}}\sim \frac{1}{3}
\ee
using the observed masses in Eqs.\ref{tcu},\ref{bsd},\ref{taumue}, and assuming 
$m_s\sim m_{\mu}$ which implies $m_e/m_d\sim 1/9$ consistent with Eq.\ref{taumue}.
We conclude that all first family masses can all be accommodated.

We may estimate the angles involved in diagonalising the quark mass matrices in Eq.\ref{Mud},
\begin{align}
\label{dmix}
\theta^d_{12}&\sim  \sqrt{\frac{m_d}{m_s}} \sim \lambda,\ \ 
\theta^d_{23}\sim \frac{m_s}{m_b}\sim \lambda^{2.5}, \ \ \theta^d_{13}\sim \frac{\sqrt{m_d m_s}}{m_b}\sim \lambda^{3.5}, 
 \\
 \label{umix}
 \theta^u_{12}&\sim  \sqrt{\frac{m_u}{m_c}} \sim \lambda^{2.1},\ \ 
\theta^u_{23}\sim \frac{m_c}{m_t}\sim \lambda^{3.3}, \ \ \theta^u_{13}\sim \frac{\sqrt{m_u m_c}}{m_t}\sim \lambda^{5.3}, 
\end{align}
assuming Eq.\ref{ude2} and 
using the observed masses in Eqs.\ref{tcu},\ref{bsd},\ref{taumue}. 
We see that the up type mixing angles are smaller than the down type mixing angles in this model, leading to 
\be
V_{us}\sim  \sqrt{\frac{m_d}{m_s}} \sim \lambda, \ \  V_{cb}\sim \frac{m_s}{m_b}\sim \lambda^{2.5}, 
\ \ V_{ub}\sim \frac{\sqrt{m_d m_s}}{m_b}\sim \lambda^{3.5}, 
\label{CKM2}
\ee
which may be compared to Eq.\ref{CKM} and includes the successful quark relation in Eq.\ref{Vus}.
These relations are encouraging, given that we have ignored the
order unity coefficients and made the symmetric approximation. We conclude that all charged fermion masses and quark mixing angles can be accommodated in the region of parameter space where there is an approximate universal texture zero in the first element of the mass matrices.

Returning to the question of the global fits of $R_{K^{(*)}}$ and $R_{D^{(*)}}$, discussed below Eq.\ref{U1op2},
we can see that the natural expectation for the mixing parameters from Eqs.\ref{dmix}, \ref{umix}, \ref{CKM2} is
\begin{align}
\beta_{s\tau}\sim \theta^d_{23}\sim  |V_{cb}|,\ \  
\beta_{b\mu}\sim  \theta^e_{23}, \ \  
\beta_{s\mu}\sim  \theta^d_{23}\theta^e_{23} \sim |V_{cb}| \theta^e_{23}
\label{betas}
\end{align}
Compared with the requirements from the global fits to the B physics anomalies \cite{Buttazzo:2017ixm},
$\beta_{s\tau}\approx 4|V_{cb}|$, with $\beta_{b\mu} < 0.5$ and $\beta_{s\mu}<  5|V_{cb}|$, it seems that the values of mixing 
in Eq.\ref{betas} are somewhat smaller than required, especially since we might expect that $\theta^e_{23}\sim m_{\mu}/m_{\tau}\sim  |V_{cb}|$.
Of course the mass matrices are not predicted to be symmetric in the present model, so it is certainly possible to choose the dimensionless coefficients in Eq.\ref{ABC} so as to enhance these parameters. However this would then imply that the up type quark mass matrix plays a significant role in the CKM mixing, which goes against the natural predictions of the model, and more generally violates the GST relation.

\subsection{Neutrino masses and mixing}
In the type I seesaw mechanism for neutrino masses~\cite{Minkowski:1977sc,Mohapatra:1979ia,Yanagida:1979as,GellMann:1980vs},
we need to consider both the Dirac mass matrix $M_{\nu}^{D}$
and the heavy Majorana mass matrix $M_{\nu}^{M}$.
We may write Dirac mass matrix $M_{\nu}^{D}$ in Eq.\ref{Mnu2} in a simplified notation as,
\begin{align}
M_{\nu}^{D} 
\equiv
\left(
\begin{array}{ccc}
0 & a & a'\\
e & b & b' \\ 
f & c & c'
\end{array}
\right)
\sim 
\left(
\begin{array}{ccc}
0 &  \bar{m}^D_{\nu_e} &  \bar{m}^D_{\nu_e} \\
m^D_{\nu_e} & m^D_{\nu_{\mu}} & m^D_{\nu_{\mu}} \\ 
m^D_{\nu_e} & m^D_{\nu_{\mu}} & m^D_{\nu_{\tau}} \end{array}
\right),
\label{MD}
\end{align}
The heavy Majorana mass matrix, follows from Eq.\ref{M},
\begin{align}
M_{\nu}^{M} \sim
\left(
\begin{array}{ccc}
 \tilde{\xi}^2 & \tilde{\xi}^5 & \tilde{\xi}^4 \\
 \tilde{\xi}^5 & \tilde{\xi}^2 &  \tilde{\xi} \\ 
 \tilde{\xi}^4 &  \tilde{\xi} & 1
\end{array}
\right)
\frac{\langle {H}' \rangle  \langle {H}' \rangle }{\Lambda}
\sim
\left(
\begin{array}{ccc}
 M^M_{1}&  0  & 0 \\
0 &  M^M_{2} &  0\\ 
0 &  0 &  M^M_{3}
\end{array}
\right)
\label{MM}
\end{align}
where we have written $ \tilde{\xi}= \langle \tilde{\xi}\rangle \ll 1$ and dropped the small off-diagonal elements with,
\be
M^M_{1}\sim M^M_{2}\sim \tilde{\xi}^2  M^M_{3}, \ \ \ \ \ \ M^M_{3}\sim \frac{\langle {H}' \rangle  \langle {H}' \rangle }{\Lambda}
\label{Mij}
\ee
Note that $M^M_{1}$ and $M^M_{2}$ are not expected to be degenerate due to the dimensionless coefficients multiplying 
each element of Eq.~\ref{MM} which we have dropped.
We shall first give a short qualitative discussion of the neutrino mass and mixing, and the scales involved, before constructing the physical neutrino mass matrix using the seesaw formula.

We assume that the first right-handed neutrino $\nu^c_1$ dominates the seesaw mechanism, as in single right-handed neutrino dominance (SRHND)~\cite{King:1998jw,King:1999cm}. Although the second right-handed neutrino mass has a similar scale, we shall assume that it is several times larger than the first, which is not unreasonable given that the higher dimensional operators in Eq.\ref{MM} may result from
a product of several Yukawa couplings, each of which may differ by a small factor.
Ignoring the other right-handed neutrinos,
then, we have just a single right-handed neutrino $\nu^c_1$ with couplings given by the first column of the Dirac mass matrix in Eq.\ref{MD},
where there is a texture zero, and the second and third elements having similar entries due to $L_2$ and $L_3$
being indistinguishable under the $Z_6$ symmetry.
Thus the dominant right-handed neutrino couples as $\nu^c_1(\nu_{\mu}+\nu_{\tau})$,
with similar couplings to $\nu_{\mu}$ and $\nu_{\tau}$, and a zero coupling to $\nu_e$ due to the texture zero,
naturally leading to large atmospheric neutrino mixing. After the single right-handed neutrino $\nu^c_1$ is integrated out 
(i.e. applying the seesaw mechanism) there is only one massive 
neutrino $\nu_3\sim \nu_{\mu}+\nu_{\tau}$ with light Majorana mass, 
\be
m_{\nu_3}\sim 2 \frac{(m^D_{\nu_e})^2}{M^M_{1}}\sim 0.05 \ {\mathrm eV}
\label{m3}
\ee
while $\nu_e$ and the orthogonal linear combination $\nu_3\sim \nu_{\mu}- \nu_{\tau}$ remain massless.
This scheme will therefore predict a normal mass hierarchy when the other smaller neutrino masses are included.
The lightest right-handed neutrino mass may be estimated by assuming $m^D_{\nu_e}\sim  \sqrt{m_u m_c}/3$,
motivated by the up quark matrix in the previous subsection, hence
\be
M^M_{1}\sim 3\times 10^7 \ {\mathrm GeV}
\label{1}
\ee

The condition for the
heaviest right-handed neutrino to decouple from the seesaw mechanism is
\be
\frac{(m^D_{\nu_{\tau}})^2}{M^M_3}\ll m_3 \sim 0.05 \ {\mathrm eV}\rightarrow M^M_3\sim M^M  \gg  6\times 10^{14} \ {\mathrm GeV}
\label{2}
\ee
assuming that $m^D_{\nu_{\tau}}\sim m_t$, as motivated in the previous subsection.
The high value of $M^M_3 \sim M^M$ in Eq.\ref{2} suggests from Eq.\ref{Mij} that the VEV $\langle H'\rangle $ should be close to the conventional scale of Grand Unified Theories (GUTs), $M_{GUT}$,
which sets the high symmetry breaking scale of the twin PS theory $M_{high}$ in Eq.\ref{breaking}.
A set of possible scales is,
\be
M^M_3\sim M^M \sim 3\times 10^{15}\ {\mathrm GeV}, \ \ 
 \langle H'\rangle  \sim M_{GUT}\sim 3\times 10^{16}\ {\mathrm GeV}, \ \ 
\Lambda \sim 3\times 10^{17}\ {\mathrm GeV}
\label{scales}
\ee
This leads to a characteristic spectrum of right-handed neutrino masses in which the lightest right-handed neutrino
has a mass 
from Eq.\ref{1} of about 30 PeV, the second one being several times heavier,
while the heaviest right-handed neutrinos $\nu^c_3$ has masses from Eq.\ref{scales}
an order of magnitude below the GUT scale.
The extreme hierarchy of right-handed neutrino masses, of order $10^{-8}$,
fixes $\tilde{\xi} \sim 10^{-4}$,
from Eqs.\ref{Mij}, \ref{1} and \ref{scales}.
Note that such a pattern of right-handed neutrino masses is typical of models based on
family symmetry and Pati-Salam~\cite{King:2003rf,King:2014iia}. Leptogenesis in this model will be highly non-standard and deserves a separate study.

The light physical effective Majorana neutrino mass matrix follows from the type I seesaw formula~\cite{Minkowski:1977sc,Mohapatra:1979ia,Yanagida:1979as,GellMann:1980vs},
\be
m_{\nu}=M_{\nu}^{D} (M_{\nu}^{M})^{-1}(M_{\nu}^{D})^T
\label{I}
\ee
In the SRHND approximation, the low energy neutrino mass matrix takes the form,
\be
m_{\nu}\approx 
\left(
\begin{array}{ccc}
 0 &  0  & 0 \\
0 & e^2  &  ef\\ 
0 &  ef &  f^2
\end{array}
\right)\frac{1}{M^M_{1}} 
\label{mnu}
\ee
with a vanishing sub-determinant and hence only one non-zero eigenvalue and a large atmospheric 
neutrino mixing angle $\theta_{23}$
~\cite{King:1998jw,King:1999cm,King:1999mb,King:2002nf,King:2002qh},
\be
m_{\nu_3}\sim \frac{e^2+f^2}{M^M_{11}}, \ \ \tan \theta_{23}\sim \frac{e}{f} \sim \frac{x^{\psi}_{25}}{x^{\psi}_{35}} \sim 1
\label{m3results}
\ee
where atmospheric neutrino mixing is expected to be large since it is given by a ratio of dimensionless coefficients of order unity.

The subdominant contribution to the seesaw mechanism comes from the second right-handed neutrino which has a similar mass to the lightest right-handed neutrino, and couples to the second column of the Dirac mass matrix in Eq.\ref{MD}.
Including also the contribution from the third right-handed neutrino, the seesaw formula Eq.\ref{I} including all three right-handed neutrinos with Eqs.\ref{MD},\ref{MM} leads
to the neutrino mass matrix, 
\be
m_{\nu}\approx 
\left(
\begin{array}{ccc}
 0 &  0  & 0 \\
0 & e^2  &  ef\\ 
0 &  ef &  f^2
\end{array}
\right)\frac{1}{M^M_{1}} 
+\left(
\begin{array}{ccc}
 a^2 &  ab  & ac \\
ab & b^2  &  bc\\ 
ac &  bc &  c^2
\end{array}
\right)\frac{1}{M^M_{2}} 
+\left(
\begin{array}{ccc}
0 &  0 & 0 \\
0 & 0  &  0\\ 
0 &  0 &  c'^2
\end{array}
\right)\frac{1}{M^M_{3}} 
\label{mnu2}
\ee
where each of the three matrices is responsible for a particular neutrino mass, yielding a normal ordered mass pattern described by 
Eq.\ref{m3results} plus the additional sequential dominance (SD) results
~\cite{King:1998jw,King:1999cm,King:1999mb,King:2002nf,King:2002qh},
\be
m_{\nu_1}\sim \frac{c'^2}{M^M_{3}} , \ \  m_{\nu_2}  \sim \frac{a^2}{M^M_2s_{12}^2} ,  \ \ 
\tan \theta_{12} \sim 
\frac{\sqrt{2}a}
{b-c}, \ \ 
\theta_{13} \lesssim \frac{m_{\nu_2}}{m_{\nu_3}}
\label{m2results}
\ee
To achieve the observed solar mixing in Eq.\ref{PMNS} we need $a\sim (b-c)/2$, where from Eqs.\ref{MD}, \ref{nu2}
and the previous assumptions,
\be
a\sim \bar{m}^D_{\nu_e}\sim m^D_{\nu_e}\sim \frac{ \sqrt{m_u m_c}}{3}, \ \ \ \ 
b\sim c\sim m^D_{\nu_{\mu}}\sim \frac{ \langle \overline{\phi}_1 \rangle\delta_u }{ \langle \overline{\phi}_3 \rangle\beta_u } m_c
\ee
which suggests that we need the pre-factor $ \frac{ \langle \overline{\phi}_1 \rangle\delta_u }{ \langle \overline{\phi}_3 \rangle\beta_u }\ll 1 $.
The partial cancellation between $b$ and $c$ in 
Eq.\ref{m2results} can also help to achieve the desired value of $\tan \theta_{12}$.

\section{Conclusions}
\label{conclusion}

The main motivation for the present work was find a realistic model with the correct ingredients for 
explaining the $B$ anomalies, as well as providing a theory quark and lepton (including neutrino) masses and mixings. 
Indeed the two endeavours have a natural synergy, since on the one hand theories which only attempt to explain the quark and lepton masses and mixings are far from unique and cannot be readily tested, while on the other hand theories which only attempt to explain the $B$ anomalies, although testable, inevitably involve input parameters which depend on the unknown quark and lepton mass matrices. 
The anomalies provide a stimulus for novel model building approaches to the flavour problem,
while upgrading the low energy phenomenological models of $B$ physics anomalies to include a realistic explanation of the quark and lepton masses provides welcome constraints on the input parameters.
Therefore searching for a realistic model of quark and lepton masses and mixings, with the correct ingredients to explain the $B$ anomalies, in an all-encompassing theory of flavour seems to be very well justified. 

In this paper we have proposed a twin PS theory of flavour broken to the $G_{4321}$ gauge group at high energies, then to the Standard Model at low energies, as in Fig.~\ref{model} and Eq.\ref{breaking}. 
The motivation for a theory of this particular kind was to 
yield a TeV scale vector leptoquark $U^{\mu}_1(3,1,2/3)$ which enables the $R_{K^{(*)}}$ and $R_{D^{(*)}}$ anomalies in $B$ decays to be addressed simultaneously, where the couplings of such a vector leptoquark could be predicted by the same theory which also explains the quark and lepton masses and mixings. In the present model we found that the twin PS theory of flavour successfully accounts for all quark and lepton (including neutrino) masses and mixings, and predicts a dominant coupling of 
$U^{\mu}_1(3,1,2/3)$ to the third family left-handed doublets, which generates flavour changing due to CKM-like mixing.
However the predicted mass matrices are not consistent with the single vector leptoquark
solution to the $B$ anomalies, given the current value of $R_{D^{(*)}}$.

It is worth emphasising that the predicted mass matrices satisfy rather generic conditions found in many models of quark and lepton masses, for example they involve a texture zero in the first entry of the mass matrices, and most of the CKM mixing comes from the down quark sector, where both features are consistent with the phenomenologically successful GST relation. This reinforces the view that the single vector leptoquark combined explanation of the $R_{K^{(*)}}$ and $R_{D^{(*)}}$ anomalies in $B$ decays, which involves regions of parameter space where the (2,3) mixings required greatly exceed $|V_{cb}|$, are not well motivated from the point of view of more general flavour models, not just the considered model. 

The twin PS theory of flavour, as an ultraviolet completion of the low energy 4321 theories, addresses 
the question of the origin of quark and lepton masses and mixings, and predicts a much richer
low energy spectrum, beyond the heavy gauge bosons, including many extra scalars and fermions. 
Therefore the twin PS theory here and the low energy 4321 models are easily distinguishable experimentally.
However the precise predictions will depend on whether the personal Higgs fields are retained or replaced by the 2HDM and the associated fields.

Although the personal Higgs doublets for the second and third families are suggested by the twin PS structure,
they come with the challenges of Higgs mixing and alignment, which depend on the Higgs potential which we have not considered in this paper.
In the low energy theory with the personal Higgs, there is a rich spectrum of scalar fields including 10 Higgs EW doublets, 4 colour octet scalar EW doublets, and 8 scalar EW doublets identified as leptoquarks $S_2$ and $\tilde{S}_2$. We have shown that the leptoquarks $S_2$ do not contribute significantly to $R_{D^{(*)}}$.

In Appendix~\ref{2HDM}
we considered replacing the personal Higgs model by a type II 2HDM where the Higgs potential is well studied.
In such a 2HDM version of the model, a plethora of scalar EW singlets and triplets are predicted, including colour octets and 
additional leptoquarks $S_3$. 
It would be interesting, in a future publication, to study in detail the phenomenology of the scalars in either version of the model, in particular the scalar leptoquarks, which could also contribute to the $B$ anomalies along with the vector leptoquark. It would also be interesting
to study the lightest VL fermion doublets and singlets with TeV scale masses accessible to colliders in a simplified model framework.

In conclusion, we have proposed a twin PS theory of flavour with a $Z_6$ family symmetry,
capable of describing the quark and lepton masses and mixing, while addressing the $B$ physics anomalies. 
It is also possible to consider twin PS models based on other discrete or continuous  
Abelian or non-Abelian family symmetries. The general approach is
to generate fermion masses by the same mixing with the VL fermions as 
that which controls the effective vector leptoquark couplings to quarks and leptons, providing a predictive framework.
In the present model, the single vector leptoquark approach to the $R_{K^{(*)}}$ and $R_{D^{(*)}}$ anomalies in $B$ decays, constrained by the mixing parameters from the predicted quark and lepton mass matrices, assuming
natural values of the parameters, cannot easily satisfy the global fits, given the current value of $R_{D^{(*)}}$.

\subsection*{Acknowledgements}
The author acknowledges the STFC Consolidated Grant ST/L000296/1 and the European Union's Horizon 2020 Research and Innovation programme under Marie Sk\l{}odowska-Curie grant agreement HIDDeN European ITN project (H2020-MSCA-ITN-2019//860881-HIDDeN).

\appendix

\section{Mixing angle formalism}
\label{A}

Since the top quark Yukawa coupling is order unity, strictly speaking we need to return to the full mass matrix in Eq.\ref{M^psi_an_1}.
For present purposes (i.e. extracting the quark and lepton mass matrices) it is not necessary to diagonalise the full mass matrix in 
Eq.\ref{M^psi_an_1}. It is sufficient to remove the largest off-diagonal elements, namely the $\phi$ and $\overline{\phi}$ terms whose VEVs are much larger than the Higgs $H,\overline{H}$ VEVs. After this is done, 
the remaining transformations required to block diagonalise the mass matrix, so that only the upper $3\times 3$ block is off-diagonal,
will only involve small angles of order $v/M^{\psi}_4$, 
or less, where $v$ is the SM Higgs VEV, which we ignore here.

The large off-diagonal $\phi$ terms in Eq.\ref{M^psi_an_1}
may be removed by the following large angle 
transformation~\cite{King:2017anf,King:2018fcg}, 
\be
\pmatr{
\psi'_3\\
\psi'_4
}
=
\pmatr{
c^{\psi}_{34}&s^{\psi}_{34}\\
-s^{\psi}_{34}&c^{\psi}_{34}
}
\pmatr{
\psi_3\\
\psi_4
}
\label{34matrix}
\ee
This large angle transformation is an important step towards diagonalising the matrix in Eq.\ref{M^psi_an_1}, replacing
the off-diagonal $\phi$ term by a zero, where the fields with primes are in the original basis~\cite{King:2021iah}. 
Such large mixing will not induce any flavour violation in the SM $W$ and $Z$ couplings since
$\psi_3$ and $\psi_4$ will have the same quantum numbers when decomposed under the SM gauge group (see later).

Beyond the mass insertion approximation, 
the couplings in the first matrix in Eq.\ref{Yuk_mass_insertion_1} should then be replaced by the above large mixing angle as follows,
\be
\frac{x^{\psi}_{34} {\phi} }{M^{\psi}_{4}}  
\rightarrow s^{\psi}_{34}=\frac{x^{\psi}_{34}\phi}{\sqrt{(x^{\psi}_{34}\phi)^2+(M^{\psi}_4)^2}}
\label{largeangle}
\ee
where $s^{\psi}_{34}=\sin \theta^{\psi}_{34}$, $c^{\psi}_{34}=\cos \theta^{\psi}_{34}$.
Similarly, we can remove 
the large (compared to the Higgs VEVs)
off-diagonal $\overline{\phi}$ terms in Eq.\ref{M^psi_an_1}, replacing them by zeros by the following approximate
transformations,\cite{King:2017anf,King:2018fcg}, 
\be
\pmatr{
\psi^{c'}_1\\
\psi^{c'}_2\\
\psi^{c'}_3\\
\psi^{c'}_4
}
=
\pmatr{
1&0&0&0\\
0&1&0&\theta^{\psi^c}_{42}\\
0&0&1&\theta^{\psi^c}_{43}\\
0&-\theta^{\psi^c}_{42}&-\theta^{\psi^c}_{43}&1
}
\pmatr{
\psi^{c}_1\\
\psi^{c}_2\\
\psi^{c}_3\\
\psi^{c}_4
}
\ee
where,
\be
\theta^{\psi^c}_{42}\approx \frac{x^{\psi^c}_{42}\overline{\phi} }{M^{\psi^c}_{4}}, \ \  \theta^{\psi^c}_{43}\approx \frac{x^{\psi^c}_{43}\overline{\phi} }{M^{\psi^c}_{4}}
\ee
which are just the combinations of couplings which appear in the second matrix in Eq.\ref{Yuk_mass_insertion_1}.
Thus the small angle approximation is equivalent to the mass insertion approximation in this case, being valid for the second family 
Yukawa couplings 
due to the hierarchy in Eq.\ref{hierarchy}.
Hence for the second family, and first family discussed below,
we may continue to use the mass insertion approximation.
Even for the third family, we shall continue to use the mass insertion approximation in the main body of the paper, since it has a simple
diagrammatic interpretation, bearing in mind that we can readily use the more exact results here if required using the replacement in 
Eq.~\ref{largeangle}.

\section{From Personal Higgs to the 2HDM}
\label{2HDM}
In this appendix we show that the personal Higgs model of the main body of the paper can be recast 
as a conventional type II 2HDM \cite{Branco:2011iw} involving only the Higgs $h$ already introduced Table~\ref{twinPS}, together with extra scalars and fermions.
This avoids the possible FCNCs due to having multiple Higgs doublets, and hence the discussion about the Higgs basis in Section~\ref{EWSB} can be avoided.

\begin{figure}[ht]
\centering
	\includegraphics[scale=0.11]{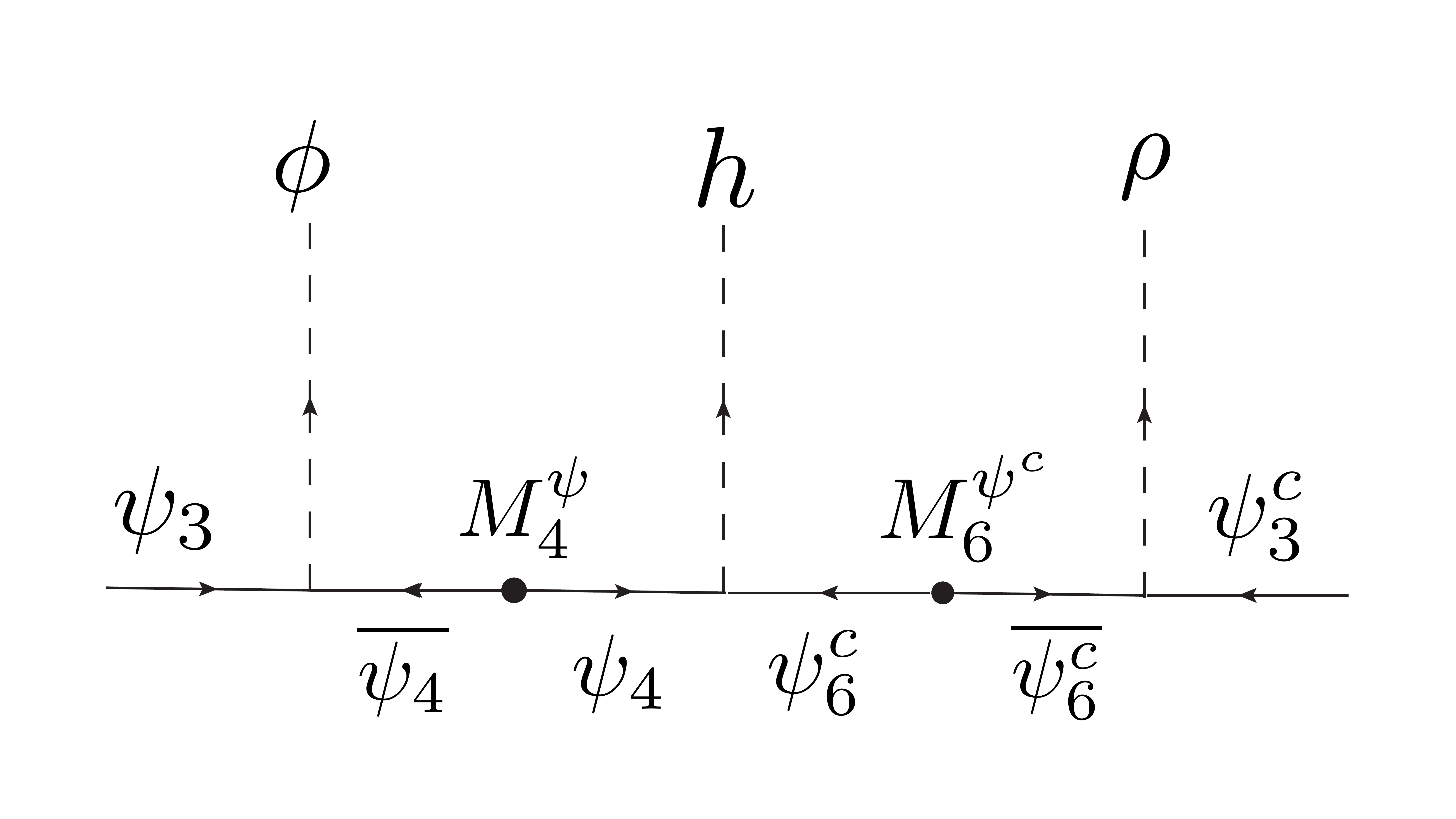}
\hspace*{1ex}
	\includegraphics[scale=0.11]{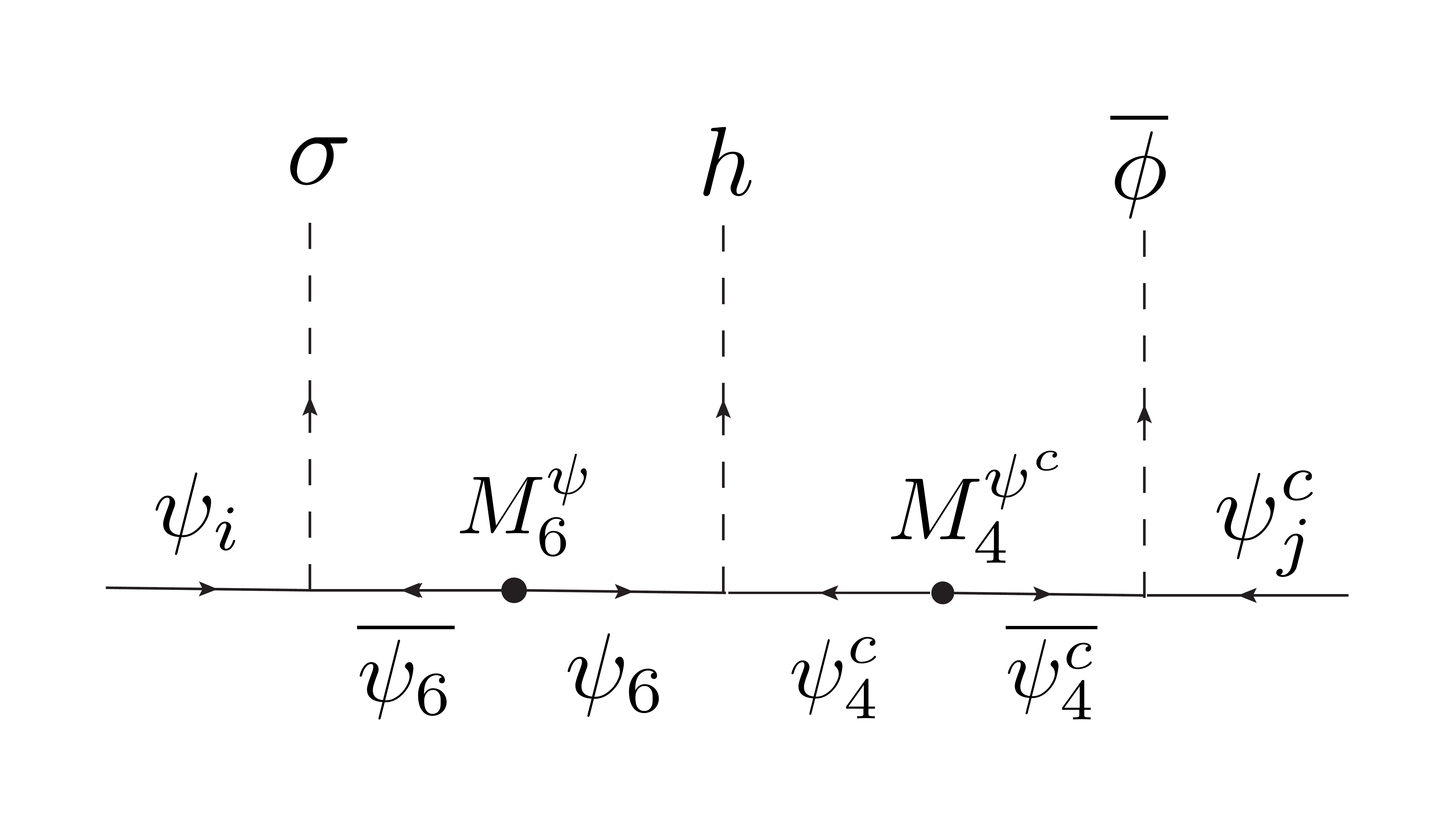}
\caption{Diagrams in the 2HDM version which lead to the effective Yukawa couplings of the third family (left panel) and second family (right panel) where $i,j=2,3$ are the only non-zero values. These arise from Fig.~\ref{Fig1}, with $H$ replaced by $(h\otimes \rho)$, and 
$\overline{H}$ replaced by $(h\otimes \sigma) $.}
\label{Fig5}
\end{figure}

In order to do this, the $H$ and $\overline{H}$ fields are removed from Table~\ref{twinPS} and replaced by two new scalar fields
$\rho$ and $\sigma$, which transform under $G_{422}^I \times G_{422}^{II} \times Z_6$ as,
\be
\rho(\overline{4},1,1;4,1,1)_{\alpha^2}, \ \ \sigma(4,2,2;\overline{4},\overline{2},\overline{2})_{\alpha^2}
\label{rhosigma}
\ee
together with new VL fermions which transform as,
\begin{align}
\psi_6(4, 2, 2; 1,1,\overline{2})_{\alpha^2} , \ \  \overline{\psi_6}(\overline{4},\overline{2},\overline{2};1,1,2)_{\alpha^4},
\label{sixth}\\
\psi^c_6(\overline{4}, 1, 1; 1,1,\overline{2})_{\alpha^2} , \ \  \overline{\psi^c_6}(4,1,1;1,1,2)_{\alpha^4}
\label{sixthc}
\end{align}
The new diagrams responsible for the third and second family fermion masses are then shown 
in Fig.~\ref{Fig5}, which replace those in Fig.~\ref{Fig1}. 

By comparing Fig.~\ref{Fig5} to Fig.~\ref{Fig1}, it is apparent that the effect of the original $H,\overline{H}$ fields is 
reproduced by combining the quantum numbers of $\rho,\sigma$ with the original $h$ field,
\begin{align}
H(\overline{4},\overline{2},1;4,1,2)_1\equiv h(1, \overline{2}, 1; 1,1,2)_{\alpha^4}\otimes  \tilde{\rho}(\overline{4},1,1;4,1,1)_{\alpha^2}, 
\label{rhoh}
\\  \overline{H}(4,1,2;\overline{4},\overline{2},1)_1\equiv 
h(1, \overline{2}, 1; 1,1,2)_{\alpha^4} \otimes \tilde{\sigma}(4,2,2;\overline{4},\overline{2},\overline{2})_{\alpha^2}
\label{hsigma}
\end{align}
where $\tilde{\rho}=\rho/M^{\psi}_6$ and $\tilde{\sigma}=\sigma/M^{\psi^c}_6$ are the fields in Eq.\ref{rhosigma}, scaled by the masses 
$M^{\psi}_6$ and $M^{\psi^c}_6$ of the heavy VL fermions in Eqs.\ref{sixth} and \ref{sixthc} which mediate Fig.~\ref{Fig5}.

Now that the Higgs fields $H,\overline{H}$ are no longer present, being replaced by 
combinations of fields in Eqs.\ref{rhoh},\ref{hsigma}, the only Higgs doublets required are those contained in $h$,
which are the same Higgs fields responsible for the first family masses, since we
retain the same mechanism for first family masses as before, as in Fig.~\ref{Fig2}.
Fermions of a given charge 
receive mass from the same Higgs doublet, either $h_u$ or $h_d$, as in the type II 2HDM.

To see how this works in detail, we must consider the decomposition of the fields under the various symmetry breakings,
as follows, noting that the extra fields which we have introduced to replace $H,\overline{H}$ in Eqs.\ref{rhoh},\ref{hsigma},\ref{sixth},\ref{sixthc}  are summarised in Table~\ref{twinPS2}, and their decompositions under
$G_{4422}$ and $G_{4321}$ are shown in Tables~\ref{tab:funfields3} and \ref{tab:funfields4}.

(i) Under $G_{4422}$ the Higgs equivalences in Eqs.\ref{rhoh},\ref{hsigma} become,
\begin{align}
H(\overline{4},4,\overline{2},2)_1 & \equiv h(1,1, \overline{2},2)_{\alpha^4} \otimes  \tilde{\rho}(\overline{4},4,1,1)_{\alpha^2}, 
\label{rhoh2}
\\  \overline{H}(4,\overline{4},\overline{2},2)_1 & \equiv 
h(1,1, \overline{2},2)_{\alpha^4} \otimes  \tilde{\sigma}(4,\overline{4},1,1)_{\alpha^2}
\label{hsigma2}
\end{align}
where in the $\sigma$ decomposition we have shown only the singlet parts of $SU(2)_L^{I+II}$ and $SU(2)_R^{I+II}$ for simplicity, bearing in mind that the triplet parts of $\sigma$ can also appear and give rise to $SU(2)_R^{I+II}$ splitting effects after the VEVs appear in the triplet components.

(ii) Under $G_{4321}$ the personal Higgs in Eqs.\ref{Ht}-\ref{Hnumu} have equivalences given from the decompositions of 
Eqs.\ref{rhoh2},\ref{hsigma2}, dropping the $Z_6$ assignments for simplicity,
\begin{align}
\label{Ht2}
{H}_{t}(\bar{4},{3},\bar{2},2/3) &\equiv  h_u(1, 1,\overline{2}, 1/2) \otimes \tilde{\rho}_q(\overline{4},3,1,1/6), 
\\
\label{Hb2}
{H}_b(\bar{4},{3},\bar{2},-1/3) &\equiv  h_d(1, 1,\overline{2}, -1/2) \otimes  \tilde{\rho}_q(\overline{4},3,1,1/6), 
\\
\label{Htau2}
{H}_{\tau}(\bar{4},1,\bar{2},-1) &\equiv h_d(1, 1,\overline{2}, -1/2) \otimes  \tilde{\rho}_l(\overline{4},1,1,-1/2), 
 \\
\label{Hnutau2}
{H}_{\nu_{\tau}}(\bar{4},1,\bar{2},0) &\equiv h_u(1, 1,\overline{2}, 1/2) \otimes \tilde{\rho}_l(\overline{4},1,1,-1/2) , 
\\
\label{Hc2}
 {H}_{c}(4,\bar{3},\bar{2},1/3) &\equiv h_u(1, 1,\overline{2}, 1/2) \otimes  \tilde{\sigma}_q(4,\overline{3},1,-1/6),
\\
\label{Hs2}
{H}_s(4,\bar{3},\bar{2},-2/3) &\equiv h_d(1, 1,\overline{2}, -1/2) \otimes  \tilde{\sigma}_q(4,\overline{3},1,-1/6),
\\
\label{Hmu2}
{H}_{\mu}(4,1,\bar{2},0) &\equiv h_d(1, 1,\overline{2}, -1/2) \otimes  \tilde{\sigma}_l(4,1,1,1/2),
 \\
\label{Hnumu2}
 {H}_{\nu_{\mu}}(4,1,\bar{2},1) &\equiv h_u(1, 1,\overline{2}, 1/2) \otimes  \tilde{\sigma}_l(4,1,1,1/2)
\end{align}

(iii) Under the breaking $G_{4321}\rightarrow G_{321}$ 
to the SM gauge group the $\rho,\sigma$ scalar fields above decompose as,
\begin{align}
\rho_q(\overline{4},3,1,{1}/{6})\rightarrow & \rho_q(1,1,0)+ \rho_q(8,1,0)+ \rho_q(3,1,{2}/{3})
\\
\rho_l(\overline{4},1,1,-{1}/{2})\rightarrow & \rho_l(1,1,0)+ \rho_l(\overline{3},1,-{2}/{3})
\\
\sigma_q(4,\overline{3},1,-{1}/{6})\rightarrow & \sigma_q(1,1,0)+\sigma_q(8,1,0)+\sigma_q(\overline{3},1,-{2}/{3})
\\
\sigma_l(4,1,1,{1}/{2})\rightarrow & \sigma_l(1,1,0)+\sigma_l({3},1,{2}/{3})
\end{align}
where each field gets a VEV in their SM singlet components.
These SM singlet VEVs reproduce the effective personal Higgs doublets once 
Eqs.\ref{Ht2}-\ref{Hnumu2} are decomposed under the symmetry breaking $G_{4321}\rightarrow G_{321}$ 
to the SM gauge group,
\begin{align}
\label{Ht3}
{H}_{t}(1,\bar{2},1/2)&\equiv  \langle \tilde{\rho}_q \rangle h_u(1,\overline{2}, 1/2), 
\\
\label{Hb3}
{H}_b(1,\bar{2},-1/2)&\equiv  \langle  \tilde{\rho}_q  \rangle h_d(1,\overline{2}, -1/2), 
\\
\label{Htau3}
{H}_{\tau}(1,\bar{2},-1/2) &\equiv   \langle  \tilde{\rho}_l  \rangle h_d(1,\overline{2}, -1/2), 
 \\
\label{Hnutau3}
{H}_{\nu_{\tau}}(1,\bar{2},1/2)&\equiv   \langle  \tilde{\rho}_l  \rangle h_u(1,\overline{2}, 1/2), 
\\
\label{Hc3}
{H}_{c}(1,\bar{2},1/2)&\equiv \langle  \tilde{\sigma}_u  \rangle h_u(1,\overline{2}, 1/2),
\\
\label{Hs3}
{H}_s(1,\bar{2},-1/2)&\equiv  \langle \tilde{\sigma}_d  \rangle h_d(1,\overline{2}, -1/2),
\\
\label{Hmu3}
{H}_{\mu}(1,\bar{2},-1/2) &\equiv \langle  \tilde{\sigma}_e \rangle h_d(1,\overline{2}, -1/2) ,
\\
\label{Hnumu3}
{H}_{\nu_{\mu}}(1,\bar{2},1/2)& \equiv \langle \tilde{\sigma}_{\nu}\rangle  h_u(1,\overline{2}, 1/2)
\end{align}
where we have written the subscripts ${u,d}$ and  ${e,\nu}$ on $\tilde{\sigma}$
to remind us that 
the $SU(2)_R^{I+II}$ triplet parts of $\sigma$ can also appear and give rise to splitting effects between the $c,s$ quark masses
and $\mu,\nu_{\mu}$ lepton masses. 

Since the VEVs of the $\rho$ and $\sigma$ fields break the $SU(4)^I_{PS}$, this means that they must have low scale values,
which in turn implies that at least some of the sixth family of VL fermions, in particular the EW singlets $\psi^c_6$ associated with the third family 
fermions in the left-hand panel of Fig.~\ref{Fig5}, along with the EW doublets $\psi_4$,
must have masses around the TeV scale. We note that the combination of the EW doublets from $\psi_4$ and the EW singlets from $\psi^c_6$, might resemble a complete VL family of fermions near the TeV scale, but in this model they would originate from different VL families, with different couplings to quarks and leptons. The prediction of such VL fermions near the TeV scale is a crucial prediction of this model and deserves a dedicated phenomenological study, along the lines of the simplified model framework of \cite{King:2021iah}.

Comparing Eqs.\ref{Ht3}-\ref{Hnumu3} to Eqs.\ref{Ht30}-\ref{Hnumu30}, we identify $h_u\equiv H_u$, $h_d\equiv H_d$, 
and the coefficients,
\begin{align}
\alpha_u &= \alpha_d \equiv \langle \tilde{\rho}_q \rangle, \ \ 
\gamma_u = \gamma_d \equiv \langle \tilde{\rho}_l \rangle, \label{r1}\\
\beta_u &\equiv \langle \tilde{\sigma}_u \rangle, \ \ 
\beta_d \equiv \langle \tilde{\sigma}_d \rangle, \ \ 
\delta_u \equiv \langle\tilde{\sigma}_{\nu} \rangle,  \label{r2}\\
\varepsilon_u&=\varepsilon_d\equiv 1  \label{r3}
\end{align}
The interpretation of the coefficients is quite different however: they are no longer elements of a unitary matrix, instead they represent scaled fields whose singlet components get VEVs, apart from $\varepsilon_u$ and $\varepsilon_d$ which are simply set equal to unity
once we identify $h_u\equiv H_u$, $h_d\equiv H_d$ as the light Higgs doublets.

The discussion of the quark and lepton masses and mixings then follows
that given below Eqs.\ref{tcu2},\ref{bsd2},\ref{taumue2}, with the identifications in Eqs.\ref{r1},\ref{r2},\ref{r3}, so we do not need to repeat it.
The relation $\alpha_u = \alpha_d $ implies that $m_t/m_b = v_u/v_d =\tan \beta$ which implies large $\tan \beta$.
Otherwise the discussion is the same as given previously, including the neutrino masses and mixing.

We emphasise that since the SM fermions of a given charge couple to the same Higgs doublet,
there is natural flavour conservation, as in the type II 2HDM, without any FCNCs from the Higgs doublet sector.
The key observation is that the personal Higgs doublets involved in the second and third family masses are replaced in
Eqs.\ref{Ht3}-\ref{Hnumu3} by the same two Higgs doublets, namely $h_u$ and $h_d$, involved in the first family masses.

In addition the $SU(2)_L$ triplet  $\sigma$ scalar fields have similar decompositions,
\begin{align}
\label{sigmaq}
\sigma_q(4,\overline{3},3,-{1}/{6})\rightarrow & \sigma_q(1,3,0)+\sigma_q(8,3,0)+\sigma_q(\overline{3},3,-2/3)
\\
\label{sigmal}
\sigma_l(4,1,3,{1}/{2})\rightarrow & \sigma_l(1,3,0)+\sigma_l({3},3,{2}/{3})
\end{align}
The $\sigma'_q$ and $\sigma'_l$ scalar fields have the same decompositions as the $\sigma_q$ and $\sigma_l$ scalar fields 
in Eqs.\ref{sigmaq},\ref{sigmal},
but with the additional hypercharges $\Delta Y$,
\begin{align}
\label{sigmapq}
\sigma'_q(4,\overline{3},3,-{1}/{6}+\Delta Y)\rightarrow & \sigma_q(1,3,\Delta Y)+\sigma_q(8,3,\Delta Y)+\sigma_q(\overline{3},3,-2/3+\Delta Y)
\\
\label{sigmpl}
\sigma'_l(4,1,3,{1}/{2}+\Delta Y)\rightarrow & \sigma_l(1,3,+\Delta Y)+\sigma_l({3},3,{2}/{3}+\Delta Y)
\end{align}
where $\Delta Y =(1,0,-1)$ corresponds to the $SU(2)_R$ triplet, plus similar decompositions for the $SU(2)_L$ singlets. 
There is clearly a rich spectrum of $\rho,\sigma$ scalar fields, which, like the personal Higgs, can also lead to FCNCs.
However, unlike the personal Higgs fields, these scalars are associated with larger VEVs at least an order of magnitude larger than the EW scale, therefore we naturally expect these scalars to have masses in the multi-TeV region. Indeed, they can lead to an interesting flavour changing phenomenology, for example 
the $\sigma'_q(\overline{3},3,1/3)$ scalar leptoquark, in Eq.\ref{sigmapq} with $\Delta Y =1$,
identified as $S_3(\overline{3},3,1/3)$ \cite{Hiller:2017bzc,deMedeirosVarzielas:2018bcy,deMedeirosVarzielas:2019okf},
could contribute a left-handed operator of the correct form for $R_{K^{(*)}}$, without violating the bounds on $B_s$ mixing or 
$\tau\rightarrow \mu \mu \mu$. However the phenomenology of such scalar leptoquarks is beyond the scope of the present paper.

\begin{table}
\centering
\begin{tabular}{| l | c  c c | c c c| c|}
\hline
Field & $SU(4)^I_{PS}$ & $SU(2)^{I}_L$ & $SU(2)^{I}_R$ &  $SU(4)^{II}_{PS}$ & $SU(2)^{II}_L$ & $SU(2)^{II}_R$ & $Z_6$\\ 
\hline
\hline
 $\rho$   & ${\overline{\bf 4}}$ & ${{\bf 1}}$ &  ${\bf 1}$ &  ${\bf 4}$   &  ${\bf 1}$  &  ${\bf 1}$ & ${\alpha^2}$ \\
$\sigma$   & ${\bf 4}$ &  ${\bf 2}$  & ${\bf 2}$    & ${\overline{\bf 4}}$  &  ${\overline{\bf 2}}$ & ${\overline{\bf 2}}$ & ${\alpha^2}$ \\
\hline
$\psi_{6}$ 		& ${\bf 4}$ & ${\bf 2}$ & ${\bf 2}$ & ${\bf 1}$ & ${\bf 1}$ & ${\overline{\bf 2}}$ & ${\alpha^2}$ \\
$\overline{\psi_{6}}$ 		& ${\overline{\bf 4}}$   & ${\overline{\bf 2}}$ & ${\overline{\bf 2}}$ & ${\bf 1}$ & ${\bf 1}$ & ${\bf 2}$  & ${\alpha^4}$\\
$\psi^c_{6}$ 		 & ${\overline{\bf 4}}$ & ${\bf 1}$  & ${\bf 1}$ & ${\bf 1}$ & ${\bf 1}$ & ${\overline{\bf 2}}$ & ${\alpha^2}$\\
$\overline{\psi^c_{6}}$ 		& ${\bf 4}$  & ${\bf 1}$  & ${\bf 1}$ & ${\bf 1}$ & ${\bf 1}$ & ${\bf 2}$ & ${\alpha^4}$\\
\hline
\hline
\end{tabular}
\caption{The additional fields which replace the personal Higgs $H,\overline{H}$ in Table~\ref{twinPS}.
All the other fields are required above the double lines in Table~\ref{twinPS}.}
\label{twinPS2}
\end{table}

\begin{table}
\centering
\begin{tabular}{| l | c c c c | c |}
\hline
Field & $SU(4)^I_{PS}$ & $SU(4)^{II}_{PS}$ & $SU(2)^{I+II}_L$ & $SU(2)^{I+II}_R$ &  $Z_6$\\ 
\hline
\hline
 $\rho$ & ${\overline{\bf 4}}$  &  ${\bf 4}$ &  ${\bf 1}$  &  ${\bf 1}$  & ${\alpha^2}$ \\
$\sigma$    & ${\bf 4}$ & ${\overline{\bf 4}}$  &  ${\bf 1+3}$  & ${\bf 1+3}$  & ${\alpha^2}$\\
\hline
$\psi_{6}$ 		& ${\bf 4}$ & ${\bf 1}$ & ${\bf 2}$ & ${\bf 1+3}$ & ${\alpha^2}$ \\
$\overline{\psi_{6}}$ 		& ${\overline{\bf 4}}$ & ${\bf 1}$   & ${\overline{\bf 2}}$ & ${\bf 1+3}$ & ${\alpha^4}$\\
$\psi^c_{6}$ 		 & ${\overline{\bf 4}}$ & ${\bf 1}$ & ${\bf 1}$ & ${\overline{\bf 2}}$& ${\alpha^2}$\\
$\overline{\psi^c_{6}}$ 		& ${\bf 4}$  & ${\bf 1}$ & ${\bf 1}$ & ${\bf 2}$ & ${\alpha^6}$ \\
\hline
\hline
\end{tabular}
\caption{The extra fields in Table~\ref{twinPS2} decompose under $G_{4422}$ as shown.
}
\label{tab:funfields3}
\end{table}

\begin{table}
\vspace{-0.5in}
\centering
\begin{tabular}{| l | c c c c | c |}
\hline
Field & $SU(4)_{PS}^I$ & $SU(3)_{c}^{II}$ & $SU(2)^{I+II}_L$ & $U(1)_{Y'}$ & $Z_6$ \\ 
\hline
\hline
$\rho_q$    & $\overline{\bf 4}$ & ${{\bf 3}}$  & ${\bf 1}$  & $\frac{1}{6}$ & ${\alpha^2}$ \\
$\rho_l$    & $\overline{\bf 4}$ & ${\bf 1}$   & ${\bf 1}$  & $-\frac{1}{2}$ & ${\alpha^2}$ \\
$\sigma_q$    & ${\bf 4}$ & ${\overline{\bf 3}}$  &${\bf 1+3}$  & $-\frac{1}{6}$ & ${\alpha^2}$ \\
$\sigma'_q$    & ${\bf 4}$ & ${\overline{\bf 3}}$  & ${\bf 1+3}$  & $(\frac{5}{6},-\frac{1}{6},-\frac{7}{6})$ & ${\alpha^2}$ \\
$\sigma_l$    & ${\bf 4}$ & ${\bf 1}$   & ${\bf 1+3}$  & $\frac{1}{2}$ & ${\alpha^2}$ \\
$\sigma'_l$    & ${\bf 4}$ & ${\bf 1}$   & ${\bf 1+3}$  & $(\frac{3}{2},\frac{1}{2},-\frac{1}{2})$ & ${\alpha^2}$ \\
\hline
$\psi_{6}$ 		& ${\bf 4}$ & ${\bf 1}$ & ${\bf 2}$ & $0$& ${\alpha^2}$ \\
$\overline{\psi}_{6}$ 		& ${\overline{\bf 4}}$ & ${\bf 1}$   & ${\overline{\bf 2}}$ & $0$& ${\alpha^4}$\\
$\psi'_{6}$ 		& ${\bf 4}$ & ${\bf 1}$ & ${\bf 2}$ & $(1,0,-1)$& ${\alpha^2}$ \\
$\overline{\psi}'_{6}$ 		& ${\overline{\bf 4}}$ & ${\bf 1}$   & ${\overline{\bf 2}}$ & $(1,0,-1)$& ${\alpha^4}$\\
$\psi^c_{6u\nu}$ 		 & ${\overline{\bf 4}}$ & ${\bf 1}$ & ${\bf 1}$ & $-\frac{1}{2}$& ${\alpha^2}$\\
$\psi^c_{6de}$ 		 & ${\overline{\bf 4}}$ & ${\bf 1}$ & ${\bf 1}$ & $\frac{1}{2}$& ${\alpha^2}$\\
$\overline{\psi^c}_{6u\nu}$ 		& ${\bf 4}$  & ${\bf 1}$ & ${\bf 1}$ & $\frac{1}{2}$  & ${\alpha^4}$\\
$\overline{\psi^c}_{6ed}$ 		& ${\bf 4}$  & ${\bf 1}$ & ${\bf 1}$ & $-\frac{1}{2}$  & ${\alpha^4}$\\
\hline
\hline
\end{tabular}
\caption{Under the subgroup $G_{4321}$, the extra fields of Tables~\ref{twinPS2},\ref{tab:funfields3} decompose as shown.
}
\label{tab:funfields4}
\end{table}

\end{document}